\documentclass[12pt]{article}

\usepackage[margin=1in]{geometry}  
\usepackage{amsmath,amssymb,amsfonts}
\usepackage{amsthm}              
\usepackage{graphicx}              
\usepackage[round]{natbib}            
\usepackage{setspace}          
\usepackage{url}                  
\usepackage{hyperref}          
\usepackage{bm}
\usepackage{xcolor}%
\usepackage{booktabs}
\usepackage{subcaption}
\usepackage{setspace}
\usepackage[english]{babel}
\usepackage[autostyle, english = american]{csquotes}
\MakeOuterQuote{"}

\newtheorem{theorem}{Theorem}

\newtheorem{proposition}{Proposition}

\theoremstyle{definition}
\newtheorem{definition}{Definition}[section]

\theoremstyle{remark}

\theoremstyle{definition}
\newtheorem{assumption}{Assumption}

\title{Modelling Skewed and Heavy-Tailed Errors in Bayesian Mediation Analysis}
\author{
  Zongyu Li \\
  \small University of Notre Dame \\
  \small \texttt{zli37@nd.edu}
  \and
  Mark Steel \\
  \small University of Warwick \\
  \small \texttt{m.steel@warwick.ac.uk}
  \and
  Zhiyong Zhang \\
  \small University of Notre Dame \\
  \small \texttt{zzhang4@nd.edu}
}
\date{\today}


\begin{document}
\doublespacing  
\maketitle

\begin{abstract}
\noindent
Traditional mediation models in both the frequentist and Bayesian frameworks typically assume normality of the error terms. Violations of this assumption can impair the estimation and hypothesis testing of the mediation effect in conventional approaches. This study addresses the non-normality issue by explicitly modelling skewed and heavy-tailed error terms within the Bayesian mediation framework. Building on the work of \citet{fernandez1998}, this study introduces a novel family of distributions, termed the \textbf{C}entred \textbf{T}wo-\textbf{P}iece Student \(\bm{t}\) Distribution (CTPT). The new distribution incorporates a skewness parameter into the Student \(t\) distribution and centres it to have a mean of zero, enabling flexible modelling of error terms in Bayesian regression and mediation analysis. A class of standard improper priors is employed, and conditions for the existence of the posterior distribution and posterior moments are established, while enabling inference on both skewness and tail parameters. Simulation studies are conducted to examine parameter recovery accuracy and statistical power in testing mediation effects. Compared to traditional Bayesian and frequentist methods, particularly bootstrap-based approaches, our method gives greater statistical power when correctly specified, while maintaining robustness against model misspecification. The application of the proposed approach is illustrated through real data analysis. Additionally, we have developed an R package \texttt{FlexBayesMed} to implement our methods in linear regression and mediation analysis, available at \url{https://github.com/Zongyu-Li/FlexBayesMed}.
\end{abstract}


\vspace{-0.5cm}
\section{Introduction}\label{sec:1}
\vspace{-0.25cm}
Mediation analysis assesses how an independent variable influences an outcome variable through one or more mediators, helping to disentangle direct and indirect effects. This method provides insights into the mechanisms underlying observed relationships and is widely applied in psychology, epidemiology, and educational sciences (\citealp{rucker2011}; \citealp{agler2017}; \citealp{santini2020}).

The basic mediation framework involves three variables: an independent variable \(X\), a mediator \(M\), and an outcome variable \(Y\). This framework is commonly represented by the following three equations (\citealp{mackinnon2007}; \citealp{mackinnon2012}):
\vspace{-0.5cm}
\begin{align}
  Y &= \beta_1 + \tau X + e_1,                 \label{eq:total}   \\
  M &= \beta_2 + \alpha X + e_2,               \label{eq:mediator}\\
  Y &= \beta_3 + \beta M + \tau' X + e_3,      \label{eq:direct}
\end{align}

\vspace{-0.75cm}
\noindent
where \(\tau\) denotes the total effect of the independent variable on the outcome variable, and \(\tau'\) denotes the direct effect after accounting for the mediator. The parameter \(\beta\) captures the mediator's effect on the outcome variable, adjusting for the independent variable, while \(\alpha\) measures the effect of the independent variable on the mediator. The error terms \(e_1\), \(e_2\), and \(e_3\) are typically assumed to be normally distributed with mean zero and finite variances \(\sigma^2_1\), \(\sigma^2_2\), and \(\sigma^2_3\), respectively (\citealp{yuan2009}; \citealp{mackinnon2012}). Additionally, $e_2$ and $e_3$ are typically assumed to be independent. 

Equations  (\ref{eq:mediator}) and (\ref{eq:direct}) allow us to disentangle the total effect $\tau$ into a direct effect $\tau'$ and a mediated (indirect) effect. Substituting (\ref{eq:mediator}) into (\ref{eq:direct}) makes clear that this decomposition is \(\tau=\tau'+\alpha\beta\). Thus, a widely used measure of the mediation effect is the product \(\alpha\beta\) (\citealp{mackinnon2012}). 
A non-zero value (\(\alpha\beta \neq 0\)) indicates mediation, while \(\alpha\beta = 0\) suggests its absence. A non-zero \(\alpha\beta\) implies that both \(\alpha\) and \(\beta\) are non-zero, indicating that the independent variable affects the mediator, which subsequently influences the outcome variable. An equivalent measure is \(\tau - \tau'\), which represents the indirect effect as the difference between the total and direct effects of \(X\) on \(Y\). We focus on \(\alpha\beta\) due to its widespread use in psychology and education.

In the frequentist framework, for the mediation model in Equations (\ref{eq:mediator}) and (\ref{eq:direct}), Ordinary Least Squares (OLS) is typically used to estimate \(\alpha\) and \(\beta\), denoted as \(\hat{\alpha}\) and \(\hat{\beta}\), which can be obtained separately from the last two equations. The mediation effect is then estimated as the product \(\hat{\alpha}\hat{\beta}\). For confidence intervals and hypothesis tests of \(H_0: \alpha\beta = 0\) against \(H_1: \alpha\beta \neq 0\), many methods rely on the asymptotic normality of \(\hat{\alpha}\hat{\beta}\) (\citealp{sobel1982}; \citealp{bollen1990}; \citealp{mackinnon2002}; \citealp{shrout2002}). These asymptotic methods perform well with large sample sizes but become unreliable in small-sample cases. 

Alternatively, assuming normally distributed error terms, the OLS estimates \(\hat{\alpha}\) and \(\hat{\beta}\) follow normal distributions. Based on the distribution of the product of two normally distributed variables (\citealp{meeker1981}), the theoretical distribution of certain test statistics can be derived under the null hypothesis \(H_0: \{\alpha = 0, \beta = 0\} \text{ or } \{\alpha \neq 0, \beta = 0\} \text{ or } \{\alpha = 0, \beta \neq 0\}\), enabling confidence interval construction and hypothesis testing for the mediation effect without relying on asymptotic results (\citealp{mackinnon2002}).

We note that many asymptotic methods also rely on the normality assumption of error terms when deriving the asymptotic variances of \(\hat{\alpha}\hat{\beta}\), as in \citet{aroian1947}. However, this assumption often fails in educational and psychological datasets, where heavy-tailed and skewed distributions are common (\citealp{micceri1989}; \citealp{cain2017}). Simulation studies show that non-normal error terms can reduce the accuracy of estimation and the power of hypothesis tests for the mediation effect when using these methods (\citealp{yuan2014}; \citealp{ng2016}; \citealp{alfons2022}).

To address non-normal error terms, researchers may apply transformations to make the error terms conform more closely to a normal distribution, such as the Box-Cox transformation (\citealp{boxcox1964}; \citealp{sakia1992}). However, these transformations often prove ineffective in practice (\citealp{osborne2002}; \citealp{fink2009}). More importantly, in fields like the social and behavioural sciences, where variables often lack predefined metrics, transforming the data is challenging and can reduce the interpretability of results.

For methods that do not rely on transformations, robust and nonparametric approaches offer viable alternatives. \citet{yuan2014} propose a robust mediation analysis method based on median regression, demonstrating that it outperforms other methods in both single-level and multilevel mediation analyses. \citet{wang2025} develop a robust regression approach for mediation analysis that minimises the Huber loss. Another popular approach to handling non-normal errors is the bootstrap (\citealp{efron1994}; \citealp{diciccio1996}). Bootstrap is a nonparametric method that avoids distributional assumptions by estimating the distributions of \(\hat{\alpha}\), \(\hat{\beta}\), or \(\hat{\alpha}\hat{\beta}\) through random sampling with replacement from the observed data. In particular, Alfons, Ateş and Groenen (2022a) combine robust statistics with bootstrap for mediation analysis, evaluating its performance under various departures from normality. Their approach uses robust MM-estimators (\citealp{yohai1987}; \citealp{salibian2006}) for \(\alpha\) and \(\beta\) instead of standard OLS estimators, and applies the fast-and-robust bootstrap method (\citealp{salibian2002}; \citealp{salibian2008}) rather than the standard bootstrap. Their results show that this robust bootstrap strategy improves the estimation and hypothesis testing of the mediation effect compared with other methods. Additionally, they provide the R package \textit{Robmed} to implement their approach (\citealp{alfons2022b}).

In the Bayesian framework, under the normality assumption of error terms, \citet{yuan2009} treat Equations (\ref{eq:mediator}) and (\ref{eq:direct}) as two separate Bayesian linear regression problems, assigning independent priors to the parameters \(\beta_2, \alpha, \sigma^2_2, \beta_3, \beta, \tau', \sigma^2_3\). The joint posterior distributions \(p(\beta_2, \alpha, \sigma^2_2 \mid X, M)\) and \(p(\beta_3, \beta, \tau', \sigma^2_3 \mid X, M, Y)\) are then obtained. Markov Chain Monte Carlo (MCMC) algorithms, such as the Gibbs sampler, generate posterior samples of size \(T\) from these distributions, denoted as \(\{\beta_2^{(i)}, \alpha^{(i)}, \sigma_2^{2\,(i)}\}_{i=1}^{T}\) and \(\{\beta_3^{(i)}, \beta^{(i)}, \tau'^{(i)}, \sigma_3^{2\,(i)}\}_{i=1}^{T}\). A point estimate of the mediation effect \(\alpha\beta\) is computed as \(\widehat{\alpha\beta} = \frac{1}{T}\sum_{i=1}^{T} \alpha^{(i)} \beta^{(i)}\), and the sample \(\{\alpha^{(i)} \beta^{(i)}\}_{i=1}^{T}\) represents the posterior distribution of the mediation effect. The \(100(1-\alpha)\%\) highest posterior density (HPD) interval can be estimated from this MCMC sample. Additionally, Bayesian statistics provides a formal framework for model comparison (hypothesis testing) using Bayes factors (\citealp{kass1995}). Methods for hypothesis testing \(H_0: \alpha\beta = 0\) against \(H_1: \alpha\beta \neq 0\) using Bayes factors are also available  (\citealp{liu2023,liu2024}) and are discussed here in Section~\ref{sec:4}.

Current Bayesian approaches generally perform inference and hypothesis testing under the assumption of normally distributed error terms, which makes them unsuitable when the error terms deviate from normality, for instance, when they are skewed and/or heavy-tailed. Although various bootstrap methods exist in the frequentist framework, they are essentially nonparametric and may lack power for testing the mediation effect. Therefore, there is a need for methods that can flexibly model skewed and/or heavy-tailed error terms in mediation analysis. Correctly specifying the model is likely to improve the accuracy of mediation effect estimation and increase the power of hypothesis testing. The Bayesian framework provides a suitable approach for this, allowing error terms to follow distributions that account for skewness and heavy tails.

To address this issue, in Section~\ref{sec:2}, we propose the Centred Two-Piece Student \(t\) Distribution (CTPT), 
which incorporates a skewness parameter into the Student \(t\) distribution and centres it to achieve a zero mean, enabling flexible modelling of error terms in Bayesian mediation analysis. In Section~\ref{sec:3}, we develop a theory for a single regression model, providing conditions for the existence of the posterior distribution and its moments, using standard improper priors to enable inference on skewness and tail parameters. In Section~\ref{sec:4}, we describe the implementation of the proposed distribution in Bayesian mediation analysis. In Section~\ref{sec:5}, we conduct parameter recovery studies to examine model identifiability and perform several simulation studies to evaluate statistical power in testing the mediation effect, comparing our method with traditional Bayesian mediation methods and bootstrap approaches. A case study in Section~\ref{sec:6} illustrates the practical application of the proposed method.

\vspace{-0.5cm}
\section{The Centred Two-Piece Student \(t\) Distribution} \label{sec:2}
\vspace{-0.25cm}
From this point forward, we focus on a reparameterisation of the mediation model. Consider a dataset of size \(n\), where the independent variable is denoted by \(X_i\) (\(i=1,2,\dots,n\)),
the mediator by \(M_i \in \mathbb{R}\), and the outcome by \(Y_i \in \mathbb{R}\). The model is expressed as:
\vspace{-0.5cm}
\begin{align}
    M_i &= \beta_0^{(M)} + \alpha X_i + \sigma^{(M)}\varepsilon^{(M)}_i,  \label{eq:my_med_1}\\
    Y_i &= \beta_0^{(Y)} + \beta M_i + \tau X_i + \sigma^{(Y)}\varepsilon^{(Y)}_i, \label{eq:my_med_2}
\end{align}

\vspace{-0.75cm}
\noindent
where the error terms \(\{\varepsilon^{(M)}_i\}_{i=1}^n\) and \(\{\varepsilon^{(Y)}_i\}_{i=1}^n\) satisfy the following assumptions:

\begin{assumption}[Independent and Identically Distributed]
  \label{assump:A1}
  \(\{\varepsilon^{(M)}_i\}_{i=1}^n\) are independent and identically distributed (i.i.d.) according to a distribution class parameterised by \(\boldsymbol{\theta}^{(M)}\). Similarly, \(\{\varepsilon^{(Y)}_i\}_{i=1}^n\) are i.i.d. according to a distribution class parameterised by \(\boldsymbol{\theta}^{(Y)}\). Furthermore, \(\{\varepsilon^{(M)}_i\}_{i=1}^n\) and \(\{\varepsilon^{(Y)}_i\}_{i=1}^n\) are independent of each other.
\end{assumption}

\begin{assumption}[Zero Means]
  \label{assump:A2}
  \(\mathbb{E}[\varepsilon^{(M)}_i|\boldsymbol{\theta}^{(M)}]=0\) and \(\mathbb{E}[\varepsilon^{(Y)}_i|\boldsymbol{\theta}^{(Y)}]=0\).
\end{assumption}

\begin{assumption}[Constant and Finite Variances]
  \label{assump:A3}
  \(\text{Var}(\varepsilon^{(M)}_i|\boldsymbol{\theta}^{(M)})< \infty\) and \(\text{Var}(\varepsilon^{(Y)}_i|\boldsymbol{\theta}^{(Y)})  < \infty\). 
\end{assumption}

\vspace{-0.5cm}
We relax the normality assumption of the error terms. 
Our objective is to capture potential skewness and heavy tails in the error distribution, which requires specifying a distribution class that accommodates these characteristics. 

There are several classes of univariate distributions that can model skewed and/or heavy-tailed error terms, such as the Student \(t\) distribution (\citealp{zhang2016}), Azzalini's skew-normal distribution (\citealp{azzalini1985}), and various skew-\(t\) distributions (\citealp{theodossiou1998}; \citealp{jones2003}; \citealp{li2020}). Among these, the family of two-piece distributions (\citealp{fernandez1998}; \citealp{rubio2014,rubio2020}) offers an elegant, simple and interpretable option for modelling skewed and heavy-tailed data in the Bayesian framework. \citet{fernandez1998} conduct Bayesian inference under a linear regression model using two-piece distributions for the error terms. As a special case, they introduce a skewed Student \(t\) distribution with two scalar parameters that separately control skewness and tail-heaviness. They investigate the use of standard improper priors, which can be interpreted as default choices, to enable inference on the regression coefficients as well as the skewness and tail parameters. \citet{rubio2014} explore the use of Jeffreys priors and other priors in this context. These works provide a solid foundation for applying two-piece distributions to model skewed and heavy-tailed data within the Bayesian framework, particularly in linear regression problems.

A special case of the class of two-piece distributions has the following density:
\begin{align}
    p(\varepsilon|\gamma,\nu) = \frac{2}{\gamma + \frac{1}{\gamma}} \left[ f_\nu\left(\frac{\varepsilon}{\gamma}\right) I_{[0,+\infty]}(\varepsilon) + f_\nu(\gamma \varepsilon) I_{(-\infty,0)}(\varepsilon) \right],\;\gamma,\,\nu>0,  \label{eq:tpd}
\end{align}
where \(I_A(\cdot)\) is the indicator function of a set \(A\), and \(f_\nu\) is the Student \(t\) distribution with \(\nu\) degrees of freedom, defined as:
\begin{align}
    f_\nu(x) = \frac{\Gamma\left(\frac{\nu+1}{2}\right)}{\sqrt{\pi \nu}\Gamma\left(\frac{\nu}{2}\right)} \left(1 + \frac{x^2}{\nu}\right)^{-\frac{\nu+1}{2}}. \label{eq:t}
\end{align}

Here, \(\gamma\) serves as the skewness parameter, introducing skewness to the Student \(t\) distribution by scaling the distribution on one side of zero by a factor of \(\gamma\) while scaling it on the other side by \(\frac{1}{\gamma}\), keeping the mode unchanged at zero. The parameter \(\nu\) controls tail-heaviness, with smaller values of \(\nu\) leading to heavier tails. In the limit as \(\nu \to \infty\), the distribution becomes a skew-normal distribution, which simplifies to a standard normal  when \(\gamma=1\).

However, the class of distributions given by (\ref{eq:tpd}) does not generally have a zero mean or finite variance, which violates Assumption~\ref{assump:A2} and Assumption~\ref{assump:A3}. The mean and variance of this distribution are given by the following propositions.

\begin{proposition}
    The mean of the distribution defined by (\ref{eq:tpd}) is 
    \begin{align}
        m(\gamma,\nu) := \frac{2\nu\Gamma\left(\frac{\nu+1}{2}\right)}{\sqrt{\pi \nu}(\nu-1)\Gamma\left(\frac{\nu}{2}\right)}\left(\gamma - \frac{1}{\gamma}\right), \quad \gamma \in \mathbb{R}_+, \; \nu > 1,
    \end{align}
    and as \(\nu \to \infty\) (where \(f_\nu(x)\) approaches a standard normal density), the mean approaches \(m^{\text{normal}}(\gamma) = \sqrt{\frac{2}{\pi}}\left(\gamma - \frac{1}{\gamma}\right)\). If \(0<\nu \leq 1\), the mean is undefined.
\end{proposition}

\begin{proposition}
    The variance of (\ref{eq:tpd}) is defined only if \(\nu > 2\), and is given by:
    \begin{align}
        \text{Var}(\varepsilon|\gamma,\nu) = \frac{\nu}{\nu-2}\left(\gamma^2 - 1 + \frac{1}{\gamma^2}\right) - m(\gamma,\nu)^2, \quad \gamma \in \mathbb{R}_+, \; \nu > 2,
    \end{align}
    and as \(\nu \to \infty\), the variance approaches \(\text{Var}^{\text{normal}}(\varepsilon|\gamma) = \gamma^2 - 1 + \frac{1}{\gamma^2} - m^{\text{normal}}(\gamma)^2\).
\end{proposition}

\begin{figure}[h]
    \centering
    \includegraphics[width=0.45\textwidth]{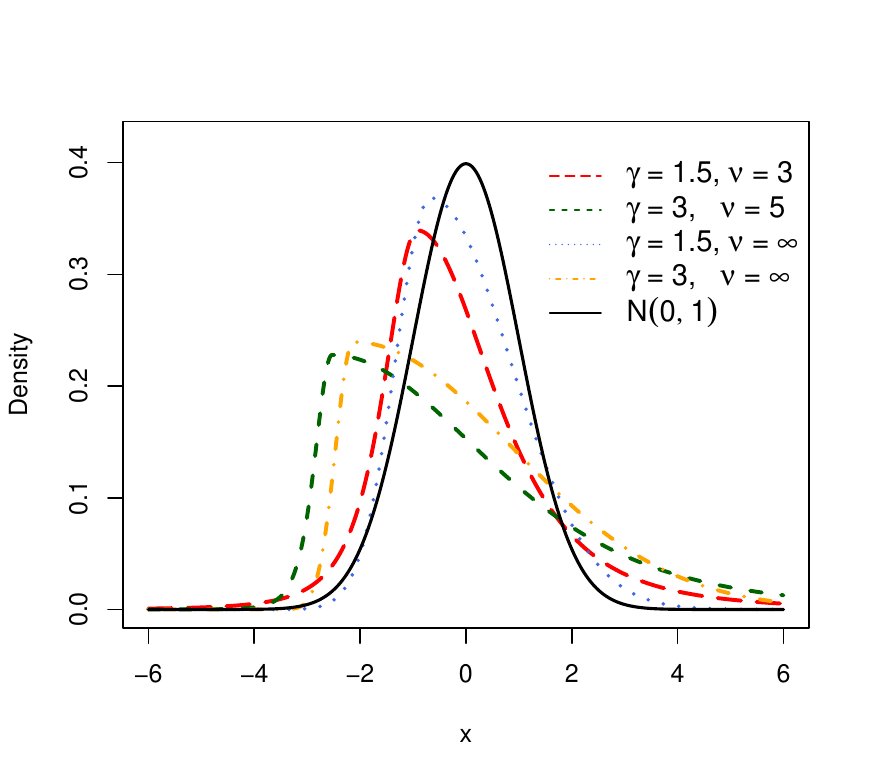}
    \includegraphics[width=0.45\textwidth]{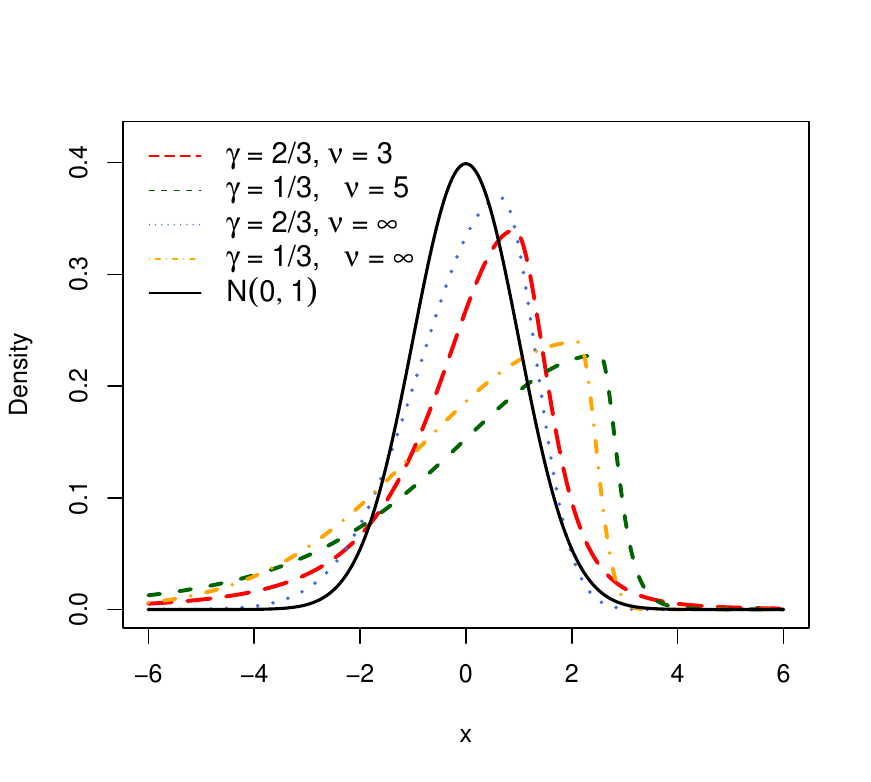}
    \caption{Density functions of the centred two-piece Student \(t\) distribution defined in (\ref{eq:pdfCPTP}).}
    \label{fig:two-side-by-side}
\end{figure}

To ensure that the error terms have zero mean and finite variance, we propose the Centred Two-Piece Student \(t\) Distribution (CTPT), denoted as \(\varepsilon\sim\text{ctpt}(\gamma,\nu)\). The distribution is characterised by the following probability density function:
\begin{equation}\label{eq:pdfCPTP}
    \begin{aligned}
        p(\varepsilon|\gamma,\nu) = \frac{2}{\gamma + \frac{1}{\gamma}} \Bigg[& f_\nu\left(\frac{\varepsilon + m(\gamma,\nu)}{\gamma}\right) I_{[-m(\gamma,\nu),+\infty)}(\varepsilon) \;+ \\
        &f_\nu\left(\gamma\big(\varepsilon + m(\gamma,\nu)\big)\right) I_{(-\infty,-m(\gamma,\nu))}(\varepsilon) \Bigg], \quad \gamma \in \mathbb{R}_+, \; \nu > 2,
    \end{aligned}
\end{equation}
where, as explained earlier, the parameter \(\gamma\) controls skewness, and the degrees of freedom \(\nu\) control the tail-heaviness on both sides of the distribution. Figure~\ref{fig:two-side-by-side} presents the density functions of CTPT under different values of \(\gamma\) and \(\nu\). In contrast to (\ref{eq:tpd}), the mean is now fixed at zero, while the mode varies. Similar distributions have been used by \cite{BauwensLaurent} in frequentist inference on multivariate financial data.

For regression modelling with the two-piece distribution defined in (\ref{eq:tpd}) and (\ref{eq:t}) with its mode at zero, it may not be a major issue if the mean of the error term is non-zero, because "modal regression" can be carried out.  
However, in the context of mediation analysis, this is awkward because most previous studies assume that the error terms have zero mean. Modelling the error terms without this zero-mean assumption can make the interpretation of the mediation effect inconsistent with previous research (see Section~\ref{sec:3.3} for a detailed discussion of the (non-)equivalence between centred and uncentred two-piece Student \(t\)-distributed errors in Bayesian regression models, and the advantages of using the centred version). This motivates our proposal of the centred two-piece Student \(t\) distribution for the error terms in regression problems.

Note that the restriction \(\nu > 2\) is not required if the assumption of finite variance is dropped. One can set \(\nu > 1\) to model heavier tails if desired (we still need \(\nu>1\) so that the mean \(m(\gamma,\nu)\) is well-defined). In our study, however, we retain the assumption of finite variance and therefore require \(\nu > 2\).

One way to characterise distributional skewness is through Fisher's moment coefficient of skewness. The following result gives the conditions for its existence and provides its closed-form expression.

\begin{proposition}
    Let \(X\sim\text{ctpt}(\gamma,\nu)\) be a random variable, and let \(r\) be a positive integer. The \(r\)-th moment of \(X\) exists if \(\nu>r\) and is given by 
    \begin{align}
        \mathbb{E}\left[X^r\right] = \sum^r_{k=0}{r\choose k} \left[-m(\gamma,\nu)\right]^{r-k}\frac{1}{\gamma+\frac{1}{\gamma}}\left[\gamma^{k+1}+(-1)^k\gamma^{-k-1}\right]\frac{\nu^{\frac{k}{2}}}{\sqrt{\pi}}\frac{\Gamma\left(\frac{k+1}{2}\right)\Gamma\left(\frac{\nu-k}{2}\right)}{\Gamma\left(\frac{\nu}{2}\right)},
    \end{align}
    and as \(\nu \to \infty\), \(\mathbb{E}\left[X^r\right]\rightarrow\sum^r_{k=0}{r\choose k} \left[-m^{\text{normal}}(\gamma)\right]^{r-k}\frac{1}{\gamma+\frac{1}{\gamma}}\left[\gamma^{k+1}+(-1)^k\gamma^{-k-1}\right]\frac{2^{\frac{k}{2}}}{\sqrt{\pi}}\Gamma\left(\frac{k+1}{2}\right)\).

    In particular, Fisher's moment coefficient of skewness exists if \(\nu >3\), and is given by:
    \begin{align}
        \text{SK}(X):=\frac{\mathbb{E}\left[X^3\right]}{\left(\mathbb{E}\left[X^2\right]\right)^{\frac{3}{2}}}=\frac{\frac{\nu^{\frac{3}{2}}}{\sqrt{\pi}}\frac{\Gamma\left(\frac{\nu-3}{2}\right)}{\Gamma\left(\frac{\nu}{2}\right)}\frac{\gamma^4-\gamma^{-4}}{\gamma+\frac{1}{\gamma}} - 3m(\gamma,\nu)\frac{\nu}{\nu-2}\left(\gamma^2-1+\frac{1}{\gamma^2}\right)+2[m(\gamma,\nu)]^3}{\left[\frac{\nu}{\nu-2}\left(\gamma^2-1+\frac{1}{\gamma^2}\right)-[m(\gamma,\nu)]^2\right]^{\frac{3}{2}}},
    \end{align}
    and as \(\nu\rightarrow \infty\), \(\text{SK}(X)\rightarrow \frac{2\sqrt{\frac{2}{\pi}}\frac{\gamma^4-\gamma^{-4}}{\gamma+\frac{1}{\gamma}}-3m^{\text{normal}}(\gamma)\left(\gamma^2-1+\frac{1}{\gamma^2}\right)+2[m^\text{normal}(\gamma)]^3}{\left[\left(\gamma^2-1+\frac{1}{\gamma^2}\right)-\left[m^{\text{normal}}(\gamma)\right]^2\right]^{\frac{3}{2}}}\).
\end{proposition}

As with the uncentred two-piece distribution obtained from Student \(t\)-sampling, the \text{ctpt}\((\gamma,\nu)\) distribution is right-skewed when \(\gamma>1\), left-skewed when \(\gamma<1\), and symmetric when \(\gamma=1\); in this case it reduces to the Student \(t\) distribution with \(\nu\) degrees of freedom. Fisher's moment coefficient of skewness, \(\text{SK}(X\mid\gamma,\nu)\), is a strictly increasing function of \(\gamma\) and satisfies \(
\text{SK}(X\mid\gamma,\nu) = -\text{SK}\!\left(X\mid \frac{1}{\gamma},\nu\right).\)

For any fixed \(\nu>3\), \(\text{SK}(X\mid\gamma,\nu)\) converges to a finite limit as \(\gamma\rightarrow\infty\), and this limit depends on \(\nu\). Figure~\ref{fig:skewness} in Supplementary Material~\ref{sm:skewness_plot} plots the coefficient as a function of \(\gamma\) for \(\nu = 5, 10\), and \(\infty\), where \(\nu = \infty\) is used, for convenience, to denote the normal base distribution. This property of Fisher's moment coefficient of skewness for the \text{ctpt}\((\gamma,\nu)\) distribution makes it difficult to compare skewness across different values of \(\nu\). A more suitable measure may be the one introduced by \citet{arnold1995}, defined as one minus twice the probability mass to the left of the mode:
\begin{align}
  \text{SK}_{\text{A\&G}}(X \mid \gamma,\nu)=\frac{\gamma^{2}-1}{\gamma^{2}+1},
\end{align}
which is independent of \(\nu\). This index is strictly increasing in \(\gamma\), equals \(0\) when \(\gamma = 1\), and takes values in \((-1,1)\). Moreover, \(\text{SK}_{\text{A\&G}}(X \mid \gamma,\nu) = -\,\text{SK}_{\text{A\&G}}\!\left(X \mid \frac{1}{\gamma},\nu\right).\)

\vspace{-0.5cm}
\section{A Single Regression Model with Skewed and Heavy-Tailed Errors}\label{sec:3}
\vspace{-0.25cm}
\subsection{Bayesian General Regression Models}\label{sec:3.1}

Although the mediation model comprises two linear regression sub-models, we begin by exploring the impact of using the centred two-piece Student \(t\) distribution for the error term within a general regression model in the Bayesian framework. Following the approach of \citet{fernandez1998}, we specify a broad class of improper priors and establish conditions under which the posterior distribution exists. We also demonstrate that these conditions are not restrictive, offering substantial flexibility for practical applications.

In the context of a general regression model, we assume that the error terms \(\varepsilon_i, i=1,2,\dots,n\) are independent and identically distributed according to \(\text{ctpt}(\gamma,\nu)\) with \(\nu > 2\). Given known measurable functions \(g_i: \mathbb{R}^k \rightarrow \mathbb{R}\) for \(i=1,2,\dots,n\), we assume that the responses \(y_i \in \mathbb{R}\) for \(i=1,2,\dots,n\) are generated by
\vspace{-0.25cm}
\begin{align}
    y_i = g_i(\bm{\beta}) + \sigma \varepsilon_i, \label{eq:general_reg}
\end{align}
\vspace{-0.25cm}
where \(\bm{\beta} = (\beta_1, \beta_2, \dots, \beta_k)^T \in \mathbb{R}^k\) is the location parameter specifying the mean of the responses, and \(\sigma \in \mathbb{R}_+\) is the scale parameter which, together with \(\gamma\) and \(\nu\), controls the variance of the responses. The likelihood is then given by:
\vspace{-0.25cm}
\begin{flalign}
  & \begin{aligned}
    p(y_1, y_2, &\dots, y_n|\bm{\beta}, \sigma, \gamma, \nu) = \\
    &\frac{2\sigma^{-n}}{\left(\gamma + \frac{1}{\gamma}\right)^n} \prod_{i=1}^n \Bigg[f_\nu\left(\frac{y_i - g_i(\bm{\beta}) + \sigma m(\gamma,\nu)}{\sigma \gamma}\right) I_{[0,+\infty)}\left(\frac{y_i - g_i(\bm{\beta}) + \sigma m(\gamma,\nu)}{\sigma}\right) + \\
    & f_\nu\left(\frac{y_i - g_i(\bm{\beta}) + \sigma m(\gamma,\nu)}{\sigma/\gamma}\right) I_{(-\infty,0)}\left(\frac{y_i - g_i(\bm{\beta}) + \sigma m(\gamma,\nu)}{\sigma}\right)\Bigg], \quad \gamma \in \mathbb{R}_+; \; \nu > 2. \label{eq:ctpt}
  \end{aligned} 
\end{flalign}\label{eq:general_reg_lik}
\vspace{-0.75cm}

We specify the class of prior distributions as follows:
\vspace{-0.25cm}
\begin{align}
    P_{\{\bm{\beta}, \sigma, \gamma, \nu\}} = P_{\bm{\beta}} \times P_{\sigma} \times P_{\gamma} \times P_\nu, \label{eq:prior_general}
\end{align}
\vspace{-0.25cm}
where
\begin{itemize}
    \item[1.] \(P_{\sigma}\) is a non-informative prior with the improper density \(p(\sigma) \propto \sigma^{-1}\).
    \item[2.] \(P_{\bm{\beta}}\) is any \(\sigma\)-finite measure defined on \(\mathbb{R}^k\).
    \item[3.] \(P_{\gamma}\) is any proper distribution defined on \([a,b]\), where \(a\) and \(b\) are fixed real constants such that \(0 < a \leq 1\) and \(1 \leq b < \infty\).
    \item[4.] \(P_\nu\) is any proper distribution defined on \((2, \infty)\).
\end{itemize}

Note that the non-informative prior for \(\sigma\) is improper, and the prior for \(\bm{\beta}\) can also be specified as improper. For example, one can adopt \(p(\bm{\beta},\sigma)\propto \sigma^{-1}\), which corresponds to a standard non-informative prior
for linear regression (\citealp{gelman2013}). When using improper priors for inference, it is necessary to prove the properness of the posterior distribution. While one may argue that proper priors can be used to avoid issues with improper posteriors, we believe that allowing the use of improper priors is important in practice, particularly in cases where specifying reasonable proper priors is challenging. Additionally, when improper priors originate from a common class of non-informative priors for \(\bm{\beta}\) and \(\sigma\), they can serve as a default choice in Bayesian analysis, with desirable properties.
 
The class of prior distributions in (\ref{eq:prior_general}) offers substantial flexibility in incorporating our (lack of) subjective knowledge about \(\bm{\beta}\), \(\gamma\), and \(\nu\):

\begin{itemize}
    \item[1.] For the location parameter \(\bm{\beta}\), we can select \(P_{\bm{\beta}}\) such that the joint prior \(P_{\bm{\beta}} \times P_{\sigma}\) has the density \(p(\bm{\beta}, \sigma) \propto \sigma^{-1}\), as mentioned earlier. This is suitable when we do not have substantive prior information on the regression and scale parameters. In addition, this prior has important invariance properties with respect to location and scale transformations of the data as well as other desirable ``objective'' qualities \citep{Bayarri_12}. Alternatively, other distributions, such as a normal distribution with a specified mean and variance, can be used to incorporate prior knowledge about \(\bm{\beta}\).

    \item[2.] For the skewness parameter \(\gamma\), we can set \(a \approx 0\) and \(b\) to be sufficiently large, specifying a flat distribution with its mode at 1 to avoid imposing strong prior information. Alternatively, setting \(a = b = 1\) results in \(P_{\gamma}\) being a Dirac measure at 1, implying that the error term distribution is symmetric around zero. In this case, we denote \(\gamma = 1\) for convenience.

    \item[3.] For the tail parameter \(\nu\), there is considerable flexibility in specifying a prior. A convenient choice is the shifted exponential distribution with density \(p(\nu)=\lambda e^{-\lambda(\nu-2)}, \;\nu>2.\) If heavy tails are expected, a large value of \(\lambda\) brings \(\nu\) closer to \(2\); when no strong preference exists, a smaller \(\lambda\) may be adopted. Alternatively, if tails are anticipated to be as light as those of the normal distribution, \(f_\nu\) can be replaced by the standard normal density (see (\ref{eq:ctpt_normal})). With a slight abuse of notation, this corresponds to assigning the prior \(\nu=\infty\).
\end{itemize}

\vspace{-0.25cm}
As mentioned earlier, we adopt a class of improper priors for our model, and it is essential to prove the existence of the posterior distribution to ensure valid statistical inference. Additionally, it is helpful to know up to which order the posterior moments of \(\bm{\beta}\) and \(\sigma\) exist, as this informs us on the posterior tails and helps us to present our inference. First, the following theorem (based on Theorem 1 from \cite{fernandez1998}) demonstrates that the existence of the posterior distribution and posterior moments is unaffected by incorporating the skewness parameter \(\gamma\).

\vspace{-0.25cm}
\begin{theorem}\label{thm:1}
   Given the prior distribution in (\ref{eq:prior_general}), for \(n\) independent samples from the sampling distribution specified in (\ref{eq:ctpt}) and with $r_j\in \Re, j=1,\dots,k$ and $r\geq0$, we have that
    \begin{align}
        \mathbb{E}\left[\sigma^r \prod_{j=1}^k |\beta_j|^{r_j} \Big| y_1, \dots, y_n\right] < \infty
    \end{align}
    if and only if the same holds under \(\gamma = 1\).
\end{theorem}
\vspace{-0.25cm}

Applying Theorem~\ref{thm:1} with \(r = 0\) and \(r_j = 0\) for \(j = 1, \dots, k\), we conclude that incorporating skewness in the Student \(t\)-sampling distribution does not affect the properness of the posterior distribution.

Next, we consider a similar model where \(f_\nu\), defined in (\ref{eq:t}), is replaced by the density of the standard normal distribution. In this case, the sampling distribution becomes

\vspace{-0.75cm}
\begin{flalign}
  & \begin{aligned}
    p(y_1, y_2, &\dots, y_n|\bm{\beta}, \sigma, \gamma) = \\
    &\frac{2\sigma^{-n}}{\left(\gamma + \frac{1}{\gamma}\right)^n} \prod_{i=1}^n \Bigg[ f\left(\frac{y_i - g_i(\bm{\beta}) + \sigma m^{\text{normal}}(\gamma)}{\sigma \gamma}\right) I_{[0,+\infty)}\left(\frac{y_i - g_i(\bm{\beta}) + \sigma m^{\text{normal}}(\gamma)}{\sigma}\right) + \\
    & f\left(\frac{y_i - g_i(\bm{\beta}) + \sigma m^{\text{normal}}(\gamma)}{\sigma/\gamma}\right) I_{(-\infty,0)}\left(\frac{y_i - g_i(\bm{\beta}) + \sigma m^{\text{normal}}(\gamma)}{\sigma}\right) \Bigg], 
\label{eq:ctpt_normal}
  \end{aligned} 
\end{flalign}
\vspace{-1cm}

\noindent
where \(f(x) = \frac{1}{\sqrt{2\pi}} \exp\left(-\frac{x^2}{2}\right)\).

For the sampling distribution in (\ref{eq:ctpt_normal}), we assume the following class of prior distributions:
\begin{align}
    P_{\{\bm{\beta}, \sigma, \gamma\}} = P_{\bm{\beta}} \times P_{\sigma} \times P_{\gamma}, \label{eq:prior_normal}
\end{align}
where \(P_{\bm{\beta}}, P_{\sigma}\) and \(P_{\gamma}\) are the same as in (\ref{eq:prior_general}). For the normal sampling model defined in (\ref{eq:general_reg}), (\ref{eq:ctpt_normal}), and (\ref{eq:prior_normal}), we also find that the existence of the posterior distribution and posterior moments is unaffected by uncertainty in the skewness parameter \(\gamma\).

\vspace{-0.25cm}
\begin{theorem}\label{thm:2}
    Given the prior distribution in (\ref{eq:prior_normal}), for \(n\) independent samples from the sampling distribution specified in (\ref{eq:ctpt_normal}) and with $r_j\in \Re, j=1,\dots,k$ and $r\geq0$, we have that
    \begin{align}
        \mathbb{E}\left[\sigma^r \prod_{j=1}^k |\beta_j|^{r_j} \Big| y_1, \dots, y_n\right] < \infty
    \end{align}
    if and only if the same holds under \(\gamma = 1\).
\end{theorem}

\vspace{-0.5cm}
\subsection{Bayesian Linear Regression Model}\label{sec:3.2}

Theorems~\ref{thm:1} and \ref{thm:2} state that the presence of the skewness parameter \(\gamma\) does not affect the existence of the posterior distribution and moments for the two models in Section~\ref{sec:3.1}. However, these theorems do not imply the properness of the posterior distribution. Additionally, while Section~\ref{sec:3.1} considers a general regression model, linear regression models are more commonly used in mediation analysis. In this section, we examine a specific case: a linear regression model with a slightly restricted class of prior distributions.

\begin{itemize}
    \item[1.] Given the predictors \(\bm{x}_i\), 
    let \(g_i(\bm{\beta}) = \bm{x}_i^T \bm{\beta}\). We assume that the design matrix \(\bm{X} = [\bm{x}_1, \dots, \bm{x}_n]^T\) with \(n\) observations has full column rank \(k\), and the response vector \(\bm{y}=(y_1,\dots,y_n)^T\) does not lie in the column space of \(\bm{X}\).

    \item[2.] The prior for \(\bm{\beta}\) is chosen such that the joint prior \(P_{\bm{\beta}} \times P_{\sigma}\) has the density \(p(\bm{\beta}, \sigma) \propto \sigma^{-1}\).

    \item[3.] \(P_{\gamma}\) is any proper distribution defined on \([a,b]\), where \(a\) and \(b\) are fixed real constants such that \(0 < a \leq 1\) and \(1 \leq b < \infty\).

    \item[4.] \(P_\nu\) is any proper distribution defined on \((2, \infty)\).
\end{itemize}

For the normal sampling model defined in (\ref{eq:general_reg}), (\ref{eq:ctpt_normal}), and (\ref{eq:prior_normal}), under the assumptions given in the first three points above, it is well known that the posterior distribution will be proper when \(\gamma = 1\) if and only if the number of observations exceeds the number of predictors, i.e., \(n > k\) (\citealp{gelman2013}). Thus, by Theorem~\ref{thm:2}, the skewed normal model leads to a proper posterior if and only if \(n>k\). In addition, posterior moments of regression coefficients and $\sigma$ exist up to order $n-k$. 

For the Student \(t\)-sampling model defined in (\ref{eq:general_reg}), (\ref{eq:ctpt}), and (\ref{eq:prior_general}), the same result for posterior existence holds. The following theorem provides the necessary and sufficient condition for the properness of the posterior distribution.

\vspace{-0.25cm}
\begin{theorem}
     Given the prior distribution specified above, for \(n\) independent samples from the sampling distribution specified in (\ref{eq:ctpt}) with \(g_i(\bm{\beta}) = \bm{x}_i^T \bm{\beta}\), the posterior distribution is proper if and only if \(n > k\).
\end{theorem}
\vspace{-0.25cm}

We can also derive a sufficient condition for the existence of marginal posterior moments of the components of \(\bm{\beta}\) for the Student \(t\)-sampling model. To do so, we first present a definition, taken from \citet{fernandez1998}, which is useful for characterising the design matrix \(\bm{X}\).

\begin{definition}[Singularity Index for Column \(j\); Definition 1 in \citet{fernandez1998}]
    Consider an \(n \times k\) matrix \(\bm{X}\) with full column rank. The singularity index for column \(j = 1, \dots, k\) is defined as the largest integer \(p_j\) \((0 < p_j < n-k)\) for which there exists a sub-matrix of \(\bm{X}\) with dimensions \((k-1+p_j) \times k\) and rank \(k-1\). This rank \(k-1\) must be preserved even after removing the \(j\)th column of the sub-matrix.
\end{definition}

\vspace{-0.25cm}
Using the singularity index for column \(j\), we can establish the following sufficient condition for the existence of the marginal posterior moments of \(\bm{\beta}\).

\begin{theorem}
    Given the prior distribution specified above, for \(n\) independent samples from the sampling distribution specified in (\ref{eq:ctpt}) with \(g_i(\bm{\beta}) = \bm{x}_i^T \bm{\beta}\), if \(r \geq n-k\), then \(\mathbb{E}(|\beta_j|^r \mid y_1, \dots, y_n) = \infty\). If \(\,0 \leq r < \min\{n-k, n-k-p_j+2\}\), then \(\mathbb{E}(|\beta_j|^r \mid y_1, \dots, y_n) < \infty\).
\end{theorem}

Following the argument of \citet{fernandez1998}, we can also derive the necessary and sufficient condition for the existence of the posterior moments of \(\sigma\).

\begin{theorem}
    Given the prior distribution specified above, for \(n\) independent samples from the sampling distribution specified in (\ref{eq:ctpt}) with \(g_i(\bm{\beta}) = \bm{x}_i^T \bm{\beta}\), \(\mathbb{E}(\sigma^r \mid y_1, \dots, y_n) < \infty\) if and only if \(r < n-k\).
\end{theorem}

This fully establishes the conditions for the existence of the posterior distribution and the posterior moments of \(\sigma\) and \(\beta_j\) for \(j = 1, \dots, k\) using the CTPT linear regression sampling model and its submodel with symmetry with the prior distributions specified in Section~\ref{sec:3.2}.

\vspace{-0.25cm}
\subsection{(Non-)Equivalence Between Bayesian Regression Models With Centred and Uncentred Two-Piece Student \(t\) Errors}\label{sec:3.3}

The centred two-piece Student \(t\) distribution is obtained by centring the original two-piece Student \(t\) distribution. We now investigate under what circumstances and in what sense the centred and uncentred forms are equivalent, when they are used for the error terms in the general regression framework of Section~\ref{sec:3.1} and in the linear regression setting of Section~\ref{sec:3.2}. The following result provides sufficient conditions for equivalence in the general regression setting of Section~\ref{sec:3.1}.

\begin{theorem}\label{thm:6}
In the general regression model defined in Section~\ref{sec:3.1}, suppose that  
\begin{enumerate}\item[(1)] the model contains a free intercept \(\beta_0\); that is, 
\vspace{-0.25cm}
\begin{align}
  y_i = g_i(\bm{\beta}) + \sigma \varepsilon_i
  \quad \Longleftrightarrow \quad
  y_i = \beta_0 + \tilde{g}_i(\bm{\beta}^*) + \sigma \varepsilon_i,
  \qquad i = 1,\dots,n, \label{eq:general_reg_free_intercept}
\end{align}

\vspace{-0.25cm}
where \(\bm{\beta}^*\) excludes \(\beta_0\); and  
\item[(2)] the prior \(P_{\bm{\beta}}\) is shift-invariant in \(\beta_0\); that is, 
\vspace{-0.5cm}
\begin{align}
  p(\beta_0,\bm{\beta}^*) \;\equiv\; p(\beta_0 + c,\bm{\beta}^*), \qquad \forall c \in \mathbb{R};
\end{align}
\end{enumerate}
\vspace{-0.25cm}
then such a regression model that uses the centred two-piece Student \(t\) distribution for the errors is equivalent to one that uses the uncentred two-piece Student \(t\) distribution, in the sense that, under the same prior specification, they induce identical posterior distributions up to a one-to-one reparameterisation of the intercept.
\end{theorem}

By “one-to-one reparameterisation” we mean that the model in~\eqref{eq:general_reg_free_intercept} can be reparameterised as
\vspace{-0.75cm}
\begin{align}
  y_i = \beta_0^{*} + \tilde{g}_i(\bm{\beta}^{*}) + \sigma \varepsilon^{*}_{i}, 
  \qquad i = 1,\dots,n,
\end{align}

\vspace{-0.5cm}
\noindent
where \(\beta_0^{*} \equiv \beta_0 - \sigma m(\gamma,\nu)\) and 
\(\varepsilon^{*}_{i} = \varepsilon_{i} + m(\gamma,\nu)\) 
The transformed errors \(\varepsilon^{*}_{i}\) follow the original uncentred two-piece Student \(t\) distribution defined in~\eqref{eq:tpd}. In other words, the regression model using the uncentred two-piece Student \(t\) distribution for the error term can be reparameterised so that the intercept \(\beta_0\) absorbs the non-zero mean of the error terms, yielding exactly the same posterior distribution for all parameters —except the intercept—under the same prior specification. Thus, within this framework, the centred and uncentred versions of the two-piece \(t\) distribution are identical up to a relabelling of the intercept.

In particular, Section~\ref{sec:3.2} considers the linear regression case with the prior \(p(\bm{\beta},\sigma)\propto 1/\sigma\). By Theorem~\ref{thm:6}, if the model includes an intercept, the specification with centred errors is equivalent to that with uncentred errors. This is advantageous because, in the original two-piece Student \(t\) distribution, the skewness parameter \(\gamma\) and the tail parameter \(\nu\) have clear interpretations. Equivalence therefore implies that \(\gamma\) and \(\nu\), together with all regression coefficients except the intercept, share the same posterior distributions --- and hence the same inferential meaning --- in the linear regression model with uncentred two-piece errors, a formulation already widely used in practice. This provides another rationale for using the non‐informative prior \(p(\bm{\beta},\sigma)\propto 1/\sigma\).

In Supplementary Material~\ref{sm:equiv_model}, we present parameter recovery results for the model obtained with uncentred errors, adopting the same prior specification described in Sections~\ref{sec:3.2} and Section~\ref{sec:4}, and the simulation settings described later in Section~\ref{sec:5.1}. The posterior summaries are very similar to those from the model with centred errors, empirically supporting the equivalence of the two models.

It is also worth noting the circumstances under which the two formulations are generally \emph{not} equivalent. The key instance arises when the prior \(P_{\bm{\beta}}\) is \emph{not} shift‐invariant in \(\beta_0\); this occurs typically when one wishes to place a proper prior on the intercept. In linear regression, the error term is usually assumed to have zero mean (e.g., the widely used OLS regression), and the intercept \(\beta_0\) represents the conditional mean of the response when all predictors are set to zero—often interpreted as an average baseline level. When previous studies have modelled errors with zero mean or when we have prior beliefs about the average baseline level, adopting the centred two‐piece Student \(t\) distribution makes it easier to borrow information from earlier work and construct informative priors, thereby improving estimation accuracy and hypothesis-testing power, as frequently desired in the Bayesian mediation analysis literature (\citealp{miocevic2017}; \citealp{miocevic2022}). By contrast, in the uncentred specification the errors satisfy \(\mathbb{E}[\varepsilon_i^*]=m(\gamma,\nu)\neq0\), so any proper prior on the intercept must, by construction, depend on \(\gamma\) and \(\nu\), whose values are usually uninformed from previous studies that do not model skewed and/or heavy-tailed errors. This issue complicates prior elicitation and risks importing unintended information about \(\gamma\) and \(\nu\). Thus, we believe it is advantageous to use the centred two-piece Student \(t\) distribution to model the error terms in linear regression and in mediation analysis.

\vspace{-0.5cm}
\section{Testing Mediation Effects}\label{sec:4}
\vspace{-0.25cm}
Since the mediation model in Equations (\ref{eq:my_med_1}) and (\ref{eq:my_med_2}) comprises two linear regression sub-models, we can apply the Bayesian linear regression models discussed earlier to incorporate skewness and heavy tails in the error terms during mediation analysis. Additionally, we assume that the path coefficients \(\alpha\) and \(\beta\) are independent a priori, allowing us to treat (\ref{eq:my_med_1}) and (\ref{eq:my_med_2}) separately by fitting each to a linear regression model. We then obtain \(T\) posterior draws of \(\alpha\) and \(\beta\) separately, denoted as \(\{\alpha^{(i)}\}_{i=1}^T\) and \(\{\beta^{(i)}\}_{i=1}^T\), respectively. This enables us to generate \(T\) posterior draws of the mediation effect \(\alpha\beta\), denoted as \(\{\alpha^{(i)}\beta^{(i)}\}_{i=1}^T\).

Here, we provide a brief overview of Bayesian hypothesis testing for the mediation effect using the Bayes factor. \citet{nuijten2015} first tried to develop a Bayesian hypothesis test for the mediation effect based on the Bayes factor. When testing the mediation effect, the null hypothesis asserts that there is no mediation effect, formalised as \(H_0: \alpha\beta = 0\), or equivalently, \(H_0: \{\alpha = 0, \beta = 0\} \text{ or } \{\alpha \neq 0, \beta = 0\} \text{ or } \{\alpha = 0, \beta \neq 0\}\). The alternative hypothesis is \(H_1: \alpha\beta \neq 0\), or equivalently, \(H_1: \{\alpha \neq 0, \beta \neq 0\}\).

\citet{nuijten2015} assume that the path parameters \(\alpha\) and \(\beta\) are independent a priori and formalise the test of the mediation effect by separately testing for the existence of path \(\alpha\) and the evidence for path \(\beta\). The test for path \(\alpha\) is framed as a model comparison problem between the following two models:
\vspace{-0.25cm}
\begin{align}
    &H_{\alpha=0}: M = \beta_0^{(M)} + \sigma^{(M)}\varepsilon^{(M)}, \\
    &H_{\alpha \neq 0}: M = \beta_0^{(M)} + \alpha X + \sigma^{(M)}\varepsilon^{(M)}.
\end{align}

\vspace{-0.25cm}
The Bayes factor for path \(\alpha\) is then calculated as:
\vspace{-0.25cm}
\begin{align}
    BF^{\alpha} = \frac{p(\text{data} \mid H_{\alpha \neq 0})}{p(\text{data} \mid H_{\alpha=0})}.
\end{align}

\vspace{-0.25cm}
Similarly, the Bayes factor for path \(\beta\), denoted as \(BF^{\beta}\), can be calculated in the same way. Under the assumption of independent paths, \citet{nuijten2015} attempt to derive the formula for the Bayes factor of the mediation effect based on \(BF^{\alpha}\) and \(BF^{\beta}\). However, errors in their calculation violate the assumption of independent paths. These errors are corrected by \citet{liu2023}, who provide the correct Bayes factor for the mediation effect under the assumption of independent paths \(\alpha\) and \(\beta\):
\vspace{-0.25cm}
\begin{align}
    BF^{\text{med}} = \frac{BF^{\alpha}BF^{\beta}}{q_{00|0} + q_{01|0}BF^{\beta} + q_{10|0}BF^{\alpha}}, \label{eq:bf}
\end{align}
where \(q_{00|0} := \mathbb{P}(\alpha = 0, \beta = 0 \mid \alpha\beta = 0)\), \(q_{01|0} := \mathbb{P}(\alpha = 0, \beta = 1 \mid \alpha\beta = 0)\), and \(q_{10|0} := \mathbb{P}(\alpha = 1, \beta = 0 \mid \alpha\beta = 0)\).

\citet{liu2023} also point out that under the assumption of independent paths \(\alpha\) and \(\beta\), the prior odds of the mediation effect must satisfy:
\vspace{-0.25cm}
\begin{align}
    PriorOdds^{\text{med}} = \frac{PriorOdds^{\alpha} \times PriorOdds^{\beta}}{1 + PriorOdds^{\beta} + PriorOdds^{\alpha}},
\end{align}
where \(PriorOdds^{\text{med}} := \frac{\mathbb{P}(H_1: \alpha\beta \neq 0)}{\mathbb{P}(H_0: \alpha\beta = 0)}\), \(PriorOdds^{\alpha} := \frac{\mathbb{P}(H_{\alpha \neq 0})}{\mathbb{P}(H_{\alpha = 0})}\), and \(PriorOdds^{\beta} := \frac{\mathbb{P}(H_{\beta \neq 0})}{\mathbb{P}(H_{\beta = 0})}\).

Under the assumption of independent paths \(\alpha\) and \(\beta\), the Bayes factor for the mediation effect in (\ref{eq:bf}) can be expressed in terms of the prior odds and Bayes factors of \(\alpha\) and \(\beta\) as:
\vspace{-0.25cm}
\begin{align}
    BF^{\text{med}} = \frac{(1 + PriorOdds^{\beta} + PriorOdds^{\alpha})BF^{\alpha}BF^{\beta}}{1 + PriorOdds^{\beta}BF^{\beta} + PriorOdds^{\alpha}BF^{\alpha}}.
\end{align}

\vspace{-0.25cm}
\citet{liu2024} further evaluate the performance of the Bayes factor in testing the mediation effect. They find that the true size of the mediation effect and the prior specification, including the prior odds for the presence of each path, can influence both false positive and true positive rates in detecting mediation.

\vspace{-0.5cm}
\section{Simulation Study}\label{sec:5}
\vspace{-0.3cm}

In Section~\ref{sec:3}, we discussed a broad class of prior distributions. For the simulations and the later case study, we specify the prior distributions for \(\gamma\) and \(\nu\) as follows:
\begin{itemize}
    \item[1.] The prior distribution of \(\gamma\) is a truncated gamma distribution on \((0.05, 20)\):
    \vspace{-0.25cm}
    \begin{align}
        p(\gamma) \propto \frac{b^{a}}{\Gamma(a)} \gamma^{a-1} e^{-b \gamma},\; \gamma \in (0.05, 20) \label{eq:gamma_prior}
    \end{align}
    and we set \(a = b = 2\) for subsequent studies. This choice gives a prior mean \(\mathbb{E}[\gamma \mid a, b] \approx 1\) and a prior variance \(\text{Var}(\gamma \mid a, b) \approx 0.5\). We believe that the support interval \((0.05, 20)\) is sufficiently wide, and the prior mean and variance are reasonable for practical applications.

    \item[2.] The prior distribution of \(\nu\) is a right-shifted exponential distribution on \((2, \infty)\):
    \vspace{-0.25cm}
    \begin{align}
        p(\nu) \propto d\,e^{-d(\nu-2)},\; \nu \in (2, \infty) \label{eq:nu_prior}
    \end{align}
    and we set \(d = 0.01\) for subsequent studies. This choice results in a prior mean \(\mathbb{E}[\nu \mid d] = 12\) and a prior variance \(\text{Var}(\nu \mid d) = 100\). This specification ensures that the density is relatively flat, thereby avoiding strong prior assumptions on \(\nu\).
\end{itemize}

To draw posterior samples \(\{\alpha^{(i)}\}_{i=1}^T\) and \(\{\beta^{(i)}\}_{i=1}^T\), we use the No-U-Turn Sampler implemented in RStan (\citealp{carpenter2017}). For estimating the marginal likelihood and Bayes factors for \(\alpha\) and \(\beta\) separately from the two linear regression models, we employ the bridge sampler from the R package \textit{bridgesampling} (\citealp{gronau2020}) and then apply (\ref{eq:bf}) to calculate the Bayes factor for the mediation effect. 

We now evaluate the performance of our proposed method under various conditions of skewed and/or heavy-tailed error distributions. The primary objectives are to assess the parameter recovery—focusing on parameters \(\gamma\), \(\nu\), and the mediation effect \(\alpha\beta\)—and to evaluate the statistical power of the method in detecting mediation effects under skewed and/or heavy-tailed errors. Additionally, the model's performance in hypothesis testing is benchmarked against three sub-models—the \(\gamma\)-Only or Skew-Normal model (where we fix \(\nu=\infty\)), the \(\nu\)-Only or Student $t$ model (where we fix \(\gamma=1\)), and the Normal model (where we fix both \(\gamma=1\) and \(\nu=\infty\))—as well as popular frequentist approaches.

\vspace{-0.5cm}
\subsection{Parameter Recovery}\label{sec:5.1}

The data for the parameter recovery studies are generated as follows. The skewness parameter \(\gamma\) takes the values \(0.33\) (highly left‑skewed), \(0.5\) (moderately left‑skewed), \(1\) (symmetric), \(2\) (moderately right‑skewed), and \(3\) (highly right‑skewed), whereas the tail parameter \(\nu\) takes the values \(3\) (very heavy tails), \(10\) (moderately heavy tails), and \(\infty\) (normal tails). These combinations yield fifteen distinct error distributions.

In the first series of simulations, for convenience, the error terms in the two equations, both \(\{\varepsilon_i^{(M)}\}_{i=1}^{n}\) and \(\{\varepsilon_i^{(Y)}\}_{i=1}^{n}\), are independently drawn from the same distribution. We must, however, examine whether this assumption is critical. For example, one might expect that if \(\varepsilon^{(M)}\) is left‑skewed while \(\varepsilon^{(Y)}\) is right‑skewed, inference about mediation may differ from the situation in which both errors share the same skewness. To explore this possibility, we consider four additional scenarios: (1) \(\varepsilon^{(M)}\sim\operatorname{ctpt}(0.33,3)\), \(\varepsilon^{(Y)}\sim\operatorname{ctpt}(2,10)\); (2) \(\varepsilon^{(M)}\sim\operatorname{ctpt}(0.33,10)\), \(\varepsilon^{(Y)}\sim\operatorname{ctpt}(2,3)\); (3) \(\varepsilon^{(M)}\sim\operatorname{ctpt}(2,3)\), \(\varepsilon^{(Y)}\sim\operatorname{ctpt}(0.33,10)\); and (4) \(\varepsilon^{(M)}\sim\operatorname{ctpt}(2,10)\), \(\varepsilon^{(Y)}\sim\operatorname{ctpt}(0.33,3)\).  

The intercepts are fixed at \(\beta_0^{(M)} = \beta_0^{(Y)} = 0\). For the path coefficients, we set \(\alpha\in\{0.4,0.7\}\) and \(\beta\in\{0.4,0.7\}\), yielding four combinations. The direct effect is fixed at \(\tau = 0.2\), and the scale parameters \(\sigma^{(M)}\) and \(\sigma^{(Y)}\) are both fixed at \(1\).

Under the null hypothesis of no mediation effect, the prior probabilities of the three possible scenarios are set equal: 
\vspace{-0.25cm}
\begin{align}
    \mathbb{P}(\alpha = 0, \beta = 0 \mid H_0) 
= \mathbb{P}(\alpha = 0, \beta \neq 0 \mid H_0) 
= \mathbb{P}(\alpha \neq 0, \beta = 0 \mid H_0) 
= \frac{1}{3}.
\end{align}

\vspace{-0.25cm}
The sample size is fixed at \(n = 50\), and \(R = 1\,000\) independent experiments are conducted to evaluate the model's performance. The performance in estimating the mediation effect \(\alpha\beta\), as well as the skewness parameter \(\gamma\) and the tail parameter \(\nu\), is evaluated across each simulation scenario. For convenience, the model that incorporates both \(\gamma\) and \(\nu\) is referred to as the "Full Model." A single Markov chain of length \(L = 30\,000\) is run, with the first \(20\%\) discarded as burn-in samples. 

First, we consider the case where \(\{\varepsilon_i^{(M)}\}_{i=1}^{n}\) and \(\{\varepsilon_i^{(Y)}\}_{i=1}^{n}\) are drawn from the same distribution. Table~\ref{tab:pr_med} presents the means ($\pm$ one standard deviation) 
of the posterior mean, posterior mode, the posterior 2.5th, 25th, 50th (median), 75th, and 97.5th percentiles, the length of the \(95\%\) credible intervals (based on the posterior 2.5th and 97.5th percentiles) for \(\alpha\beta\) and their empirical coverage rate of the true value across the \(1\,000\) experiments with \(\alpha=\beta=0.4\), where \(\varepsilon^{(M)}\) and \(\varepsilon^{(Y)}\) share the same distribution. The results for the remaining three combinations of \(\alpha\) and \(\beta\) are presented in Supplementary Material~\ref{sm:med_diff}. The posterior means of \(\alpha\beta\) align well with the true values, confirming the model's ability to recover the mediation effect under normal, skewed, and/or heavy-tailed error distributions. The empirical coverage rates of the 95\% credible intervals are slightly higher than the nominal 95 percent when skewness is presented. We note that the prior of \(\gamma\) is weakly-informative with its mean close to \(1\), which may lead to slightly conservative credible intervals when skewness is present. In Supplementary Material~\ref{sm:alpha_beta}, parameter recovery results for \(\alpha\) and \(\beta\) are presented in Table~\ref{tab:pr_alpha} and Table~\ref{tab:pr_beta}, respectively.

\setlength{\tabcolsep}{2.1pt}
\begin{table}[t]
\centering
\fontsize{6}{10}\selectfont
\caption{\textbf{Parameter recovery results for the mediation effect.} The "True \(\alpha\beta\)" is the product of \(\alpha=0.4\) and \(\beta=0.4\), with \(\varepsilon^{(M)}\) and \(\varepsilon^{(Y)}\) drawn from the same distribution. The first two columns specify the skewness and tail parameters in the error distribution. It presents the means ($\pm$ one standard deviation) of posterior means, modes, 2.5th, 25th, 50th, 75th, and 97.5th percentiles, the length of the \(95\%\) credible intervals for \(\alpha\beta\), and the coverage rate of the true value by the credible intervals across \(1\,000\) experiments with sample size \(n = 50\).}

\begin{tabular}{*{12}{c}}
    \toprule
    \textbf{\(\gamma\)} & \textbf{\(\nu\)} & \textbf{True \(\alpha\beta\)} & \textbf{Mean} & \textbf{Mode} & \textbf{2.5th} & \textbf{25th} & \textbf{50th} & \textbf{75th} & \textbf{97.5th} & \textbf{Length} & \textbf{Rate} \\
    \midrule
    0.33 & 3  & 0.16 & \(0.165 \pm 0.102\) & \(0.151 \pm 0.100\) & \(-0.030 \pm 0.104\) & \(0.096 \pm 0.094\) & \(0.160 \pm 0.101\) & \(0.229 \pm 0.112\) & \(0.384 \pm 0.146\) & \(0.413 \pm 0.129\) & .966 \\
    0.33 & 10 & 0.16 & \(0.161 \pm 0.087\) & \(0.143 \pm 0.086\) & \(-0.015 \pm 0.086\) & \(0.097 \pm 0.079\) & \(0.155 \pm 0.086\) & \(0.220 \pm 0.097\) & \(0.371 \pm 0.128\) & \(0.386 \pm 0.109\) & .970 \\
    0.33 & \(\infty\) & 0.16 & \(0.161 \pm 0.088\) & \(0.142 \pm 0.084\) & \(-0.008 \pm 0.081\) & \(0.098 \pm 0.077\) & \(0.155 \pm 0.086\) & \(0.218 \pm 0.099\) & \(0.367 \pm 0.133\) & \(0.375 \pm 0.110\) & .961 \\
    0.5  & 3  & 0.16 & \(0.156 \pm 0.093\) & \(0.138 \pm 0.092\) & \(-0.018 \pm 0.084\) & \(0.091 \pm 0.083\) & \(0.150 \pm 0.092\) & \(0.216 \pm 0.105\) & \(0.364 \pm 0.137\) & \(0.382 \pm 0.113\) & .960 \\
    0.5  & 10 & 0.16 & \(0.162 \pm 0.091\) & \(0.141 \pm 0.090\) & \(0.003 \pm 0.072\) & \(0.100 \pm 0.078\) & \(0.155 \pm 0.090\) & \(0.217 \pm 0.104\) & \(0.360 \pm 0.136\) & \(0.357 \pm 0.104\) & .945 \\
    0.5  & \(\infty\) & 0.16 & \(0.167 \pm 0.087\) & \(0.146 \pm 0.086\) & \(0.014 \pm 0.065\) & \(0.107 \pm 0.075\) & \(0.160 \pm 0.086\) & \(0.220 \pm 0.100\) & \(0.360 \pm 0.129\) & \(0.346 \pm 0.094\) & .939 \\
    1 & 3  & 0.16 & \(0.157 \pm 0.096\) & \(0.131 \pm 0.092\) & \(-0.016 \pm 0.079\) & \(0.089 \pm 0.082\) & \(0.148 \pm 0.094\) & \(0.217 \pm 0.109\) & \(0.377 \pm 0.144\) & \(0.393 \pm 0.108\) & .949 \\
    1 & 10 & 0.16 & \(0.163 \pm 0.091\) & \(0.136 \pm 0.090\) & \(0.003 \pm 0.065\) & \(0.098 \pm 0.077\) & \(0.154 \pm 0.090\) & \(0.219 \pm 0.106\) & \(0.373 \pm 0.137\) & \(0.370 \pm 0.098\) & .950 \\
    1 & \(\infty\) & 0.16 & \(0.159 \pm 0.090\) & \(0.132 \pm 0.089\) & \(0.006 \pm 0.063\) & \(0.096 \pm 0.076\) & \(0.150 \pm 0.089\) & \(0.212 \pm 0.104\) & \(0.359 \pm 0.135\) & \(0.354 \pm 0.094\) & .935 \\
    2 & 3  & 0.16 & \(0.158 \pm 0.087\) & \(0.139 \pm 0.085\) & \(-0.018 \pm 0.082\) & \(0.092 \pm 0.078\) & \(0.152 \pm 0.086\) & \(0.218 \pm 0.097\) & \(0.367 \pm 0.126\) & \(0.385 \pm 0.106\) & .974 \\
    2 & 10 & 0.16 & \(0.160 \pm 0.085\) & \(0.137 \pm 0.082\) & \(-0.001 \pm 0.069\) & \(0.097 \pm 0.073\) & \(0.153 \pm 0.083\) & \(0.216 \pm 0.096\) & \(0.363 \pm 0.127\) & \(0.364 \pm 0.097\) & .963 \\
    2 & \(\infty\) & 0.16 & \(0.161 \pm 0.085\) & \(0.138 \pm 0.083\) & \(0.006 \pm 0.065\) & \(0.099 \pm 0.072\) & \(0.153 \pm 0.084\) & \(0.215 \pm 0.097\) & \(0.358 \pm 0.127\) & \(0.351 \pm 0.093\) & .959 \\
    3 & 3  & 0.16 & \(0.161 \pm 0.100\) & \(0.144 \pm 0.094\) & \(-0.046 \pm 0.106\) & \(0.087 \pm 0.092\) & \(0.155 \pm 0.097\) & \(0.228 \pm 0.108\) & \(0.396 \pm 0.143\) & \(0.441 \pm 0.122\) & .972 \\
    3 & 10 & 0.16 & \(0.163 \pm 0.092\) & \(0.142 \pm 0.088\) & \(-0.022 \pm 0.085\) & \(0.094 \pm 0.080\) & \(0.156 \pm 0.090\) & \(0.225 \pm 0.103\) & \(0.387 \pm 0.139\) & \(0.408 \pm 0.113\) & .973 \\
    3 & \(\infty\) & 0.16 & \(0.160 \pm 0.091\) & \(0.137 \pm 0.086\) & \(-0.019 \pm 0.083\) & \(0.092 \pm 0.079\) & \(0.152 \pm 0.088\) & \(0.219 \pm 0.102\) & \(0.378 \pm 0.137\) & \(0.397 \pm 0.110\) & .971 \\
    \bottomrule
\end{tabular}
\label{tab:pr_med}
\end{table}

We also assess the model's ability to recover the parameters \(\gamma\) and \(\nu\). In Supplementary Material~\ref{sm:skew_tail}, Tables~\ref{tab:pr_gamma1} and \ref{tab:pr_gamma2} report the parameter recovery results for \(\gamma\), while Tables \ref{tab:pr_v1} and \ref{tab:pr_v2} present the results for \(\nu\), corresponding to \(\varepsilon^{(M)}\) and \(\varepsilon^{(Y)}\). The results show that the recovery of \(\gamma\) is generally good, although slight under‑estimation occurs when the error terms are highly skewed (i.e., \(\gamma = 0.33\) or \(3\)), with lower coverage rates than the nominal \(95\%\) for the credible intervals. For \(\nu\), the recovery is satisfactory when \(\nu = 3\) or \(10\), though the intervals are rather wide, which indicates the difficulty to estimate \(\nu\) with small sample size. 

Furthermore, the parameter recovery results for the mediation effect \(\alpha\beta\) are compared between the Full Model and the Normal Model, which assumes \(\gamma = 1\) and \(\nu = \infty\). Table \ref{tab:pr_norm3} in Supplementary Material~\ref{sm:pr_mis} reports the estimated mediation effect based on the Normal Model. The posterior means of \(\alpha\beta\) are close to the true values in both models. However, the credible intervals given by the Normal Model are generally much wider compared with the Full Model's results. This suggests that the Normal Model may not perform as effectively in uncertainty quantification and may lead to lower statistical power in Bayesian hypothesis testing for the mediation effect.

\setlength{\tabcolsep}{1.8pt}
\begin{table}[h!]
\centering
\fontsize{5}{10}\selectfont
\caption{\textbf{Parameter recovery results for the mediation effect with opposing error skewness and different tail-heaviness.} The four error scenarios are  (1) \(\varepsilon^{(M)}\sim\operatorname{ctpt}(0.33,3)\), \(\varepsilon^{(Y)}\sim\operatorname{ctpt}(2,10)\); (2) \(\varepsilon^{(M)}\sim\operatorname{ctpt}(0.33,10)\), \(\varepsilon^{(Y)}\sim\operatorname{ctpt}(2,3)\); (3) \(\varepsilon^{(M)}\sim\operatorname{ctpt}(2,3)\), \(\varepsilon^{(Y)}\sim\operatorname{ctpt}(0.33,10)\); (4) \(\varepsilon^{(M)}\sim\operatorname{ctpt}(2,10)\), \(\varepsilon^{(Y)}\sim\operatorname{ctpt}(0.33,3)\). The first two columns specify \((\gamma,\, \nu)\) for \(\varepsilon^{(M)}\) and \(\varepsilon^{(Y)}\). It presents the means ($\pm$ one standard deviation) of posterior means, modes, 2.5th, 25th, 50th, 75th, and 97.5th percentiles, the length of the \(95\%\) credible intervals for \(\alpha\beta\), and the coverage rate of the true value by the credible intervals across \(1\,000\) experiments with sample size \(n = 50\).}

\begin{tabular}{*{12}{c}}
    \toprule
    \textbf{\(\varepsilon^{(M)}\)} & \textbf{\(\varepsilon^{(Y)}\)} & \textbf{True \(\alpha\beta\)} & \textbf{Mean} & \textbf{Mode} & \textbf{2.5th} & \textbf{25th} & \textbf{50th} & \textbf{75th} & \textbf{97.5th} & \textbf{Length} & \textbf{Rate} \\
    \midrule
    \multicolumn{12}{c}{Path Coefficients: \(\alpha = 0.4, \beta = 0.4\)} \\
    (0.33, 3) & (2, 10) & 0.16 & \(0.165 \pm 0.101\) & \(0.156 \pm 0.100\) & \(-0.025 \pm 0.108\) & \(0.100 \pm 0.097\) & \(0.162 \pm 0.100\) & \(0.226 \pm 0.108\) & \(0.371 \pm 0.135\) & \(0.396 \pm 0.117\) & .951 \\
    (0.33, 10) & (2, 3) & 0.16 & \(0.156 \pm 0.090\) & \(0.136 \pm 0.087\) & \(-0.015 \pm 0.089\) & \(0.093 \pm 0.081\) & \(0.149 \pm 0.088\) & \(0.213 \pm 0.101\) & \(0.364 \pm 0.134\) & \(0.378 \pm 0.114\) & .960 \\
    (2, 3) & (0.33, 10) & 0.16 & \(0.156 \pm 0.091\) & \(0.137 \pm 0.086\) & \(-0.019 \pm 0.088\) & \(0.091 \pm 0.082\) & \(0.150 \pm 0.089\) & \(0.215 \pm 0.101\) & \(0.369 \pm 0.133\) & \(0.388 \pm 0.107\) & .951 \\
    (2, 10) & (0.33, 3) & 0.16 & \(0.161 \pm 0.096\) & \(0.128 \pm 0.091\) & \(-0.022 \pm 0.080\) & \(0.088 \pm 0.080\) & \(0.150 \pm 0.094\) & \(0.223 \pm 0.112\) & \(0.405 \pm 0.156\) & \(0.426 \pm 0.129\) & .971 \\
    \midrule
    \multicolumn{12}{c}{Path Coefficients: \(\alpha = 0.4, \beta = 0.7\)} \\
    (0.33, 3) & (2, 10)  & 0.28 & \(0.287 \pm 0.161\) & \(0.281 \pm 0.161\) & \(-0.043 \pm 0.186\) & \(0.178 \pm 0.162\) & \(0.285 \pm 0.160\) & \(0.394 \pm 0.167\) & \(0.629 \pm 0.198\) & \(0.672 \pm 0.189\) & .965 \\
    (0.33, 10) & (2, 3) & 0.28 & \(0.281 \pm 0.149\) & \(0.266 \pm 0.145\) & \(-0.010 \pm 0.168\) & \(0.181 \pm 0.146\) & \(0.276 \pm 0.147\) & \(0.376 \pm 0.156\) & \(0.599 \pm 0.191\) & \(0.609 \pm 0.165\) & .955 \\
    (2, 3) & (0.33, 10)  & 0.28 & \(0.272 \pm 0.155\) & \(0.258 \pm 0.152\) & \(-0.020 \pm 0.162\) & \(0.170 \pm 0.150\) & \(0.267 \pm 0.153\) & \(0.369 \pm 0.162\) & \(0.591 \pm 0.193\) & \(0.611 \pm 0.151\) & .945 \\
    (2, 10) & (0.33, 3) & 0.28 & \(0.282 \pm 0.140\) & \(0.252 \pm 0.136\) & \(0.019 \pm 0.125\) & \(0.182 \pm 0.126\) & \(0.272 \pm 0.138\) & \(0.371 \pm 0.155\) & \(0.604 \pm 0.200\) & \(0.585 \pm 0.150\) & .950 \\
    \midrule
    \multicolumn{12}{c}{Path Coefficients: \(\alpha = 0.7, \beta = 0.4\)} \\
    (0.33, 3) & (2, 10)  & 0.28 & \(0.280 \pm 0.104\) & \(0.265 \pm 0.103\) & \(0.084 \pm 0.106\) & \(0.210 \pm 0.099\) & \(0.275 \pm 0.103\) & \(0.345 \pm 0.112\) & \(0.504 \pm 0.137\) & \(0.420 \pm 0.113\) & .958 \\
    (0.33, 10) & (2, 3) & 0.28 & \(0.277 \pm 0.104\) & \(0.252 \pm 0.103\) & \(0.081 \pm 0.088\) & \(0.200 \pm 0.094\) & \(0.269 \pm 0.103\) & \(0.345 \pm 0.116\) & \(0.523 \pm 0.150\) & \(0.443 \pm 0.113\) & .969 \\
    (2, 3) & (0.33, 10)  & 0.28 & \(0.285 \pm 0.112\) & \(0.257 \pm 0.109\) & \(0.080 \pm 0.096\) & \(0.204 \pm 0.101\) & \(0.275 \pm 0.110\) & \(0.355 \pm 0.123\) & \(0.544 \pm 0.158\) & \(0.463 \pm 0.117\) & .962 \\
    (2, 10) & (0.33, 3) & 0.28 & \(0.282 \pm 0.134\) & \(0.245 \pm 0.130\) & \(0.035 \pm 0.125\) & \(0.184 \pm 0.121\) & \(0.269 \pm 0.132\) & \(0.366 \pm 0.149\) & \(0.601 \pm 0.196\) & \(0.566 \pm 0.161\) & .963 \\
    \midrule
    \multicolumn{12}{c}{Path Coefficients: \(\alpha = 0.7, \beta = 0.7\)} \\
    (0.33, 3) & (2, 10)  & 0.49 & \(0.491 \pm 0.160\) & \(0.479 \pm 0.163\) & \(0.165 \pm 0.175\) & \(0.380 \pm 0.158\) & \(0.487 \pm 0.159\) & \(0.597 \pm 0.168\) & \(0.837 \pm 0.201\) & \(0.672 \pm 0.185\) & .964 \\
    (0.33, 10) & (2, 3) & 0.49 & \(0.492 \pm 0.159\) & \(0.470 \pm 0.156\) & \(0.190 \pm 0.153\) & \(0.383 \pm 0.149\) & \(0.485 \pm 0.157\) & \(0.593 \pm 0.171\) & \(0.838 \pm 0.213\) & \(0.648 \pm 0.159\) & .961 \\
    (2, 3) & (0.33, 10)  & 0.49 & \(0.485 \pm 0.157\) & \(0.463 \pm 0.156\) & \(0.178 \pm 0.148\) & \(0.372 \pm 0.149\) & \(0.477 \pm 0.156\) & \(0.589 \pm 0.168\) & \(0.837 \pm 0.202\) & \(0.659 \pm 0.146\) & .955 \\
    (2, 10) & (0.33, 3) & 0.49 & \(0.482 \pm 0.168\) & \(0.444 \pm 0.167\) & \(0.172 \pm 0.139\) & \(0.360 \pm 0.152\) & \(0.469 \pm 0.167\) & \(0.590 \pm 0.185\) & \(0.868 \pm 0.233\) & \(0.697 \pm 0.166\) & .949 \\
    \bottomrule
\end{tabular}
\label{tab:pr_oppo_skew}
\end{table}

We next consider cases in which \(\varepsilon^{(M)}\) and \(\varepsilon^{(Y)}\) exhibit opposing skewness and different tail-heaviness. Table~\ref{tab:pr_oppo_skew} reports the recovery results for the mediation effect under four combinations of \(\alpha\) and \(\beta\). The results broadly mirror those obtained when \(\varepsilon^{(M)}\) and \(\varepsilon^{(Y)}\) share the same distribution. The posterior means match well with the true values of the mediation effect, and coverage of the \(95\%\) credible intervals is quite good.

\vspace{-0.25cm}
\subsection{Hypothesis Testing of the Mediation Effect}\label{sec:5.2}

In this section, we conduct simulation studies to evaluate our model's ability to test the mediation effect. We generate data under both the null hypothesis (i.e., no mediation effect) and the alternative hypothesis (i.e., mediation effect exists) for each simulation scenario outlined in Section~\ref{sec:5.1}. We refer to the data generated under the null hypothesis as the "null data". 

In Bayesian hypothesis testing, the Bayes factor is often reported to provide evidence for each hypothesis. Additionally, a cutoff can be chosen to accept or reject the null hypothesis. For example, a commonly used cutoff for the Bayes factor is 10, 
indicating strong evidence to reject the null hypothesis (\citealp{jeffreys1998}). However, this choice is subjective, and the distribution of the Bayes factor depends on the prior and likelihood (see \citet{rubio2018} for an example where the Bayes factor does not even depend on the data). As a result, one can set the cutoff based on the value of the false positive rate estimated from \(1\,000\) experiments on the null data for each of the models being compared.

We also compare the statistical power of our Full Model and three sub-models with two popular frequentist methods, both of which are bootstrap-based approaches widely used in mediation analysis. Specifically, we compare our models with the OLS bootstrap and the robust bootstrap, as described in \citet{alfons2022}. The bootstrap methods are implemented using the R package \textit{Robmed} (\citealp{alfons2022b}). 

We first consider simulation scenarios in which \(\varepsilon^{(M)}\) and \(\varepsilon^{(Y)}\) are symmetric or right-skewed, following the same distribution. For each setting (defined by \(\gamma\) and \(\nu\)), we compute the true positive and false positive rates of the Full Model and three sub‑models, using a Bayes factor cutoff of 
10. They are compared with the true positive rate and the false positive rate of the OLS bootstrap method and the robust bootstrap method under the nominal \(0.05\) significance level. Results are presented in Table~\ref{tab:bf_comparison_3} in Supplementary Material~\ref{sm:add_ht}. From Table \ref{tab:bf_comparison_3}, we observe that under the Bayes factor cutoff of \(10\), the true positive rates of our Full Model and three sub-models are generally much higher than the bootstrap-based methods, although their false‑positive rates are slightly higher as well. 

To allow a more direct comparison, we estimate separate cutoff values for the Full Model and its three sub-models so that their false positive rates match those obtained with the OLS bootstrap and the robust bootstrap, respectively.  For example, if the OLS bootstrap yields a false positive rate of \(0.050\), we take the 50th-largest Bayes factor among the \(1{,}000\) null data as its cutoff. In this way, we can evaluate which model gives the highest true positive rate under the same false positive rate. The results are presented in Table~\ref{tab:bf_comparison_match}, 
from which we observe that the true positive rates generally improve when the model specification aligns with the data generation mechanism. 
The Full Model generally performs the best over the three sub-models and bootstrap-based methods when the error is both skewed and heavy-tailed, and it also performs well for skewed errors with normal tails and for errors that follow either the Student \(t\) or the normal distribution.

\setlength{\tabcolsep}{7pt}
\begin{table}[h!]
    \centering
    \caption{\textbf{Comparison of true positive rates for four Bayesian models, OLS bootstrap, and robust bootstrap at matched false positive rates when the errors follow the same centred two-piece Student \(t\) distribution}: The first two columns specify the skewness and tail parameters in the error distribution. Cutoff values of the Bayes factor are estimated from null data so that the resulting false positive rate matches that of the OLS bootstrap or the robust bootstrap under comparison. Results are based on \(1\,000\) experiments with sample size \(n = 50\). \textbf{Bold} and \underline{underlined} denote the best and second-best results, respectively.}

    \fontsize{6}{8}\selectfont
    \begin{tabular}{cc|cccccc|cccccc}
        \toprule
        \multicolumn{2}{c}{Settings} & \multicolumn{2}{c}{OLS Bootstrap}  & \multicolumn{1}{c}{Full} & \multicolumn{1}{c}{\(\gamma\)-Only} & \multicolumn{1}{c}{\(\nu\)-Only} & \multicolumn{1}{c}{Normal} & \multicolumn{2}{c}{Robust Bootstrap}  & \multicolumn{1}{c}{Full} & \multicolumn{1}{c}{\(\gamma\)-Only} & \multicolumn{1}{c}{\(\nu\)-Only} & \multicolumn{1}{c}{Normal}\\
        \cmidrule(lr){1-2} \cmidrule(lr){3-4} \cmidrule(lr){9-10}
        \(\gamma\) & \(\nu\) & TPR & FPR &  &  & &  & TPR & FPR &  &  &  &  \\
        \midrule
        \multicolumn{14}{c}{Path Coefficients: \(\alpha = 0.4, \beta = 0.4\)} \\
        0.33  & 3  & .109 & .004 & \textbf{.374} & \underline{.225} & .129 & .059 & .229 & .010 & \textbf{.435} & \underline{.368} & .228 & .111 \\ 
        0.33  & 10  & .204 & .009 & \textbf{.466} & \underline{.456} & .212 & .219 & .222 & .009 & \textbf{.466} & \underline{.456} & .212 & .219  \\ 
        0.33     & \(\infty\)  & .252 & .012 & \underline{.503} & \textbf{.516} & .294 & .276 & .227 & .010 & \underline{.472} & \textbf{.500} & .248 & .272 \\ 
        0.5     & 3  & .191 & .007 & \textbf{.413} & \underline{.310} & .177 & .111 & .354 & .019 & \textbf{.571} & \underline{.452} & .427 & .205 \\ 
        0.5   & 10  & .325 & .013 & \textbf{.522} & \underline{.473} & .383 & .355 & .322 & .010 & \textbf{.469} & \underline{.431} & .362 & .316 \\ 
        0.5   & \(\infty\)  & .399 & .016 & \underline{.557} & \textbf{.568} & .388 & .394 & .043 & .013 & \underline{.533} & \textbf{.548} & .379 & .360 \\ 
        1   & 3  & .301 & .017 & \underline{.510} & .324 & \textbf{.553} & .295 & .422 & .017 &  \underline{.510} & .324 & \textbf{.553} & .295 \\ 
        1   & 10  & .511 & .019 &  \underline{.569} & .534 & \textbf{.613} & .568 & .436 & .015 &  \underline{.523} & .475 & \textbf{.586} & .506 \\ 
        1   & \(\infty\)  & .558 & .017 & .606 & .596 &  \underline{.630} & \textbf{.639} & .422 & .012 & .553 &  \underline{.577} & .571 & \textbf{.586} \\ 
        \midrule
        \multicolumn{14}{c}{Path Coefficients: \(\alpha = 0.4, \beta = 0.7\)} \\
        0.33  & 3  & .166 & .013 & \textbf{.477} &  \underline{.377} & .201 & .106 & .253 & .008 & \textbf{.463} &  \underline{.266} & .157 & .065 \\ 
        0.33  & 10  & .285 & .025 & \textbf{.564} &  \underline{.517} & .342 & .308 & .286 & .020 & \textbf{.524} &  \underline{.491} & .294 & .290 \\ 
        0.33     & \(\infty\)  & .301 & .017 & \textbf{.542} &  \underline{.518} & .342 & .326 & .302 & .018 & \textbf{.561} &  \underline{.522} & .349 & .327 \\ 
        0.5     & 3  & .299 & .019 & \textbf{.548} &  \underline{.412} & .390 & .230 & .409 & .019 & \textbf{.548} &  \underline{.412} & .390 & .230 \\ 
        0.5   & 10  & .462 & .028 & \textbf{.649} &  \underline{.644} & .530 & .485 & .430 & .021 & \textbf{.635} &  \underline{.620} & .479 & .411 \\ 
        0.5   & \(\infty\)  & .546 & .027 & \textbf{.683} &  \underline{.682} & .582 & .589 & .492 & .019 &  \underline{.622} & \textbf{.636} & .549 & .568 \\ 
        1   & 3  & .456 & .032 & \textbf{.574} & .457 &  \underline{.561} & .431 & \textbf{.574} & .032 & \textbf{.574} & .457 &  .561 & .431 \\ 
        1   & 10  & .728 & .039 &  \underline{.739} & .707 & \textbf{.768} & .732 & .634 & .027 & \textbf{.679} & .642 & .668 &  \underline{.678} \\ 
        1   & \(\infty\)  & \textbf{.774} & .037 & .740 & .726 & .744 &  \underline{.760} & .657 & .028 &  \underline{.713} & .712 & .709 & \textbf{.737} \\ 
        \midrule
        \multicolumn{14}{c}{Path Coefficients: \(\alpha = 0.7, \beta = 0.4\)} \\
        0.33  & 3  & .232 & .011 & \textbf{.799} &  \underline{.596} & .446 & .130 & .509 & .015 & \textbf{.821} &  \underline{.679} & .547 & .142 \\ 
        0.33  & 10  & .462 & .013 &  \underline{.757} & \textbf{.795} & .512 & .486 & .492 & .011 &  \underline{.713} & \textbf{.745} & .500 & .440 \\ 
        0.33     & \(\infty\)  & .531 & .012 &  \underline{.719} & \textbf{.755} & .493 & .506 & .468 & .026 &  \underline{.831} & \textbf{.855} & .626 & .593 \\ 
        0.5     & 3  & .440 & .012 & \textbf{.762} & .540 &  \underline{.608} & .307 & .677 & .023 & \textbf{.836} & .635 &  \underline{.707} & .420 \\ 
        0.5   & 10  & .636 & .020 & \textbf{.785} &  \underline{.775} & .673 & .612 & .630 & .024 & \textbf{.803} &  \underline{.786} & .733 & .659 \\ 
        0.5   & \(\infty\)  & .726 & .033 & \textbf{.866} &  \underline{.858} & .719 & .691 & .628 &  .020 & \textbf{.795} &  \underline{.787} & .640 & .606 \\ 
        1   & 3  & .578 & .026 &  \underline{.770} & .649 & \textbf{.813} & .668 & .727 & .027 &  \underline{.770} & .651 & \textbf{.824} & .669 \\ 
        1   & 10  & .740 & .022 & .772 & .724 &  \underline{.780} & \textbf{.781} & .673 & .019 & .749 & .698 & \textbf{.772} &  \underline{.771} \\ 
        1   & \(\infty\)  & .742 & .027 & .739 & .742 & \textbf{.768} &  \underline{.749} & .628 & .022 &  \underline{.728} & .703 & \textbf{.742} & .725 \\ 
        \midrule
        \multicolumn{14}{c}{Path Coefficients: \(\alpha = 0.7, \beta = 0.7\)} \\
        0.33  & 3  & .357 & .017 & \textbf{.746} &  \underline{.655} & .478 & .307 & .564 & .031 & \textbf{.838} &  \underline{.761} & .579 & .377 \\ 
        0.33  & 10  & .599 & .025 & \textbf{.876} &  \underline{.862} & .720 & .585 & .618 & .021 & \textbf{.853} &  \underline{.845} & .659 & .562 \\  
        0.33     & \(\infty\)  & .716 & .031 &  \underline{.899} & \textbf{.907} & .737 & .742 & .672 & .031 &   \underline{.899} & \textbf{.907} & .737 & .742 \\ 
        0.5     & 3  & .557 & .026 & \textbf{.903} & .764 &  \underline{.800} & .499 & .783 & .027 & \textbf{.909} & .767 &  \underline{.800} & .501 \\ 
        0.5   & 10  & .852 & .052 & \textbf{.966} &  \underline{.955} & .917 & .882 & .853 & .039 & \textbf{.959} &  \underline{.945} & .891 & .838 \\ 
        0.5   & \(\infty\)  & .935 & .042 & \textbf{.974} &  \underline{.970} & .952 & .945 & .886 & .029 & \textbf{.959} &  \underline{.954} & .935 & .931 \\ 
        1   & 3  & .776 & .049 &  \underline{.927} & .818 & \textbf{.941} & .840 & .896 & .046 &  \underline{.917} & .804 & \textbf{.937} & .831 \\ 
        1   & 10  & .975 & .053 & .982 & .972 & \textbf{.989} &  \underline{.981} & .945 & .034 & .966 & .956 & \textbf{.978} &  \underline{.974} \\ 
        1   & \(\infty\)  & .984 & .083 & .983 &  \underline{.986} & .984 & \textbf{.989} & .944 & .046 & .970 & .970 & \textbf{.973} &  \underline{.972} \\ 
        \bottomrule
    \end{tabular}
    \label{tab:bf_comparison_match}
\end{table}

We note that, when the true \(\alpha\) and \(\beta\) are small (e.g., \(\alpha=0.4\) and \(\beta=0.4\)), the false positive rate of the bootstrap-based methods is very small. If we want to estimate the cutoff of Bayes factor of the Full Model and other three sub-models to match the false discovery rates, the estimation may be inaccurate. Readers can still refer to Table \ref{tab:bf_comparison_3} for the comparison of the models' performance, which yields consistent results.

Note that, for the results in Table~\ref{tab:bf_comparison_match} and Table~\ref{tab:bf_comparison_3}, we fix the sample size at \(n = 50\) for each simulated data set. In Supplementary Material~\ref{sm:diff_sample} we also examine how sample size affects hypothesis testing for our models and for bootstrap-based approaches. We additionally test \(n = 100\) and \(200\), keeping all other simulation settings unchanged; the results are consistent with the \(n = 50\) case.

Next, like in the parameter recovery studies, we examine four additional scenarios:  (1) \(\varepsilon^{(M)}\sim\operatorname{ctpt}(0.33,3)\), \(\varepsilon^{(Y)}\sim\operatorname{ctpt}(2,10)\); (2) \(\varepsilon^{(M)}\sim\operatorname{ctpt}(0.33,10)\), \(\varepsilon^{(Y)}\sim\operatorname{ctpt}(2,3)\); (3) \(\varepsilon^{(M)}\sim\operatorname{ctpt}(2,3)\), \(\varepsilon^{(Y)}\sim\operatorname{ctpt}(0.33,10)\); and (4) \(\varepsilon^{(M)}\sim\operatorname{ctpt}(2,10)\), \(\varepsilon^{(Y)}\sim\operatorname{ctpt}(0.33,3)\), where the \(\varepsilon^{(M)}\) and \(\varepsilon^{(Y)}\) follow different centred two-piece Student \(t\) distribution with opposing skewness and different tail-heaviness. The sample size is fixed at \(n=50\) and we re-run the experiments to compare the hypothesis testing performance of our Full Model and the three sub-models with bootstrap-based methods. The results appear in Section~\ref{sm:tp_ht}. When the false discovery rate is matched, our Full Model performs the best over all scenarios.

To further demonstrate the Full Model’s robustness, we evaluate its performance when the error terms depart from the centred two-piece Student \(t\) distribution and instead arise from other unimodal, skewed, heavy‑tailed distributions. A widely used family for modelling skewness and heavy tails is the univariate Tukey \(g\)-and-\(h\) distribution (\citealp{yan2019}). A random variable \(T\) following this distribution is obtained by transforming a standard normal variable \(Z\) by
\begin{align}
    T_{g,h}(Z)=\left(\frac{e^{gZ}-1}{g}\right)e^{hZ^{2}/2},
\end{align}
where \(g\in\mathbb{R}\) and \(h\ge 0\) are parameters. The parameter \(g\) governs skewness: \(g>0\) yields right‑skewness that increases with \(g\); \(g<0\) yields left‑skewness that intensifies as \(g\) becomes more negative. When \(g=0\), the distribution is symmetric. The parameter \(h\) controls tail-heaviness; both tails become heavier as \(h\) increases. In the special case \(g=h=0\), \(T\) reduces to a standard normal random variable.

We examine the hypothesis testing performance of the Full Model and three sub-models when the error terms follow the univariate Tukey \(g\)-and-\(h\) distribution with (1) \(g = 0.2,\; h = 0\); (2) \(g = 0.2,\; h = 0.2\); (3) \(g = 0.5,\; h = 0\); and (4) \(g = 0.5,\; h = 0.2\). Errors are simulated from these distributions and centred by subtracting their theoretical means. For convenience, we assume both \(\varepsilon^{(M)}\) and \(\varepsilon^{(Y)}\) follow the same distribution. We fix the sample size at 50, and all other simulation settings remain unchanged. Similarly, we compare the true positive and false positive rates of the three sub-models, the OLS bootstrap, and the robust bootstrap. As before, we estimate the cutoff values for the Full Model and the three sub-models so that their false positive rates match those of the OLS and robust bootstraps, and we report the results in Table~\ref{tab:bf_comparison_match_tukey}. Table~\ref{tab:bf_comparison_tukey} presents the results with the Bayes factor cutoff fixed at 10.

\setlength{\tabcolsep}{7pt}
\begin{table}[h!]
    \centering
    \caption{\textbf{Comparison of true positive rates for four models at matched false positive rates when the errors follow the same Tukey \(g\)-and-\(h\) distribution}: The first two columns specify the \(g\) and \(h\) parameters in the error distribution. Cutoff values of the Bayes factor are estimated from null data so that the resulting false positive rate matches that of the OLS bootstrap or the robust bootstrap under comparison. Results are based on \(1\,000\) experiments with sample size \(n = 50\). \textbf{Bold} and \underline{underlined} denote the best and second-best results, respectively.}
    \fontsize{6}{8}\selectfont
    \begin{tabular}{cc|cccccc|cccccc}
        \toprule
        \multicolumn{2}{c}{Settings} & \multicolumn{2}{c}{OLS Bootstrap}  & \multicolumn{1}{c}{Full} & \multicolumn{1}{c}{\(\gamma\)-Only} & \multicolumn{1}{c}{\(\nu\)-Only} & \multicolumn{1}{c}{Normal} & \multicolumn{2}{c}{Robust Bootstrap}  & \multicolumn{1}{c}{Full} & \multicolumn{1}{c}{\(\gamma\)-Only} & \multicolumn{1}{c}{\(\nu\)-Only} & \multicolumn{1}{c}{Normal}\\
        \cmidrule(lr){1-2} \cmidrule(lr){3-4} \cmidrule(lr){9-10}
        \(g\) & \(h\) & TPR & FPR &  &  & &  & TPR & FPR &  &  &  &  \\
        \midrule
        \multicolumn{14}{c}{Path Coefficients: \(\alpha = 0.4, \beta = 0.4\)} \\
        0.2  & 0  & .583 & .015 & \underline{.642} & \textbf{.645} & .612 & .566 & .488 & .012 & \textbf{.610} & \underline{.598} & .580 & .548 \\ 
        0.2  & 0.2  & .364 & .011 & \textbf{.560} & .394 & \underline{.541} & .298 & .514 & .009 & \underline{.523} & .359 & \textbf{.524} & .253 \\ 
        0.5  & 0  & .485 & .013 & \underline{.782} & \textbf{.786} & .641 & .479 & .597 & .010 & \textbf{.776} & \underline{.703} & .625 & .435 \\ 
        0.5  & 0.2  & .311 & .014 & \textbf{.689} & .449 & \underline{.615} & .238 & .572 & .013 & \textbf{.687} & .414 & \underline{.609} & .238 \\ 
        \midrule
        \multicolumn{14}{c}{Path Coefficients: \(\alpha = 0.4, \beta = 0.7\)} \\
        0.2  & 0  & .746 & .033 & \textbf{.803} & \textbf{.803} & .781 & .779 & .669 & .025 & \textbf{.782} & \underline{.772} & .755 & .753 \\ 
        0.2  & 0.2  & .496 & .025 & \underline{.662} & .519 & \textbf{.689} & .515 & .610 & .023 & \underline{.649} & .503 & \textbf{.666} & .495 \\ 
        0.5  & 0  & .632 & .028 & \textbf{.866} & \underline{.859} & .750 & .603 & .723 & .028 & \textbf{.866} & \underline{.859} & .750 & .603 \\ 
        0.5  & 0.2 & .423 & .025 & \textbf{.671} & .541 & \underline{.655} & .377 & .668 & .029 & \textbf{.697} & .549 & \underline{.676} & .402 \\ 
        \midrule
        \multicolumn{14}{c}{Path Coefficients: \(\alpha = 0.7, \beta = 0.4\)} \\
        0.2  & 0  & .768 & .025 & \textbf{.796} & \underline{.780} & .779 & .753 & .671 & .020 & \textbf{.768} & .756 & \underline{.765} & .703 \\ 
        0.2  & 0.2  & .635 & .019 & \textbf{.816} & .683 & \underline{.786} & .608 & .772 & .023 & \textbf{.823} & .694 & \underline{.798} & .631 \\ 
        0.5 & 0 & .733 & .020 & \textbf{.917} & \underline{.878} & .778 & .676 & .816 & .020 & \textbf{.917} & \underline{.878} & .778 & .676 \\ 
        0.5 & 0.2  & .570 & .021 & \textbf{.911} & .720 & \underline{.837} & .471 & .854 & .028 & \textbf{.923} & .753 & \underline{.865} & .547 \\ 
        \midrule
        \multicolumn{14}{c}{Path Coefficients: \(\alpha = 0.7, \beta = 0.7\)} \\
        0.2  & 0  & .978 & .062 & .979 & .981 & \underline{.984} & \textbf{.987} & .945 & .037 & .972 & \textbf{.975} & \underline{.974} & \underline{.974} \\ 
        0.2  & 0.2  & .831 & .043 & \underline{.943} & .869 & \textbf{.957} & .868 & .919 & .040 & \underline{.942} & .864 & \textbf{.953} & .867 \\ 
        0.5 & 0  & .921 & .052 & \textbf{.993} & \underline{.991} & .977 & .944 & .959 & .044 & \textbf{.991} & \underline{.990} & .973 & .942 \\ 
        0.5 & 0.2  & .729 & .037 & \textbf{.960} & .873 & \underline{.946} & .748 & .945 & .044 & \textbf{.972} & .882 & \underline{.950} & .762 \\ 
        \bottomrule
    \end{tabular}
    \label{tab:bf_comparison_match_tukey}
\end{table}

From Tables \ref{tab:bf_comparison_match_tukey} and \ref{tab:bf_comparison_tukey}, we observe that the Full Model performs well under a variety of simulation scenarios. This suggests that the Full Model is robust to model misspecification and can be applied when the error terms follow other unimodal, skewed and/or heavy-tailed distributions. It also generally outperforms the bootstrap-based methods. As expected, each sub‑model performs best when the data‑generating mechanism is closest to its own specification. For example, when \(g = 0.2\) and \(h = 0\)—yielding a skewed distribution with moderately light tails (noting that \(g\) also influences tail-heaviness)—the \(\gamma\)-Only Model sometimes achieves the best performance. In practice, the Bayes factor can be used to select the most appropriate model from the four models, as demonstrated in the next section.

\vspace{-0.5cm}
\section{Case Study}\label{sec:6}
\vspace{-0.25cm}
As a case study, we analyse data from the National Longitudinal Survey of Youth 1979 cohort (NLSY79). The NLSY79 survey is sponsored and directed by the U.S. Bureau of Labor Statistics and managed by the Center for Human Resource Research (CHRR) at The Ohio State University. Interviews are conducted by the National Opinion Research Center (NORC) at the University of Chicago. We apply our Bayesian mediation model to investigate the mediation effect of the home environment on the relationship between mothers' education level and children's mathematical achievement.

In our dataset of 371 samples, three variables are included: mothers' education level (ME), home environment (HE), and children's mathematical achievement (MATH). The ME variable takes integer values from 3 to 18, with higher values representing higher levels of maternal education. The HE variable, derived from observational data, reflects how mothers engage with and support their children at home (\citealp{bradley1984}). In the dataset, HE takes integer values ranging from 0 to 9, where higher values indicate greater support for the children. The MATH variable measures children's mathematical achievement, taking integer values from 0 to 42, where higher values represent higher levels of achievement.

We hypothesise that mothers with higher educational attainment create more nurturing home environments—characterised by positive interactions with their children—which, in turn, foster the development of children's mathematical skills. In this mediation analysis, ME serves as the independent variable, the HE acts as the mediator, and MATH is the outcome. We examine the following two linear regressions:
\begin{align}
    &\text{HE}_i = \beta_0^{(M)} + \alpha \text{ME}_i + \sigma^{(M)}\varepsilon_i^{(M)}, \label{eq:case_1}\\
    &\text{MATH}_i = \beta_0^{(Y)} + \beta \text{HE}_i + \tau \text{ME}_i + \sigma^{(Y)}\varepsilon_i^{(Y)}. \label{eq:case_2}
\end{align}
We treat the discrete scores MATH and HE as continuous, implicitly assuming that their distributions can be well approximated by a continuous distribution. While this is defensible for MATH, the assumption is stronger for HE, which has only ten distinct values, and should therefore be interpreted with appropriate caution.

Before applying our method, we first apply the OLS method to fit the two linear regression models described in Equations (\ref{eq:case_1}) and (\ref{eq:case_2}). We then examine the residuals to assess potential skewness and heavy tails in the data. In Supplementary Material~\ref{sm:add_case}, Figure \ref{fig:case1} presents the Normal Q–Q plot of the residuals from (\ref{eq:case_1}), and Figure \ref{fig:case2} shows the corresponding plot for (\ref{eq:case_2}). We observe that the OLS residuals from (\ref{eq:case_1}) exhibit clear left skewness, while the residuals from (\ref{eq:case_2}) show right skewness. In both cases, the residuals appear unimodal. Further calculations show that the Fisher's moment coefficient of skewness and the excess kurtosis 
of the residuals from (\ref{eq:case_1}) are -0.679 and 0.097, respectively, whereas the skewness and kurtosis of the residuals from (\ref{eq:case_2}) are 1.253 and 5.423, respectively. These results suggest the need for modelling both skewness and heavy tails in the error terms.

We then examine the posterior estimates of \(\gamma\) and \(\nu\) from our Full Model in (\ref{eq:case_1}). A single Markov chain of length \(L = 30\,000\) is generated, with the first \(20\%\) discarded as burn-in samples. The posterior mean estimate of \(\gamma\) is \(0.593\), with a 95\% credible interval of \((0.488, 0.689)\). The posterior mean estimate of \(\nu\) is \(129.680\), with a 95\% credible interval of \((19.228, 397.627)\). The kernel density estimates of the posterior densities of \(\gamma\) and \(\nu\), with their prior densities, are shown in Figure \ref{fig:full1} in Supplementary Material~\ref{sm:add_case}. These results indicate that the errors are left skewed with tails not much heavier than the normal tails.

\begin{table}[t]
  \centering
  \caption{Log-transformed Bayes factors for the four candidate models under two data-generating equations. Each row corresponds to the model in the numerator.}
  \captionsetup{justification = centering}

  \begin{subtable}[t]{0.45\textwidth}
    \centering
    \caption{Equation~(\ref{eq:case_1})}
    \label{tab:case1}
    \footnotesize                               
    \begin{tabular}{lcccc}
      \toprule
              & Full & $\gamma$-Only & $\nu$-Only & Normal \\ \midrule
      Full     & 0   & -0.369 & 17.461 & 17.429 \\ 
      $\gamma$-Only &     & 0      & 17.830 & 17.798 \\ 
      $\nu$-Only    &     &        & 0      & -0.032 \\ 
      Normal        &     &        &        & 0      \\ \bottomrule
    \end{tabular}
  \end{subtable}
  \hfill
  \begin{subtable}[t]{0.45\textwidth}
    \centering
    \caption{Equation~(\ref{eq:case_2})}
    \label{tab:case2}
    \footnotesize                               
    \begin{tabular}{lcccc}
      \toprule
              & Full & $\gamma$-Only & $\nu$-Only & Normal \\ \midrule
      Full     & 0   & 12.008 & 2.296 & 22.966 \\ 
      $\gamma$-Only &     & 0      & -9.712 & 10.958 \\ 
      $\nu$-Only    &     &        & 0      & 20.671 \\ 
      Normal        &     &        &        & 0      \\ \bottomrule
    \end{tabular}
  \end{subtable}

  \label{fig:bf_tables_side_by_side}
\end{table}

Next, for Equation (\ref{eq:case_1}), we also fit the \(\gamma\)-Only Model, \(\nu\)-Only Model, and the Normal Model. A formal Bayesian model comparison was conducted using the Bayes factor computed by the bridge sampler, and the results are shown in Table \ref{tab:case1}, where the Bayes factors are log-transformed. Each row corresponds to the model in the numerator of the Bayes factor. The results indicate that the \(\gamma\)-Only Model performs best for Equation (\ref{eq:case_1}) with the Full Model a close second, while both the \(\nu\)-Only Model and the Normal Model perform substantially worse. 
This can be expected because the \(\nu\)-Only Model and the Normal Model fail to capture the strong skewness of the error terms.

\begin{figure}[t]
  \centering
  
  \begin{subfigure}[b]{0.32\textwidth}
    \centering
    \includegraphics[width=\linewidth]{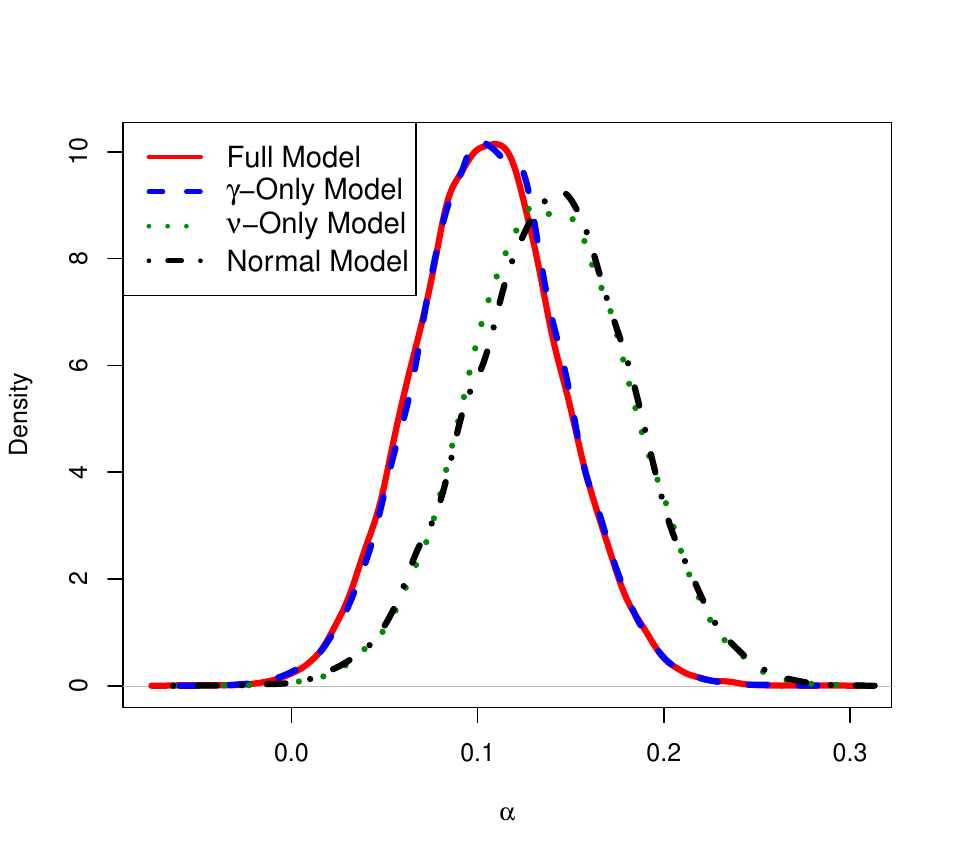}
    \caption{Posterior of $\alpha$}
    \label{fig:alpha_1}
  \end{subfigure}
  \hfill
  \begin{subfigure}[b]{0.32\textwidth}
    \centering
    \includegraphics[width=\linewidth]{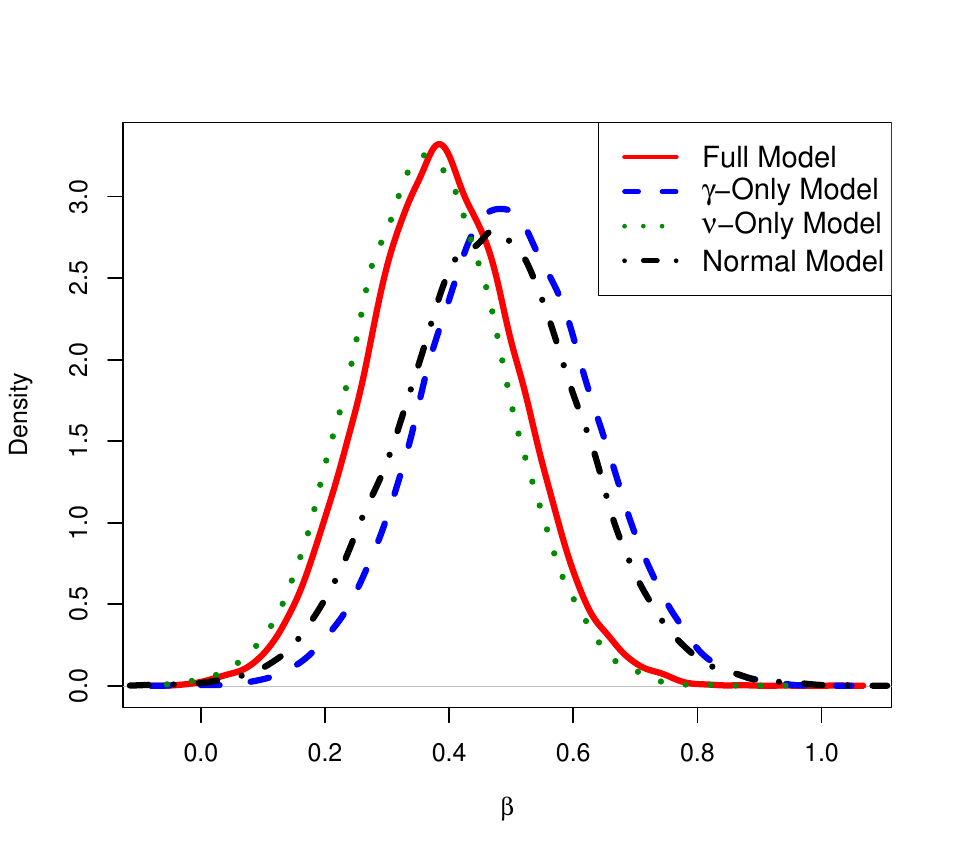}
    \caption{Posterior of $\beta$}
    \label{fig:beta}
  \end{subfigure}
  \hfill
  \begin{subfigure}[b]{0.32\textwidth}
    \centering
    \includegraphics[width=\linewidth]{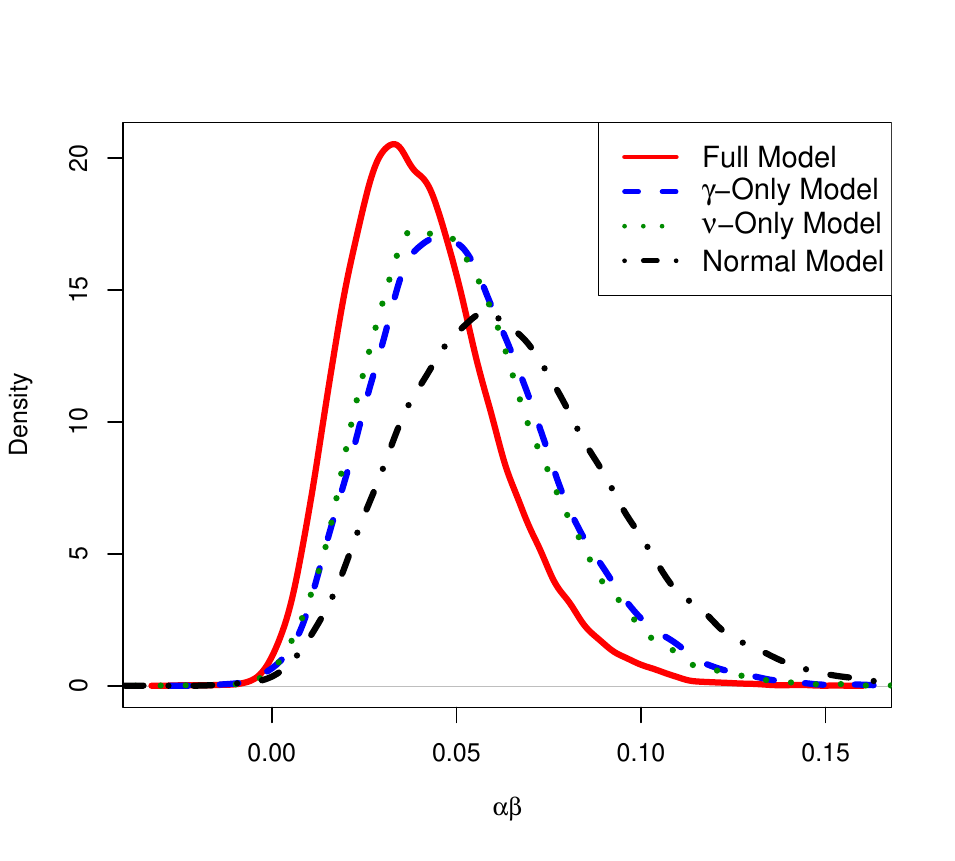}
    \caption{Posterior of mediation effect}
    \label{fig:med}
  \end{subfigure}

  \caption{Kernel density estimates of key posterior quantities.}
  \label{fig:posteriors_side_by_side}
\end{figure}

For the Full Model, the posterior mean estimate of \(\alpha\) is \(0.105\) with a 95\% credible interval of \((0.031, 0.182)\). For the \(\gamma\)-Only Model, the posterior mean estimate of \(\alpha\) is \(0.106\) with a 95\% credible interval of \((0.031, 0.181)\). Figure~\ref{fig:alpha_1} shows the kernel density estimates of the posterior distributions of $\alpha$ under the four models. The inference with the Full Model and the $\gamma$-Only Model is almost identical, which supports the robustness of the Full Model. However, the corresponding estimates from the $\nu$-Only and Normal Models deviate from the former two, since those models do not account for skewness. 

Furthermore, we examine the posterior estimates of \(\gamma\) and \(\nu\) in Equation (\ref{eq:case_2}) from our Full Model. The posterior mean estimate of \(\gamma\) is \(1.310\) with a 95\% credible interval of \((1.104, 1.546)\). The posterior mean estimate of \(\nu\) is \(6.602\) with a 95\% credible interval of \((3.719, 12.320)\). The kernel density estimates of the posterior densities of \(\gamma\) and \(\nu\) are shown in Figure \ref{fig:full2} in Supplementary Material~\ref{sm:add_case}. This results indicate that the errors are right skewed with heavy tails.

Additionally, for Equation (\ref{eq:case_2}), we fit the \(\gamma\)-Only Model, \(\nu\)-Only Model, and the Normal Model. We also conduct a formal Bayesian model comparison using the Bayes factor. Table \ref{tab:case2} shows the model comparison results, where the Bayes factors are log-transformed. Each row corresponds to the model in the numerator of the Bayes factor. The results suggest that the Full Model performs best for Equation (\ref{eq:case_2}), while both the \(\gamma\)-Only Model and the Normal Model perform considerably worse than the Full Model and the \(\nu\)-Only Model. This is  because the \(\gamma\)-Only Model and the Normal Model fail to capture the heavy tails of the error terms. As there is also evidence of skewness, the Full Model is the overall winner here. 

For the Full Model, the posterior mean estimate of \(\beta\) is \(0.388\) with a 95\% credible interval of \((0.151, 0.631)\). For the \(\nu\)-Only Model, the posterior mean estimate of \(\beta\) is \(0.365\) with a 95\% credible interval of \((0.129, 0.607)\). Figure~\ref{fig:beta} shows the kernel density estimates of the posterior distributions of $\beta$ under the four models. The two curves from the Full Model and the $\nu$-Only Model are similar, while the estimates from the $\gamma$-only and Normal models deviate from the former two, as these latter models 
do not accommodate  heavy tails.

Finally, we use our Full Model's fit for (\ref{eq:case_1}) and (\ref{eq:case_2}) to estimate and test the mediation effect, i.e., testing \(H_0:\alpha\beta=0\) against \(H_1:\alpha\beta\neq0\). The posterior mean estimate of the mediation effect \(\alpha\beta\) is \(0.041\) with a 95\% credible interval of \((0.008, 0.087)\). Figure~\ref{fig:med} presents the estimated posterior density of the mediation effect. We can see that the posterior density is more concentrated around the mode for the Full Model, which may indicate that the Full Model reduces uncertainty more effectively by modelling the skewness and heavy tails. Assuming the paths are independent, the Bayes factor for the mediation effect is \(BF^{med} = 13.021\), given by the Full Model. 
Thus, the mediation effect is very likely to exist. 

Based on the posterior estimation of \(\alpha\) and \(\beta\), we conclude that the data provide strong evidence that mothers with higher educational attainment create more nurturing home environments, which in turn improve the development of children's mathematical skills. This conclusion aligns with existing analyses from a frequentist perspective (\citealp{zhang2013}).

\vspace{-0.5cm}
\section{Discussion}\label{sec:7}
\vspace{-0.25cm}
In our study, we introduced a flexible Bayesian mediation model designed to address skewness and heavy tails in error distributions. We used the two-piece distribution of \citet{fernandez1998} as a starting point but noted that directly applying it to mediation analysis was less natural because of its non‑zero mean. Consequently, we proposed the centred two-piece Student \(t\) distribution for the error terms, exploring the conditions for the existence of the posterior distribution and its moments using a class of standard and convenient improper priors. This class of improper priors has attractive properties, is widely applicable and flexible enough to serve as a default prior. 

We conducted extensive simulation studies to evaluate our method's performance. The parameter recovery studies showed that our method effectively recovers parameters under various conditions of skewness and tail-heaviness, although they indicated that, as expected, a larger sample size may be necessary for accurate estimation of the tail parameter \(\nu\). We also ran simulations to assess the statistical power of our model in testing the mediation effect, comparing it to popular bootstrap-based methods. These results demonstrated that by formally allowing for skewness and tail-heaviness of error terms through a flexible parametric distribution, our model can provide greater statistical power to detect mediation effects. Additionally, we illustrated how to implement our model in practice through a case study examining the mediation effect of home environment on children's mathematical achievement, highlighting the model's ability to conduct meaningful inference in the context of non-normal data.

We note that, although we investigate only the simplest mediation model, our method can be naturally extended to more complicated mediation models such as the moderated mediation. We have developed the \texttt{FlexBayesMed} \textsf{R} package to implement our method. \texttt{FlexBayesMed} enables Bayesian mean regression and mediation analysis with error terms following normal distributions, Student \(t\) distributions, centred two-piece normal distributions, and the most general centred two-piece Student \(t\) distributions. The package also implements model selection and hypothesis tests via Bayes factors. \texttt{FlexBayesMed}, together with supplementary code and a user manual, is available at \url{https://github.com/Zongyu-Li/FlexBayesMed}.

The present study may have several limitations. First, in the parameter recovery and hypothesis testing simulations, we fixed the direct effect \(\tau\) 
and the scale parameters \(\sigma^{(M)}\) and \(\sigma^{(Y)}\); more extensive simulations are needed to assess how varying these values influences performance. We also considered only a limited set of distributional families in the data-generation process; future studies should explore a broader range of error terms drawn from other classes of unimodal skewed and/or heavy-tailed distributions. Second, the current model is confined to unimodal errors and is not suited to multimodal errors. Third, we note again that the current model, along with previous Bayesian mediation models which use normal distribution to model error terms, are essentially for continuous mediator and outcome variables, while in fields such as psychology, ordinal mediators and outcomes are common. Thus, the model is misspecified if either the mediator or the outcome variable is discrete, as noted in the case study. Further studies should evaluate the impact of modelling discrete mediators and outcomes in Bayesian mediation, or propose new distributions to deal with the discrete cases.  
Finally, we focused primarily on the use of improper priors, without investigating how incorporating substantive knowledge into informative priors might improve estimation and hypothesis testing. We leave this for future exploration.

Nevertheless, we believe that our approach offers both methodological novelty and considerable practical value. By explicitly modelling skewed and heavy‑tailed errors within a Bayesian mediation framework, our approach improves estimation accuracy and hypothesis-testing power. The availability of our method in the open-source \texttt{FlexBayesMed} package further enhances its accessibility, enabling applied researchers to incorporate flexible error structures into mediation analyses with minimal additional effort. 

\section*{Acknowledgments}
Zhang is supported by the US Department of Education (R305D210023) and Notre Dame Global.

\clearpage
\appendix       
\begin{center}
    \Large \bfseries
    Supplementary Material for ``Modelling Skewed and Heavy-Tailed Errors in Bayesian Mediation Analysis''
\end{center}

\setcounter{section}{0}
\renewcommand{\thesection}{\Alph{section}}
\renewcommand{\thesubsection}{\thesection.\arabic{subsection}}

\setcounter{figure}{0}
\renewcommand{\thefigure}{S.\arabic{figure}}

\setcounter{table}{0}
\renewcommand{\thetable}{S.\arabic{table}}

\setcounter{equation}{0}
\renewcommand{\theequation}{A\arabic{equation}}

\section{Proofs of Main Results} \label{sec:A}
\subsection{Proof of Propositions 1 \& 2}  \label{sec:A.1}
For \(\gamma \in \mathbb{R}_+\) and \(\nu > 0\), we have:
\begin{flalign*}
    \int_{0}^{\infty} \varepsilon f_\nu\left(\frac{\varepsilon}{\gamma}\right) d\varepsilon &= \frac{\Gamma\left(\frac{\nu+1}{2}\right)}{\sqrt{\pi \nu} \Gamma\left(\frac{\nu}{2}\right)} \int_{0}^{\infty} \varepsilon \left(1 + \frac{\varepsilon^2}{\nu\gamma^2}\right)^{-\frac{\nu+1}{2}} d\varepsilon \\
    &= \frac{\Gamma\left(\frac{\nu+1}{2}\right)}{2\sqrt{\pi \nu} \Gamma\left(\frac{\nu}{2}\right)} \int_{0}^{\infty} \left(1 + \frac{x}{\nu\gamma^2}\right)^{-\frac{\nu+1}{2}} dx \text{   (substituting } x = \varepsilon^2 \text{)} \\
    &= \frac{\Gamma\left(\frac{\nu+1}{2}\right)}{2\sqrt{\pi \nu} \Gamma\left(\frac{\nu}{2}\right)} \left[-\frac{2\nu\gamma^2}{\nu-1}\left(1 + \frac{x}{\nu\gamma^2}\right)^{-\frac{\nu-1}{2}}\right]_{0}^{\infty} \\
    &= \frac{\Gamma\left(\frac{\nu+1}{2}\right)}{\sqrt{\pi \nu} \Gamma\left(\frac{\nu}{2}\right)} \cdot \frac{\nu\gamma^2}{\nu-1}.
\end{flalign*}

Similarly, for the negative part:
\begin{flalign*}
    \int_{-\infty}^{0} \varepsilon f_\nu\left(\gamma \varepsilon\right) d\varepsilon &= \frac{\Gamma\left(\frac{\nu+1}{2}\right)}{\sqrt{\pi \nu} \Gamma\left(\frac{\nu}{2}\right)} \int_{-\infty}^{0} \varepsilon \left(1 + \frac{\gamma^2 \varepsilon^2}{\nu}\right)^{-\frac{\nu+1}{2}} d\varepsilon \\
    &= \frac{\Gamma\left(\frac{\nu+1}{2}\right)}{2\sqrt{\pi \nu} \Gamma\left(\frac{\nu}{2}\right)} \int_{-\infty}^{0} \left(1 + \frac{\gamma^2 x}{\nu}\right)^{-\frac{\nu+1}{2}} dx \text{   (substituting } x = \varepsilon^2 \text{)} \\
    &= \frac{\Gamma\left(\frac{\nu+1}{2}\right)}{2\sqrt{\pi \nu} \Gamma\left(\frac{\nu}{2}\right)} \left[-\frac{2\nu}{(\nu-1)\gamma^2}\left(1 + \frac{\gamma^2 x}{\nu}\right)^{-\frac{\nu-1}{2}}\right]_{-\infty}^{0} \\
    &= -\frac{\Gamma\left(\frac{\nu+1}{2}\right)}{\sqrt{\pi \nu} \Gamma\left(\frac{\nu}{2}\right)} \cdot \frac{\nu}{(\nu-1)\gamma^2}.
\end{flalign*}

Thus, the mean of the distributions given by (\ref{eq:tpd}) is:
\begin{flalign*}
    m(\nu,\gamma) &= \frac{2}{\gamma + \frac{1}{\gamma}} \Bigg[\int_{0}^{\infty} \varepsilon f_\nu\left(\frac{\varepsilon}{\gamma}\right) d\varepsilon + \int_{-\infty}^{0} \varepsilon f_\nu\left(\gamma \varepsilon\right) d\varepsilon \Bigg] \\
    &= \frac{2\nu\Gamma\left(\frac{\nu+1}{2}\right)}{\sqrt{\pi \nu}(\nu-1)\Gamma\left(\frac{\nu}{2}\right)}\left(\gamma - \frac{1}{\gamma}\right), \quad \gamma \in \mathbb{R}_+, \nu > 0.
\end{flalign*}

As \(\nu \rightarrow \infty\), using Stirling's approximation for the gamma function, we have:
\begin{flalign*}
    m^{\text{normal}}(\gamma) = \left(\gamma-\frac{1}{\gamma}\right)\lim_{\nu \rightarrow \infty} \frac{2\nu\Gamma\left(\frac{\nu+1}{2}\right)}{\sqrt{\pi \nu}(\nu-1)\Gamma\left(\frac{\nu}{2}\right)}=\sqrt{\frac{2}{\pi}}\left(\gamma-\frac{1}{\gamma}\right)
\end{flalign*}

When \(\nu>2\), we consider
\begin{flalign*}
    \mathbb{E}[\varepsilon^2\mid \gamma,\nu]&=\frac{2}{\gamma + \frac{1}{\gamma}}\left[\int^\infty_0\varepsilon^2f_\nu\left(\frac{\varepsilon}{\gamma}\right)d\,\varepsilon + \int^0_{-\infty}\varepsilon^2f_\nu(\gamma\varepsilon)d\,\varepsilon\right]\\
    & =\frac{1}{\gamma + \frac{1}{\gamma}}\left[\gamma^3\int^\infty_02x^2f_\nu(x)dx+\frac{1}{\gamma^3}\int^0_{-\infty}2x^2f_\nu(x)dx\right]\\
    &=\frac{1}{\gamma + \frac{1}{\gamma}}\left(\gamma^3+\frac{1}{\gamma^3}\right)\frac{\nu}{\nu-2}\\
    &= \frac{\nu}{\nu-2}\left(\gamma^2 - 1 + \frac{1}{\gamma^2}\right)
\end{flalign*}
where the third equality holds because \(\int^\infty_02x^2f_\nu(x)dx=\int^0_{-\infty}2x^2f_\nu(x)dx\) and both are equal to the variance of a Student \textit{t} distribution with degrees of freedom \(\nu\), which is \(\frac{\nu}{\nu-2}\).

Then, the variance is given by:
\begin{flalign*}
    \text{Var}(\varepsilon \mid \gamma, \nu) = \frac{\nu}{\nu-2}\left(\gamma^2 - 1 + \frac{1}{\gamma^2}\right) - m(\gamma, \nu)^2, \quad \gamma \in \mathbb{R}_+, \nu > 2.
\end{flalign*}

By letting \(\nu \rightarrow \infty\), we have
\begin{flalign*}
    \text{Var}^{\text{normal}}(\varepsilon|\gamma) &= \lim_{\nu \rightarrow \infty}\left[\frac{\nu}{\nu-2}\left(\gamma^2 - 1 + \frac{1}{\gamma^2}\right) - m(\gamma, \nu)^2\right] \\
    &= \gamma^2 - 1 + \frac{1}{\gamma^2} - m^{\text{normal}}(\gamma)^2
\end{flalign*}

\subsection{Proof of Proposition 3}
Let \(Y\) be a random variable following the uncentred two-piece distribution with the Student \textit{t} distribution as specified in (\ref{eq:tpd}) with \(\nu>1\). Then, for a random variable \(X\sim\text{ctpt}(\gamma,\nu)\), we have \(X=Y-m(\gamma,\nu)\).

Then, for a positive integer \(r\), we have
\begin{align*}
    \mathbb{E}\left[X^r\right]=\mathbb{E}\left[\left(Y-m(\gamma,\nu)\right)^r\right]=\sum^r_{k=0}{r\choose k}\left[-m(\gamma,\nu)\right]^{r-k}\mathbb{E}\left[Y^k\right]
\end{align*}
where
\begin{align*}
    \mathbb{E}\left[Y^k\right]=\frac{2}{\gamma+\frac{1}{\gamma}}\left[\int^\infty_0y^k f_\nu\left(\frac{y}{\gamma}\right)dy\;+\;\int^0_{-\infty}y^kf_\nu(\gamma y)dy\right]
\end{align*}

For the positive part, we have
\begin{align*}
    \int^\infty_0y^k f_\nu\left(\frac{y}{\gamma}\right)dy=\gamma^{k+1}\int^\infty_0u^kf_\nu(u)du
\end{align*}

For the negative part, we have
\begin{align*}
    \int^0_{-\infty}y^kf_\nu(\gamma y)dy=\gamma^{-k-1}\int^0_{-\infty}v^kf_{\nu}(v)dv=\gamma^{-k-1}(-1)^k\int^\infty_0u^kf_\nu(u)du
\end{align*}
where the second equality holds due to the change of variable \(u:=-v\) and the fact that \(f_\nu(\cdot)\) is even.

Hence,
\begin{align}
    \mathbb{E}\left[Y^k\right]&=\frac{2}{\gamma+\frac{1}{\gamma}}\left[\gamma^{k+1}+(-1)^k\gamma^{-k-1}\right]\int^\infty_0u^kf_\nu(u)du  \notag\\
    &=\frac{2}{\gamma+\frac{1}{\gamma}}\left[\gamma^{k+1}+(-1)^k\gamma^{-k-1}\right]\frac{\nu^{\frac{k}{2}}}{2\sqrt{\pi}}\frac{\Gamma\left(\frac{k+1}{2}\right)\Gamma\left(\frac{\nu-k}{2}\right)}{\Gamma\left(\frac{\nu}{2}\right)} \label{eq:ctpt_moments}
\end{align}
where the second equality is obtained from the standard beta function integrals, which requires \(\nu>r\).

Plugging (\ref{eq:ctpt_moments}) into the definition of Fisher's moment coefficient of skewness, we can obtain the expression in (\ref{eq:ctpt}). Also, the limiting case as \(\nu\rightarrow\infty\) can be obtained using the same approach as the proofs in Section~\ref{sec:A.1}.

\subsection{Proof of Theorem 1} 

The proof here follows a similar approach to the \textit{Proof of Theorem 1} in \citet{fernandez1998}.
We start by observing that \(\mathbb{E}\left[\sigma^r \prod_{j=1}^k |\beta_j|^{r_j} \mid y_1, \dots, y_n\right] < \infty\) if and only if
\begin{flalign}\label{eq:moments_gen}
    I = \int_{\mathbb{R}^k \times \mathbb{R}_+ \times (a, b) \times (2, \infty)} \left(\prod_{j=1}^k |\beta_j|^{r_j}\right) \sigma^{r-1} \left\{\prod_{i=1}^n p(y_i \mid \bm{\beta}, \sigma, \gamma, \nu)\right\} dP_{\bm{\beta}} d\sigma dP_\gamma dP_\nu 
\end{flalign}
is finite, where
\begin{flalign}
    & \begin{aligned}
    &p(y_i \mid \bm{\beta}, \sigma, \gamma, \nu) = \frac{2\sigma^{-1}}{\gamma + \frac{1}{\gamma}} \Bigg[f_\nu\left(\frac{y_i - g_i(\bm{\beta}) + \sigma m(\gamma, \nu)}{\sigma \gamma}\right) I_{[0, +\infty)}\left(\frac{y_i - g_i(\bm{\beta}) + \sigma m(\gamma, \nu)}{\sigma}\right) \\
    & \quad + f_\nu\left(\frac{y_i - g_i(\bm{\beta}) + \sigma m(\gamma, \nu)}{\sigma / \gamma}\right) I_{(-\infty, 0)}\left(\frac{y_i - g_i(\bm{\beta}) + \sigma m(\gamma, \nu)}{\sigma}\right) \Bigg], \quad \gamma \in \mathbb{R}_+,\, \nu > 2. \label{eq:2}
    \end{aligned} 
\end{flalign}

Next, we note  that \(f_\nu(s) = f_\nu(|s|)\) is decreasing in \(|s|\). Therefore, for the sampling density \(p(y_i \mid \bm{\beta}, \sigma, \gamma, \nu)\), we have the following upper and lower bounds:
\begin{flalign}
    \frac{2\sigma^{-1}}{\gamma + \frac{1}{\gamma}} f_\nu\left(\frac{|y_i - g_i(\bm{\beta}) + \sigma m(\gamma, \nu)|}{\sigma h^{\text{lower}}(\gamma)}\right)\leq p(y_i \mid \bm{\beta}, \sigma, \gamma, \nu) \leq \frac{2\sigma^{-1}}{\gamma + \frac{1}{\gamma}} f_\nu\left(\frac{|y_i - g_i(\bm{\beta}) + \sigma m(\gamma, \nu)|}{\sigma h^{\text{upper}}(\gamma)}\right), \label{eq:3}
\end{flalign}
where \(h^{\text{upper}}(\gamma) = \max\{\gamma, \frac{1}{\gamma}\}\) and  \(h^{\text{lower}}(\gamma) = \min\{\gamma, \frac{1}{\gamma}\}\).

For any positive bounded function \(h(\cdot)\), the ratio
\begin{flalign}
    R_i = \frac{f_\nu\left(\frac{|y_i - g_i(\bm{\beta}) + \sigma m(\gamma, \nu)|}{\sigma h(\gamma)}\right)}{f_\nu\left(\frac{|y_i - g_i(\bm{\beta})|}{\sigma h(\gamma)}\right)}
\end{flalign}
is bounded, due to the tail behaviour of \(f_\nu(\cdot)\) and the fact that:
\begin{flalign}
    -\frac{|m(\gamma,\nu)|}{h(\gamma)}\leq \frac{|y_i - g_i(\bm{\beta}) + \sigma m(\gamma, \nu)|}{\sigma h(\gamma)} - \frac{|y_i - g_i(\bm{\beta})|}{\sigma h(\gamma)} \leq \frac{|m(\gamma,\nu)|}{h(\gamma)}
\end{flalign}
where both \(m(\gamma,\nu)\) and \(h(\gamma)\) are bounded.
Thus, an upper bound of the sampling density (\ref{eq:2}) is:
\begin{flalign}   C_i^{\text{upper}}\frac{2\sigma^{-1}}{\gamma + \frac{1}{\gamma}} f_\nu\left(\frac{|y_i - g_i(\bm{\beta})|}{\sigma h^{\text{upper}}(\gamma)}\right),
\end{flalign}
where \(C_i^{\text{upper}}\) is a positive constant.

Similarly, a lower bound of the sampling density (\ref{eq:2}) is:
\begin{flalign}
    C_i^{\text{lower}} \frac{2\sigma^{-1}}{\gamma + \frac{1}{\gamma}} f_\nu\left(\frac{|y_i - g_i(\bm{\beta})|}{\sigma h^{\text{lower}}(\gamma)}\right),
\end{flalign}
where \(C_i^{\text{lower}}\) is a positive constant.

Substituting each of these bounds into (\ref{eq:moments_gen}), we obtain:
\begin{flalign}
    &\begin{aligned}
    \text{Constant} \times \int_{\mathbb{R}^k \times \mathbb{R}_+ \times (a, b) \times (2, \infty)} \frac{1}{\left(\gamma + \frac{1}{\gamma}\right)^n} & \left(\prod_{j=1}^k |\beta_j|^{r_j}\right) \sigma^{r-n-1} \times \\
    & \left\{\prod_{i=1}^n f_\nu\left(\frac{|y_i - g_i(\bm{\beta})|}{\sigma h(\gamma)}\right)\right\} dP_{\bm{\beta}} d\sigma dP_{\gamma} dP_\nu. 
    \end{aligned} 
\end{flalign}

Transforming from \(\sigma\) to \(\theta = \sigma h(\gamma)\) and applying Fubini's theorem, the upper and lower bounds of \(I\) are:
\begin{flalign}
    &\begin{aligned}
    \text{Constant} \times \int_{(a, b)} \frac{h(\gamma)^{n-r}}{\left(\gamma + \frac{1}{\gamma}\right)^n} dP_{\gamma} \times \int_{\mathbb{R}^k \times \mathbb{R}_+ \times (2, \infty)} & \left(\prod_{j=1}^k |\beta_j|^{r_j}\right) \theta^{r-1-n} \times  \\
    & \left\{\prod_{i=1}^n f_\nu\left(\frac{|y_i - g_i(\bm{\beta})|}{\theta}\right)\right\} dP_{\bm{\beta}} d\theta dP_\nu, \label{eq:10}
    \end{aligned} 
\end{flalign}
with \(h(\gamma)\) as defined in (\ref{eq:3}).

For the first integral in (\ref{eq:10}), it is strictly positive if \(h(\gamma) = \min\{\gamma, \frac{1}{\gamma}\}\). When \(h(\gamma) = \max\{\gamma, \frac{1}{\gamma}\}\), this integral is less than 1.

For the second integral in (\ref{eq:10}), it is finite if and only if \(\mathbb{E}\left[\sigma^r \prod_{j=1}^k |\beta_j|^{r_j} \mid y_1, \dots, y_n\right]\) \( < \infty\) holds under \(\gamma = 1\).

Therefore, Theorem 1 holds.

\subsection{Proof of Theorem 2}
The proof follows a similar approach to the proof of Theorem 1. All we need to show is that the ratio
\begin{flalign*}
    R_i = \frac{f\left(\frac{|y_i - g_i(\bm{\beta}) + \sigma m^{\text{normal}}(\gamma)|}{\sigma h(\gamma)}\right)}{f\left(\frac{|y_i - g_i(\bm{\beta})|}{\sigma h(\gamma) }\right)}, \quad \text{where } f(x) = \frac{1}{\sqrt{2\pi}} \exp\left(-\frac{x^2}{2}\right),
\end{flalign*}
is bounded, with \(h(\gamma)\) as defined in (3).

We have:
\begin{flalign*}
    -\frac{|m^{\text{normal}}(\gamma)|}{h(\gamma)}\leq \frac{|y_i - g_i(\bm{\beta}) + \sigma m^{\text{normal}}(\gamma)|}{\sigma h(\gamma)} - \frac{|y_i - g_i(\bm{\beta})|}{\sigma h(\gamma)} \leq \frac{|m^{\text{normal}}(\gamma)|}{h(\gamma)}
\end{flalign*}
which is bounded  since both \(h(\gamma)\) and \(m^{\text{normal}}(\gamma)\) are bounded for \(\gamma \in (a,b)\), where \(0<a\leq 1\) and \(1\leq b < \infty\). 

Therefore, Theorem 2 holds.

\subsection{Proof of Theorems 3, 4 and 5}
To prove Theorem 3, we only need to consider the case where \(\gamma = 1\) based on Theorem 1. The sampling distribution for a single observation becomes:
\begin{flalign*}
    p(y_i \mid \bm{\beta}, \sigma, \nu) = \sigma^{-1} \Bigg[ f_\nu\left(\frac{y_i - g_i(\bm{\beta})}{\sigma}\right) & I_{[0, +\infty)}\left(\frac{y_i - g_i(\bm{\beta})}{\sigma}\right) +  \\
    & f_\nu\left(\frac{y_i - g_i(\bm{\beta})}{\sigma}\right) I_{(-\infty, 0)}\left(\frac{y_i - g_i(\bm{\beta})}{\sigma}\right) \Bigg], \quad \nu > 2 
\end{flalign*}

The remainder of the proof follows exactly as in the \textit{Proof of Theorem 2} in \citet{fernandez1998}.

To prove Theorem 4 and 5, similarly, we only need to consider the case where \(\gamma = 1\). The remainder of the proofs follows exactly as \textit{Proof of Theorem 3(b)} and \textit{Proof of Theorem 4}, respectively, in \citet{fernandez1998}

\subsection{Proof of Theorem 6 with Discussions on the Equivalence}
Assume that the general regression model in~\eqref{eq:general_reg} includes a free intercept \(\beta_0\)
\begin{align}
    y_i = \beta_0 + \tilde{g}_i(\bm{\beta}^*) + \sigma\varepsilon_i,\quad i=1,\dots,n \label{eq:appendix_centred_model}
\end{align}
where the parameter vector \(\bm{\beta}^*\) excludes \(\beta_0\) and \(\varepsilon_i \stackrel{\text{i.i.d.}}{\sim} \mathrm{ctpt}(\gamma,\nu)\). This model gives the likelihood
\begin{flalign}
  & \begin{aligned}
    &p(y_1, y_2, \dots, y_n|\beta_0,\bm{\beta}^*, \sigma, \gamma, \nu) = \\
    &\frac{2\sigma^{-n}}{\left(\gamma + \frac{1}{\gamma}\right)^n} \prod_{i=1}^n \Bigg[f_\nu\left(\frac{y_i - \beta_0- \tilde{g}_i(\bm{\beta}^*) + \sigma m(\gamma,\nu)}{\sigma \gamma}\right) I_{[0,+\infty)}\left(\frac{y_i - \beta_0- \tilde{g}_i(\bm{\beta}^*) + \sigma m(\gamma,\nu)}{\sigma}\right) + \\
    & f_\nu\left(\frac{y_i - \beta_0- \tilde{g}_i(\bm{\beta}^*) + \sigma m(\gamma,\nu)}{\sigma/\gamma}\right) I_{(-\infty,0)}\left(\frac{y_i - \beta_0- \tilde{g}_i(\bm{\beta}^*) + \sigma m(\gamma,\nu)}{\sigma}\right)\Bigg], \quad \gamma \in \mathbb{R}_+; \; \nu > 2. 
  \end{aligned} 
\end{flalign}

Under the prior specification in~\eqref{eq:prior_general}, and imposing that the prior \(p(\beta_0,\bm{\beta}^*)\) is shift-invariant in \(\beta_0\), the full posterior distribution is
\begin{align}
    p(\beta_0,\bm{\beta}^*, \sigma, \gamma, \nu\mid y_1, y_2, \dots, y_n)\propto p(y_1, y_2, \dots, y_n|\beta_0,\bm{\beta}^*, \sigma, \gamma, \nu)p(\beta_0,\bm{\beta}^*)p(\sigma)p(\gamma)p(\nu)
\end{align}

Next, perform the change of variable \(\beta_0^{*} = \beta_0 - \sigma m(\gamma,\nu)\). The Jacobian of this transformation equals \(1\), so the posterior becomes
\begin{align}
    p(\beta_0^*,\bm{\beta}^*, \sigma, \gamma, \nu\mid y_1, y_2, \dots, y_n)&\propto \tilde{p}(y_1, y_2, \dots, y_n|\beta_0^*,\bm{\beta}^*, \sigma, \gamma, \nu)p(\beta_0^*+\sigma m(\gamma,\nu),\bm{\beta}^*)p(\sigma)p(\gamma)p(\nu)\\
    &= \tilde{p}(y_1, y_2, \dots, y_n|\beta_0^*,\bm{\beta}^*, \sigma, \gamma, \nu)p(\beta_0^*,\bm{\beta}^*)p(\sigma)p(\gamma)p(\nu) \label{eq:equivalent_posterior}
\end{align}
where the equality follows from the shift invariance of \(p(\beta_0,\bm{\beta}^*)\) in \(\beta_0\). The resulting likelihood in (\ref{eq:equivalent_posterior}) is:
\begin{flalign}
  & \begin{aligned}
    \tilde{p}(y_1, y_2, &\dots, y_n|\beta_0^*,\bm{\beta}^*, \sigma, \gamma, \nu) = \\
    &\frac{2\sigma^{-n}}{\left(\gamma + \frac{1}{\gamma}\right)^n} \prod_{i=1}^n \Bigg[f_\nu\left(\frac{y_i - \beta_0^*- \tilde{g}_i(\bm{\beta}^*)}{\sigma \gamma}\right) I_{[0,+\infty)}\left(\frac{y_i - \beta_0^*- \tilde{g}_i(\bm{\beta}^*)}{\sigma}\right) + \\
    & f_\nu\left(\frac{y_i - \beta_0^* - \tilde{g}_i(\bm{\beta}^*)}{\sigma/\gamma}\right) I_{(-\infty,0)}\left(\frac{y_i - \beta^*_0- \tilde{g}_i(\bm{\beta}^*)}{\sigma}\right)\Bigg], \quad \gamma \in \mathbb{R}_+; \; \nu > 2. 
  \end{aligned} 
\end{flalign}
and this is exactly the likelihood of the model given by
\begin{align}
    y_i = \beta_0^* + \tilde{g}_i(\bm{\beta}^*) + \sigma\varepsilon^*_i,\quad i= 1,\dots,n \label{eq:appendix_uncentred_model}
\end{align}
where \(\beta^*_0 \equiv \beta_0 - \sigma m(\gamma,\nu)\), and \(\varepsilon_i^*=\varepsilon_i+m(\gamma,\nu)\), which follows the original (uncentred) two-piece Student \(t\) distribution introduced in (\ref{eq:tpd}).

Thus, under the common prior specification
\(P_{\{\bm{\beta},\sigma,\gamma,\nu\}}
      = P_{\bm{\beta}}\times P_{\sigma}\times P_{\gamma}\times P_{\nu}\) in~\eqref{eq:prior_general},
the centred model in~\eqref{eq:appendix_centred_model} and the uncentred model in~\eqref{eq:appendix_uncentred_model} induce posterior distributions differing only by a one-to-one reparameterisation of the intercept.

\newpage
\section{Additional Figures and Tables} \label{sm:additional_plots}

\subsection{Fisher’s Moment Coefficient of Skewness for CTPT} \label{sm:skewness_plot}

\begin{figure}[h!]
    \centering
    \includegraphics[width=0.65\textwidth]{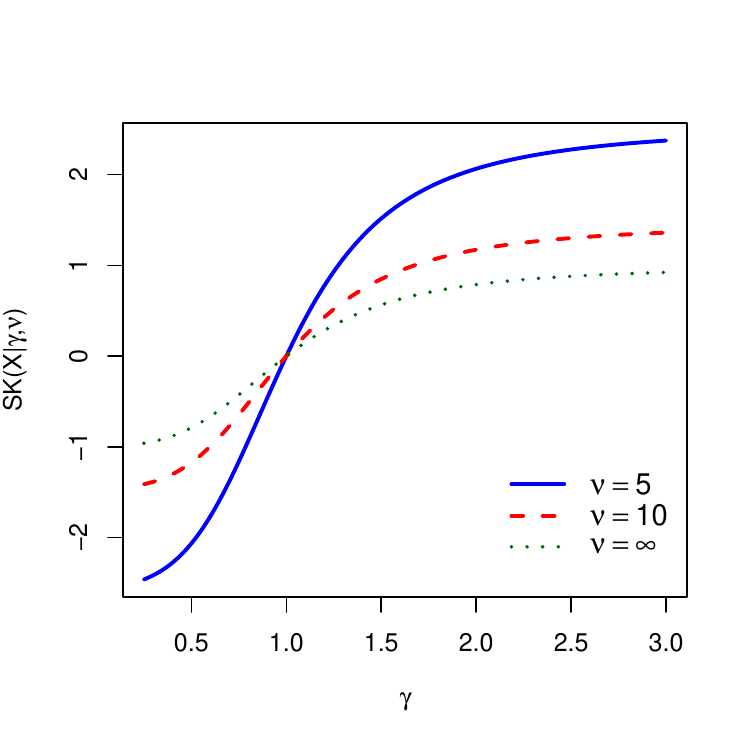}
    \caption{Fisher’s moment coefficient of skewness for the centred two-piece Student $t$ distribution as a function of the skewness parameter $\gamma$ at $\nu = 5,\;10,$ and $\infty$.}
    \label{fig:skewness}
\end{figure}

\subsection{Additional Parameter Recovery Results}

\subsubsection{Mediation Effect under Different \(\alpha\) and \(\beta\)} \label{sm:med_diff}

Table~\ref{tab:pr_med} presents the parameter recovery result of the mediation effect under \(\alpha=\beta=0.4\), with \(\varepsilon^{(M)}\) and \(\varepsilon^{(Y)}\) sampled from the same distribution. Similarly, Table~\ref{tab:pr_med_diff} here presents the parameter recovery result of the mediation effect under (1) \(\alpha=0.4\) and \(\beta=0.7\); (2) \(\alpha=0.7\) and \(\beta=0.4\); (3) \(\alpha=0.7\) and \(\beta=0.7\), again with errors \(\varepsilon^{(M)}\) and \(\varepsilon^{(Y)}\) sampled from the same distribution.

\setlength{\tabcolsep}{2.1pt}
\begin{table}[t]
\centering
\fontsize{6}{10}\selectfont
\caption{\textbf{Additional parameter recovery results for the mediation effect.} The "True \(\alpha\beta\)" is the product of (1) \(\alpha=0.4\) and \(\beta=0.7\); (2) \(\alpha=0.7\) and \(\beta=0.4\); (3) \(\alpha=0.7\) and \(\beta=0.7\). It presents the means ($\pm$ one standard deviation) of posterior means, modes, 2.5th, 25th, 50th, 75th, and 97.5th percentiles, the length of the \(95\%\) credible intervals for \(\alpha\beta\), and the coverage rate of the true value by the credible intervals across \(1\,000\) experiments with sample size \(n = 50\).}

\begin{tabular}{*{12}{c}}
    \toprule
    \textbf{\(\gamma\)} & \textbf{\(\nu\)} & \textbf{True \(\alpha\beta\)} & \textbf{Mean} & \textbf{Mode} & \textbf{2.5th} & \textbf{25th} & \textbf{50th} & \textbf{75th} & \textbf{97.5th} & \textbf{Length} & \textbf{Rate} \\
    \midrule
    \multicolumn{12}{c}{Path Coefficients: \(\alpha = 0.4, \beta = 0.7\)} \\
    0.33 & 3  & 0.28 & \(0.280 \pm 0.151\) & \(0.269 \pm 0.148\) & \(-0.048 \pm 0.178\) & \(0.170 \pm 0.149\) & \(0.276 \pm 0.148\) & \(0.386 \pm 0.157\) & \(0.625 \pm 0.196\) & \(0.673 \pm 0.191\) & .966 \\
    0.33 & 10 & 0.28 & \(0.281 \pm 0.148\) & \(0.268 \pm 0.145\) & \(-0.008 \pm 0.160\) & \(0.182 \pm 0.142\) & \(0.277 \pm 0.146\) & \(0.376 \pm 0.156\) & \(0.595 \pm 0.193\) & \(0.604 \pm 0.166\) & .949 \\
    0.33 & \(\infty\) & 0.28 & \(0.275 \pm 0.140\) & \(0.260 \pm 0.135\) & \(0.000 \pm 0.153\) & \(0.180 \pm 0.134\) & \(0.270 \pm 0.137\) & \(0.365 \pm 0.148\) & \(0.577 \pm 0.185\) & \(0.576 \pm 0.155\) & .955 \\
    0.5 & 3  & 0.28 & \(0.283 \pm 0.143\) & \(0.268 \pm 0.143\) & \(-0.004 \pm 0.149\) & \(0.182 \pm 0.137\) & \(0.278 \pm 0.142\) & \(0.379 \pm 0.152\) & \(0.596 \pm 0.183\) & \(0.599 \pm 0.154\) & .961 \\
    0.5 & 10 & 0.28 & \(0.282 \pm 0.134\) & \(0.266 \pm 0.133\) & \(0.032 \pm 0.130\) & \(0.192 \pm 0.126\) & \(0.277 \pm 0.133\) & \(0.367 \pm 0.144\) & \(0.562 \pm 0.172\) & \(0.530 \pm 0.130\) & .942 \\
    0.5 & \(\infty\) & 0.28 & \(0.281 \pm 0.124\) & \(0.261 \pm 0.123\) & \(0.046 \pm 0.114\) & \(0.194 \pm 0.113\) & \(0.274 \pm 0.123\) & \(0.361 \pm 0.136\) & \(0.552 \pm 0.169\) & \(0.506 \pm 0.130\) & .949 \\
    1 & 3  & 0.28 & \(0.284 \pm 0.148\) & \(0.262 \pm 0.144\) & \(0.011 \pm 0.141\) & \(0.183 \pm 0.138\) & \(0.277 \pm 0.146\) & \(0.378 \pm 0.159\) & \(0.599 \pm 0.192\) & \(0.589 \pm 0.133\) & .959 \\
    1 & 10 & 0.28 & \(0.291 \pm 0.133\) & \(0.266 \pm 0.131\) & \(0.053 \pm 0.115\) & \(0.200 \pm 0.121\) & \(0.283 \pm 0.132\) & \(0.374 \pm 0.145\) & \(0.577 \pm 0.174\) & \(0.524 \pm 0.110\) & .954 \\
    1 & \(\infty\) & 0.28 & \(0.278 \pm 0.125\) & \(0.253 \pm 0.124\) & \(0.058 \pm 0.100\) & \(0.192 \pm 0.112\) & \(0.270 \pm 0.124\) & \(0.355 \pm 0.138\) & \(0.544 \pm 0.168\) & \(0.486 \pm 0.107\) & .943 \\
    2 & 3  & 0.28 & \(0.281 \pm 0.149\) & \(0.267 \pm 0.146\) & \(-0.013 \pm 0.154\) & \(0.178 \pm 0.142\) & \(0.277 \pm 0.147\) & \(0.380 \pm 0.157\) & \(0.601 \pm 0.189\) & \(0.614 \pm 0.152\) & .957 \\
    2 & 10 & 0.28 & \(0.279 \pm 0.135\) & \(0.261 \pm 0.133\) & \(0.022 \pm 0.129\) & \(0.186 \pm 0.126\) & \(0.273 \pm 0.134\) & \(0.366 \pm 0.146\) & \(0.571 \pm 0.176\) & \(0.548 \pm 0.125\) & .959 \\
    2 & \(\infty\) & 0.28 & \(0.280 \pm 0.123\) & \(0.260 \pm 0.120\) & \(0.039 \pm 0.117\) & \(0.191 \pm 0.113\) & \(0.273 \pm 0.121\) & \(0.362 \pm 0.133\) & \(0.559 \pm 0.164\) & \(0.520 \pm 0.122\) & .953 \\
    3 & 3  & 0.28 & \(0.283 \pm 0.164\) & \(0.272 \pm 0.159\) & \(-0.075 \pm 0.186\) & \(0.163 \pm 0.161\) & \(0.279 \pm 0.161\) & \(0.400 \pm 0.170\) & \(0.662 \pm 0.206\) & \(0.736 \pm 0.183\) & .969 \\
    3 & 10 & 0.28 & \(0.285 \pm 0.157\) & \(0.267 \pm 0.149\) & \(-0.023 \pm 0.160\) & \(0.177 \pm 0.147\) & \(0.279 \pm 0.154\) & \(0.387 \pm 0.169\) & \(0.625 \pm 0.212\) & \(0.648 \pm 0.163\) & .961 \\
    3 & \(\infty\) & 0.28 & \(0.288 \pm 0.143\) & \(0.269 \pm 0.136\) & \(-0.009 \pm 0.149\) & \(0.183 \pm 0.135\) & \(0.281 \pm 0.140\) & \(0.386 \pm 0.152\) & \(0.618 \pm 0.190\) & \(0.627 \pm 0.149\) & .971 \\
    \midrule
    \multicolumn{12}{c}{Path Coefficients: \(\alpha = 0.7, \beta = 0.4\)} \\
    0.33 & 3  & 0.28 & \(0.280 \pm 0.105\) & \(0.266 \pm 0.107\) & \(0.068 \pm 0.098\) & \(0.203 \pm 0.098\) & \(0.275 \pm 0.105\) & \(0.352 \pm 0.116\) & \(0.521 \pm 0.148\) & \(0.453 \pm 0.128\) & .967 \\
    0.33 & 10 & 0.28 & \(0.286 \pm 0.105\) & \(0.268 \pm 0.105\) & \(0.080 \pm 0.084\) & \(0.209 \pm 0.093\) & \(0.280 \pm 0.104\) & \(0.357 \pm 0.119\) & \(0.530 \pm 0.158\) & \(0.450 \pm 0.122\) & .962 \\
    0.33 & \(\infty\) & 0.28 & \(0.279 \pm 0.103\) & \(0.260 \pm 0.103\) & \(0.078 \pm 0.084\) & \(0.202 \pm 0.093\) & \(0.272 \pm 0.102\) & \(0.347 \pm 0.115\) & \(0.519 \pm 0.150\) & \(0.441 \pm 0.114\) & .966 \\
    0.5 & 3  & 0.28 & \(0.279 \pm 0.112\) & \(0.260 \pm 0.113\) & \(0.074 \pm 0.091\) & \(0.201 \pm 0.100\) & \(0.272 \pm 0.112\) & \(0.350 \pm 0.125\) & \(0.522 \pm 0.161\) & \(0.447 \pm 0.126\) & .948 \\
    0.5 & 10 & 0.28 & \(0.282 \pm 0.111\) & \(0.261 \pm 0.112\) & \(0.079 \pm 0.088\) & \(0.204 \pm 0.100\) & \(0.275 \pm 0.111\) & \(0.352 \pm 0.124\) & \(0.525 \pm 0.155\) & \(0.445 \pm 0.109\) & .947 \\
    0.5 & \(\infty\) & 0.28 & \(0.279 \pm 0.107\) & \(0.259 \pm 0.108\) & \(0.081 \pm 0.084\) & \(0.203 \pm 0.096\) & \(0.272 \pm 0.106\) & \(0.348 \pm 0.118\) & \(0.518 \pm 0.147\) & \(0.437 \pm 0.100\) & .954 \\
    1 & 3  & 0.28 & \(0.284 \pm 0.124\) & \(0.255 \pm 0.123\) & \(0.063 \pm 0.099\) & \(0.196 \pm 0.110\) & \(0.274 \pm 0.123\) & \(0.361 \pm 0.139\) & \(0.560 \pm 0.176\) & \(0.497 \pm 0.128\) & .950 \\
    1 & 10 & 0.28 & \(0.280 \pm 0.125\) & \(0.253 \pm 0.123\) & \(0.058 \pm 0.102\) & \(0.193 \pm 0.112\) & \(0.271 \pm 0.124\) & \(0.357 \pm 0.138\) & \(0.551 \pm 0.169\) & \(0.493 \pm 0.111\) & .942 \\
    1 & \(\infty\) & 0.28 & \(0.280 \pm 0.132\) & \(0.255 \pm 0.130\) & \(0.059 \pm 0.107\) & \(0.194 \pm 0.118\) & \(0.272 \pm 0.131\) & \(0.357 \pm 0.145\) & \(0.547 \pm 0.176\) & \(0.487 \pm 0.111\) & .930 \\
    2 & 3  & 0.28 & \(0.280 \pm 0.109\) & \(0.259 \pm 0.108\) & \(0.073 \pm 0.089\) & \(0.200 \pm 0.098\) & \(0.272 \pm 0.108\) & \(0.352 \pm 0.122\) & \(0.528 \pm 0.155\) & \(0.455 \pm 0.116\) & .966 \\
    2 & 10 & 0.28 & \(0.280 \pm 0.112\) & \(0.256 \pm 0.111\) & \(0.074 \pm 0.085\) & \(0.200 \pm 0.099\) & \(0.272 \pm 0.112\) & \(0.352 \pm 0.126\) & \(0.531 \pm 0.159\) & \(0.457 \pm 0.109\) & .953 \\
    2 & \(\infty\) & 0.28 & \(0.279 \pm 0.111\) & \(0.256 \pm 0.109\) & \(0.075 \pm 0.086\) & \(0.199 \pm 0.098\) & \(0.271 \pm 0.110\) & \(0.351 \pm 0.124\) & \(0.530 \pm 0.156\) & \(0.455 \pm 0.106\) & .958 \\
    3 & 3  & 0.28 & \(0.285 \pm 0.106\) & \(0.264 \pm 0.105\) & \(0.061 \pm 0.094\) & \(0.201 \pm 0.096\) & \(0.278 \pm 0.105\) & \(0.361 \pm 0.118\) & \(0.546 \pm 0.155\) & \(0.485 \pm 0.122\) & .971 \\
    3 & 10 & 0.28 & \(0.281 \pm 0.112\) & \(0.257 \pm 0.110\) & \(0.065 \pm 0.086\) & \(0.197 \pm 0.098\) & \(0.273 \pm 0.111\) & \(0.356 \pm 0.127\) & \(0.545 \pm 0.168\) & \(0.480 \pm 0.125\) & .962 \\
    3 & \(\infty\) & 0.28 & \(0.276 \pm 0.111\) & \(0.250 \pm 0.109\) & \(0.063 \pm 0.084\) & \(0.192 \pm 0.097\) & \(0.267 \pm 0.110\) & \(0.350 \pm 0.125\) & \(0.540 \pm 0.163\) & \(0.477 \pm 0.115\) & .970 \\
    \midrule
    \multicolumn{12}{c}{Path Coefficients: \(\alpha = 0.7, \beta = 0.7\)} \\
    0.33 & 3   & 0.49 & \(0.490 \pm 0.160\) & \(0.476 \pm 0.165\) & \(0.152 \pm 0.173\) & \(0.373 \pm 0.157\) & \(0.485 \pm 0.160\) & \(0.601 \pm 0.169\) & \(0.855 \pm 0.206\) & \(0.703 \pm 0.188\) & .969 \\
    0.33 & 10  & 0.49 & \(0.490 \pm 0.158\) & \(0.473 \pm 0.158\) & \(0.186 \pm 0.150\) & \(0.382 \pm 0.148\) & \(0.484 \pm 0.157\) & \(0.591 \pm 0.170\) & \(0.829 \pm 0.208\) & \(0.643 \pm 0.160\) & .957 \\
    0.33 & \(\infty\) & 0.49 & \(0.494 \pm 0.152\) & \(0.475 \pm 0.151\) & \(0.195 \pm 0.144\) & \(0.385 \pm 0.142\) & \(0.487 \pm 0.150\) & \(0.594 \pm 0.163\) & \(0.831 \pm 0.203\) & \(0.636 \pm 0.154\) & .971 \\
    0.5 & 3   & 0.49 & \(0.492 \pm 0.158\) & \(0.472 \pm 0.160\) & \(0.186 \pm 0.148\) & \(0.380 \pm 0.149\) & \(0.485 \pm 0.158\) & \(0.596 \pm 0.171\) & \(0.834 \pm 0.206\) & \(0.648 \pm 0.158\) & .959 \\
    0.5 & 10  & 0.49 & \(0.492 \pm 0.146\) & \(0.471 \pm 0.147\) & \(0.217 \pm 0.126\) & \(0.389 \pm 0.136\) & \(0.485 \pm 0.146\) & \(0.588 \pm 0.158\) & \(0.812 \pm 0.189\) & \(0.595 \pm 0.130\) & .960 \\
    0.5 & \(\infty\) & 0.49 & \(0.498 \pm 0.141\) & \(0.476 \pm 0.143\) & \(0.231 \pm 0.117\) & \(0.397 \pm 0.130\) & \(0.490 \pm 0.141\) & \(0.590 \pm 0.153\) & \(0.808 \pm 0.184\) & \(0.577 \pm 0.121\) & .957 \\
    1 & 3   & 0.49 & \(0.488 \pm 0.169\) & \(0.459 \pm 0.168\) & \(0.185 \pm 0.140\) & \(0.372 \pm 0.155\) & \(0.478 \pm 0.168\) & \(0.593 \pm 0.184\) & \(0.846 \pm 0.223\) & \(0.662 \pm 0.142\) & .948 \\
    1 & 10  & 0.49 & \(0.490 \pm 0.157\) & \(0.460 \pm 0.157\) & \(0.212 \pm 0.118\) & \(0.380 \pm 0.142\) & \(0.480 \pm 0.157\) & \(0.588 \pm 0.173\) & \(0.826 \pm 0.207\) & \(0.614 \pm 0.116\) & .949 \\
    1 & \(\infty\) & 0.49 & \(0.488 \pm 0.159\) & \(0.459 \pm 0.160\) & \(0.220 \pm 0.120\) & \(0.382 \pm 0.145\) & \(0.478 \pm 0.160\) & \(0.583 \pm 0.175\) & \(0.811 \pm 0.206\) & \(0.591 \pm 0.108\) & .937 \\
    2 & 3   & 0.49 & \(0.483 \pm 0.162\) & \(0.462 \pm 0.163\) & \(0.172 \pm 0.150\) & \(0.369 \pm 0.153\) & \(0.476 \pm 0.161\) & \(0.589 \pm 0.173\) & \(0.834 \pm 0.205\) & \(0.662 \pm 0.145\) & .956 \\
    2 & 10  & 0.49 & \(0.491 \pm 0.149\) & \(0.467 \pm 0.149\) & \(0.209 \pm 0.126\) & \(0.384 \pm 0.138\) & \(0.483 \pm 0.149\) & \(0.588 \pm 0.162\) & \(0.817 \pm 0.193\) & \(0.608 \pm 0.122\) & .954 \\
    2 & \(\infty\) & 0.49 & \(0.483 \pm 0.144\) & \(0.459 \pm 0.143\) & \(0.212 \pm 0.116\) & \(0.379 \pm 0.131\) & \(0.474 \pm 0.143\) & \(0.577 \pm 0.157\) & \(0.801 \pm 0.191\) & \(0.590 \pm 0.118\) & .954 \\
    3 & 3   & 0.49 & \(0.489 \pm 0.167\) & \(0.474 \pm 0.162\) & \(0.130 \pm 0.176\) & \(0.364 \pm 0.160\) & \(0.484 \pm 0.164\) & \(0.609 \pm 0.176\) & \(0.878 \pm 0.216\) & \(0.748 \pm 0.177\) & .968 \\
    3 & 10  & 0.49 & \(0.489 \pm 0.161\) & \(0.468 \pm 0.155\) & \(0.162 \pm 0.151\) & \(0.370 \pm 0.149\) & \(0.482 \pm 0.159\) & \(0.600 \pm 0.174\) & \(0.859 \pm 0.217\) & \(0.697 \pm 0.161\) & .973 \\
    3 & \(\infty\) & 0.49 & \(0.490 \pm 0.158\) & \(0.468 \pm 0.154\) & \(0.178 \pm 0.144\) & \(0.375 \pm 0.147\) & \(0.482 \pm 0.156\) & \(0.596 \pm 0.169\) & \(0.848 \pm 0.207\) & \(0.670 \pm 0.138\) & .964 \\
    \bottomrule
\end{tabular}
\label{tab:pr_med_diff}
\end{table}

\clearpage
\newpage

\subsubsection{Parameters \(\alpha\) and \(\beta\)} \label{sm:alpha_beta}

For simplicity, we only present the results where \(\alpha=\beta=0.4\) and \(\{\varepsilon_i^{(M)}\}_{i=1}^{n}\) and \(\{\varepsilon_i^{(Y)}\}_{i=1}^{n}\) share the same distribution. Tables~\ref{tab:pr_alpha} and \ref{tab:pr_beta} report the parameter recovery results for \(\alpha\) and \(\beta\), respectively.

\setlength{\tabcolsep}{2.3pt}
\begin{table}[h!]
\centering
\fontsize{6}{10}\selectfont
\caption{\textbf{Parameter recovery results for \(\alpha\).} Path coefficients \(\alpha\) and \(\beta\) are both fixed at 0.4, and \(\{\varepsilon_i^{(M)}\}_{i=1}^{n}\) and \(\{\varepsilon_i^{(Y)}\}_{i=1}^{n}\) are drawn from the same distribution. It presents the means ($\pm$ one standard deviation) of posterior means, modes, 2.5th, 25th, 50th, 75th, and 97.5th percentiles, the length of the \(95\%\) credible intervals for \(\alpha\), and the coverage rate of the true value by the credible intervals across \(1\,000\) experiments with sample size \(n = 50\).}
\begin{tabular}{*{12}{c}}
    \toprule
    \textbf{\(\gamma\)} & \textbf{\(\nu\)} & \textbf{True \(\alpha\)} & \textbf{Mean} & \textbf{Mode} & \textbf{2.5th} & \textbf{25th} & \textbf{50th} & \textbf{75th} & \textbf{97.5th} & \textbf{Length} & \textbf{Rate} \\
    \midrule
    0.33 & 3  & 0.4 & \(0.413 \pm 0.243\) & \(0.415 \pm 0.252\) & \(-0.055 \pm 0.269\) & \(0.262 \pm 0.244\) & \(0.413 \pm 0.243\) & \(0.564 \pm 0.252\) & \(0.879 \pm 0.288\) & \(0.934 \pm 0.255\) & .955 \\
    0.33 & 10 & 0.4 & \(0.401 \pm 0.196\) & \(0.399 \pm 0.202\) & \(-0.007 \pm 0.226\) & \(0.269 \pm 0.200\) & \(0.400 \pm 0.196\) & \(0.532 \pm 0.199\) & \(0.811 \pm 0.224\) & \(0.818 \pm 0.203\) & .957 \\
    0.33 & \(\infty\) & 0.4 & \(0.405 \pm 0.187\) & \(0.405 \pm 0.189\) & \(0.023 \pm 0.216\) & \(0.281 \pm 0.188\) & \(0.405 \pm 0.186\) & \(0.529 \pm 0.192\) & \(0.791 \pm 0.221\) & \(0.768 \pm 0.192\) & .946 \\
    0.5 & 3  & 0.4 & \(0.396 \pm 0.206\) & \(0.393 \pm 0.212\) & \(-0.014 \pm 0.224\) & \(0.259 \pm 0.207\) & \(0.395 \pm 0.207\) & \(0.532 \pm 0.212\) & \(0.808 \pm 0.232\) & \(0.822 \pm 0.199\) & .947 \\
    0.5 & 10 & 0.4 & \(0.398 \pm 0.173\) & \(0.399 \pm 0.176\) & \(0.057 \pm 0.189\) & \(0.285 \pm 0.174\) & \(0.398 \pm 0.173\) & \(0.511 \pm 0.175\) & \(0.738 \pm 0.191\) & \(0.681 \pm 0.147\) & .941 \\
    0.5 & \(\infty\) & 0.4 & \(0.408 \pm 0.163\) & \(0.407 \pm 0.165\) & \(0.088 \pm 0.179\) & \(0.302 \pm 0.165\) & \(0.408 \pm 0.163\) & \(0.514 \pm 0.164\) & \(0.728 \pm 0.177\) & \(0.640 \pm 0.138\) & .936 \\
    1 & 3  & 0.4 & \(0.396 \pm 0.193\) & \(0.396 \pm 0.193\) & \(0.016 \pm 0.204\) & \(0.268 \pm 0.193\) & \(0.396 \pm 0.193\) & \(0.523 \pm 0.196\) & \(0.775 \pm 0.212\) & \(0.759 \pm 0.155\) & .947 \\
    1 & 10 & 0.4 & \(0.403 \pm 0.160\) & \(0.402 \pm 0.164\) & \(0.089 \pm 0.167\) & \(0.296 \pm 0.161\) & \(0.403 \pm 0.161\) & \(0.509 \pm 0.162\) & \(0.718 \pm 0.168\) & \(0.630 \pm 0.106\) & .947 \\
    1 & \(\infty\) & 0.4 & \(0.399 \pm 0.154\) & \(0.399 \pm 0.157\) & \(0.107 \pm 0.160\) & \(0.300 \pm 0.155\) & \(0.399 \pm 0.154\) & \(0.498 \pm 0.155\) & \(0.693 \pm 0.161\) & \(0.586 \pm 0.097\) & .945 \\
    2 & 3  & 0.4 & \(0.396 \pm 0.194\) & \(0.395 \pm 0.195\) & \(-0.019 \pm 0.219\) & \(0.258 \pm 0.197\) & \(0.396 \pm 0.194\) & \(0.534 \pm 0.196\) & \(0.812 \pm 0.217\) & \(0.831 \pm 0.193\) & .971 \\
    2 & 10 & 0.4 & \(0.392 \pm 0.164\) & \(0.394 \pm 0.164\) & \(0.038 \pm 0.186\) & \(0.275 \pm 0.167\) & \(0.393 \pm 0.163\) & \(0.510 \pm 0.165\) & \(0.743 \pm 0.179\) & \(0.704 \pm 0.147\) & .969 \\
    2 & \(\infty\) & 0.4 & \(0.401 \pm 0.162\) & \(0.403 \pm 0.162\) & \(0.071 \pm 0.177\) & \(0.291 \pm 0.164\) & \(0.401 \pm 0.161\) & \(0.511 \pm 0.163\) & \(0.730 \pm 0.177\) & \(0.659 \pm 0.129\) & .958 \\
    3 & 3  & 0.4 & \(0.401 \pm 0.229\) & \(0.402 \pm 0.226\) & \(-0.098 \pm 0.271\) & \(0.238 \pm 0.234\) & \(0.401 \pm 0.227\) & \(0.563 \pm 0.229\) & \(0.899 \pm 0.259\) & \(0.997 \pm 0.235\) & .965 \\
    3 & 10 & 0.4 & \(0.408 \pm 0.201\) & \(0.409 \pm 0.198\) & \(-0.023 \pm 0.227\) & \(0.268 \pm 0.202\) & \(0.409 \pm 0.199\) & \(0.549 \pm 0.203\) & \(0.838 \pm 0.227\) & \(0.861 \pm 0.177\) & .964 \\
    3 & \(\infty\) & 0.4 & \(0.398 \pm 0.190\) & \(0.397 \pm 0.186\) & \(-0.010 \pm 0.215\) & \(0.264 \pm 0.191\) & \(0.397 \pm 0.188\) & \(0.531 \pm 0.192\) & \(0.807 \pm 0.216\) & \(0.817 \pm 0.168\) & .965 \\
    \bottomrule
\end{tabular}
\label{tab:pr_alpha}
\end{table}

\setlength{\tabcolsep}{2.3pt}
\begin{table}[h!]
\centering
\fontsize{6}{10}\selectfont
\caption{\textbf{Parameter recovery results for \(\beta\).} Path coefficients \(\alpha\) and \(\beta\) are both fixed at 0.4, and \(\{\varepsilon_i^{(M)}\}_{i=1}^{n}\) and \(\{\varepsilon_i^{(Y)}\}_{i=1}^{n}\) are drawn from the same distribution. It presents the means ($\pm$ one standard deviation) of posterior means, modes, 2.5th, 25th, 50th, 75th, and 97.5th percentiles, the length of the \(95\%\) credible intervals for \(\beta\), and the coverage rate of the true value by the credible intervals across \(1\,000\) experiments with sample size \(n = 50\).}
\begin{tabular}{*{12}{c}}
    \toprule
    \textbf{\(\gamma\)} & \textbf{\(\nu\)} & \textbf{True \(\beta\)} & \textbf{Mean} & \textbf{Mode} & \textbf{2.5th} & \textbf{25th} & \textbf{50th} & \textbf{75th} & \textbf{97.5th} & \textbf{Length} & \textbf{Rate} \\
    \midrule
    0.33 & 3  & 0.4 & \(0.399 \pm 0.077\) & \(0.415 \pm 0.079\) & \(0.228 \pm 0.094\) & \(0.351 \pm 0.079\) & \(0.404 \pm 0.077\) & \(0.452 \pm 0.079\) & \(0.543 \pm 0.098\) & \(0.314 \pm 0.111\) & .964 \\
    0.33 & 10 & 0.4 & \(0.405 \pm 0.092\) & \(0.417 \pm 0.094\) & \(0.199 \pm 0.107\) & \(0.343 \pm 0.094\) & \(0.409 \pm 0.092\) & \(0.469 \pm 0.094\) & \(0.590 \pm 0.110\) & \(0.391 \pm 0.103\) & .958 \\
    0.33 & \(\infty\) & 0.4 & \(0.399 \pm 0.102\) & \(0.406 \pm 0.102\) & \(0.178 \pm 0.118\) & \(0.330 \pm 0.104\) & \(0.401 \pm 0.101\) & \(0.469 \pm 0.102\) & \(0.607 \pm 0.115\) & \(0.429 \pm 0.096\) & .964 \\
    0.5 & 3  & 0.4 & \(0.395 \pm 0.100\) & \(0.406 \pm 0.103\) & \(0.192 \pm 0.124\) & \(0.333 \pm 0.104\) & \(0.399 \pm 0.100\) & \(0.460 \pm 0.102\) & \(0.577 \pm 0.118\) & \(0.385 \pm 0.133\) & .949 \\
    0.5 & 10 & 0.4 & \(0.407 \pm 0.127\) & \(0.415 \pm 0.129\) & \(0.164 \pm 0.136\) & \(0.330 \pm 0.128\) & \(0.409 \pm 0.127\) & \(0.486 \pm 0.129\) & \(0.637 \pm 0.141\) & \(0.473 \pm 0.109\) & .930 \\
    0.5 & \(\infty\) & 0.4 & \(0.406 \pm 0.121\) & \(0.412 \pm 0.124\) & \(0.161 \pm 0.132\) & \(0.328 \pm 0.123\) & \(0.408 \pm 0.122\) & \(0.487 \pm 0.123\) & \(0.643 \pm 0.134\) & \(0.482 \pm 0.107\) & .942 \\
    1 & 3  & 0.4 & \(0.398 \pm 0.132\) & \(0.399 \pm 0.134\) & \(0.142 \pm 0.149\) & \(0.313 \pm 0.134\) & \(0.398 \pm 0.133\) & \(0.483 \pm 0.135\) & \(0.652 \pm 0.150\) & \(0.510 \pm 0.141\) & .952 \\
    1 & 10 & 0.4 & \(0.404 \pm 0.150\) & \(0.404 \pm 0.151\) & \(0.117 \pm 0.158\) & \(0.307 \pm 0.151\) & \(0.404 \pm 0.150\) & \(0.501 \pm 0.151\) & \(0.692 \pm 0.158\) & \(0.575 \pm 0.108\) & .943 \\
    1 & \(\infty\) & 0.4 & \(0.395 \pm 0.152\) & \(0.396 \pm 0.154\) & \(0.102 \pm 0.158\) & \(0.296 \pm 0.153\) & \(0.395 \pm 0.153\) & \(0.494 \pm 0.154\) & \(0.688 \pm 0.160\) & \(0.587 \pm 0.100\) & .942 \\
    2 & 3  & 0.4 & \(0.399 \pm 0.091\) & \(0.411 \pm 0.091\) & \(0.196 \pm 0.112\) & \(0.337 \pm 0.094\) & \(0.403 \pm 0.091\) & \(0.464 \pm 0.094\) & \(0.583 \pm 0.115\) & \(0.387 \pm 0.130\) & .962 \\
    2 & 10 & 0.4 & \(0.407 \pm 0.120\) & \(0.413 \pm 0.121\) & \(0.163 \pm 0.132\) & \(0.329 \pm 0.122\) & \(0.409 \pm 0.120\) & \(0.487 \pm 0.121\) & \(0.639 \pm 0.132\) & \(0.476 \pm 0.103\) & .947 \\
    2 & \(\infty\) & 0.4 & \(0.400 \pm 0.120\) & \(0.404 \pm 0.120\) & \(0.147 \pm 0.130\) & \(0.318 \pm 0.122\) & \(0.402 \pm 0.120\) & \(0.484 \pm 0.121\) & \(0.645 \pm 0.131\) & \(0.498 \pm 0.092\) & .957 \\
    3 & 3  & 0.4 & \(0.399 \pm 0.081\) & \(0.412 \pm 0.079\) & \(0.217 \pm 0.097\) & \(0.346 \pm 0.081\) & \(0.403 \pm 0.080\) & \(0.456 \pm 0.084\) & \(0.558 \pm 0.106\) & \(0.341 \pm 0.114\) & .979 \\
    3 & 10 & 0.4 & \(0.400 \pm 0.101\) & \(0.409 \pm 0.099\) & \(0.180 \pm 0.113\) & \(0.332 \pm 0.102\) & \(0.403 \pm 0.100\) & \(0.471 \pm 0.103\) & \(0.605 \pm 0.117\) & \(0.424 \pm 0.094\) & .962 \\
    3 & \(\infty\) & 0.4 & \(0.402 \pm 0.111\) & \(0.409 \pm 0.110\) & \(0.171 \pm 0.123\) & \(0.330 \pm 0.112\) & \(0.405 \pm 0.111\) & \(0.477 \pm 0.113\) & \(0.622 \pm 0.127\) & \(0.451 \pm 0.097\) & .951 \\
    \bottomrule
\end{tabular}
\label{tab:pr_beta}
\end{table}

\clearpage
\newpage
\subsubsection{Skewness and Tail Parameters}\label{sm:skew_tail}

\vspace{-0.5cm}

For simplicity, we present the results only for \(\alpha=\beta=0.4\), with \(\{\varepsilon_i^{(M)}\}_{i=1}^{n}\) and \(\{\varepsilon_i^{(Y)}\}_{i=1}^{n}\) drawn from the same distribution. Tables~\ref{tab:pr_gamma1} and \ref{tab:pr_gamma2} report the results for \(\gamma\) corresponding to \(\varepsilon^{(M)}\) and \(\varepsilon^{(Y)}\). Tables~\ref{tab:pr_v1} and \ref{tab:pr_v2} present the results for \(\nu\) corresponding to \(\varepsilon^{(M)}\) and \(\varepsilon^{(Y)}\).

\setlength{\tabcolsep}{2.8pt}
\begin{table}[h!]
\centering
\fontsize{6}{9.5}\selectfont
\caption{\textbf{Parameter recovery results for \(\gamma\) corresponding to \(\varepsilon^{(M)}\).} Path coefficients \(\alpha\) and \(\beta\) are both fixed at 0.4, and \(\{\varepsilon_i^{(M)}\}_{i=1}^{n}\) and \(\{\varepsilon_i^{(Y)}\}_{i=1}^{n}\) are drawn from the same distribution. It presents the means ($\pm$ one standard deviation) of posterior means, modes, 2.5th, 25th, 50th, 75th, and 97.5th percentiles, the length of the \(95\%\) credible intervals for \(\gamma\), and the coverage rate of the true value by the credible intervals across \(1\,000\) experiments with sample size \(n = 50\).}
\begin{tabular}{*{12}{c}}
    \toprule
    \textbf{\(\gamma\)} & \textbf{\(\nu\)} & \textbf{True \(\gamma\)} & \textbf{Mean} & \textbf{Mode} & \textbf{2.5th} & \textbf{25th} & \textbf{50th} & \textbf{75th} & \textbf{97.5th} & \textbf{Length} & \textbf{Rate} \\
    \midrule
    0.33 & 3  & 0.33 & \(0.356 \pm 0.109\) & \(0.326 \pm 0.123\) & \(0.132 \pm 0.084\) & \(0.262 \pm 0.106\) & \(0.347 \pm 0.110\) & \(0.439 \pm 0.116\) & \(0.639 \pm 0.137\) & \(0.508 \pm 0.088\) & .963 \\
    0.33 & 10 & 0.33 & \(0.387 \pm 0.116\) & \(0.346 \pm 0.134\) & \(0.133 \pm 0.076\) & \(0.277 \pm 0.108\) & \(0.375 \pm 0.118\) & \(0.482 \pm 0.131\) & \(0.720 \pm 0.169\) & \(0.587 \pm 0.132\) & .969 \\
    0.33 & \(\infty\) & 0.33 & \(0.410 \pm 0.131\) & \(0.364 \pm 0.153\) & \(0.135 \pm 0.077\) & \(0.289 \pm 0.117\) & \(0.395 \pm 0.132\) & \(0.512 \pm 0.151\) & \(0.779 \pm 0.215\) & \(0.645 \pm 0.178\) & .968 \\
    0.5 & 3  & 0.5 & \(0.508 \pm 0.150\) & \(0.484 \pm 0.151\) & \(0.254 \pm 0.137\) & \(0.410 \pm 0.145\) & \(0.500 \pm 0.148\) & \(0.597 \pm 0.156\) & \(0.811 \pm 0.187\) & \(0.558 \pm 0.107\) & .936 \\
    0.5 & 10 & 0.5 & \(0.527 \pm 0.150\) & \(0.489 \pm 0.156\) & \(0.236 \pm 0.126\) & \(0.410 \pm 0.141\) & \(0.515 \pm 0.147\) & \(0.628 \pm 0.160\) & \(0.892 \pm 0.212\) & \(0.656 \pm 0.148\) & .983 \\
    0.5 & \(\infty\) & 0.5 & \(0.549 \pm 0.167\) & \(0.508 \pm 0.170\) & \(0.237 \pm 0.128\) & \(0.423 \pm 0.150\) & \(0.535 \pm 0.163\) & \(0.659 \pm 0.185\) & \(0.948 \pm 0.262\) & \(0.711 \pm 0.202\) & .971 \\
    1 & 3  & 1 & \(1.046 \pm 0.240\) & \(0.983 \pm 0.213\) & \(0.679 \pm 0.173\) & \(0.892 \pm 0.201\) & \(1.023 \pm 0.228\) & \(1.174 \pm 0.268\) & \(1.543 \pm 0.402\) & \(0.864 \pm 0.275\) & .935 \\
    1 & 10 & 1 & \(1.041 \pm 0.252\) & \(0.964 \pm 0.223\) & \(0.632 \pm 0.180\) & \(0.866 \pm 0.209\) & \(1.013 \pm 0.239\) & \(1.184 \pm 0.285\) & \(1.617 \pm 0.442\) & \(0.985 \pm 0.329\) & .960 \\
    1 & \(\infty\) & 1 & \(1.063 \pm 0.262\) & \(0.975 \pm 0.233\) & \(0.621 \pm 0.178\) & \(0.872 \pm 0.213\) & \(1.030 \pm 0.247\) & \(1.216 \pm 0.300\) & \(1.696 \pm 0.484\) & \(1.075 \pm 0.379\) & .969 \\
    2 & 3  & 2 & \(1.989 \pm 0.410\) & \(1.798 \pm 0.349\) & \(1.234 \pm 0.222\) & \(1.642 \pm 0.312\) & \(1.918 \pm 0.386\) & \(2.256 \pm 0.485\) & \(3.160 \pm 0.771\) & \(1.926 \pm 0.589\) & .952 \\
    2 & 10 & 2 & \(1.922 \pm 0.385\) & \(1.712 \pm 0.338\) & \(1.138 \pm 0.218\) & \(1.557 \pm 0.299\) & \(1.843 \pm 0.366\) & \(2.198 \pm 0.456\) & \(3.168 \pm 0.719\) & \(2.030 \pm 0.566\) & .971 \\
    2 & \(\infty\) & 2 & \(1.863 \pm 0.369\) & \(1.651 \pm 0.335\) & \(1.071 \pm 0.219\) & \(1.493 \pm 0.293\) & \(1.783 \pm 0.355\) & \(2.142 \pm 0.436\) & \(3.124 \pm 0.671\) & \(2.053 \pm 0.533\) & .970 \\
    3 & 3  & 3 & \(2.467 \pm 0.375\) & \(2.193 \pm 0.335\) & \(1.460 \pm 0.211\) & \(1.993 \pm 0.295\) & \(2.366 \pm 0.360\) & \(2.828 \pm 0.444\) & \(4.054 \pm 0.677\) & \(2.594 \pm 0.525\) & .922 \\
    3 & 10 & 3 & \(2.316 \pm 0.365\) & \(2.035 \pm 0.340\) & \(1.324 \pm 0.217\) & \(1.846 \pm 0.295\) & \(2.213 \pm 0.354\) & \(2.671 \pm 0.431\) & \(3.898 \pm 0.628\) & \(2.574 \pm 0.478\) & .904 \\
    3 & \(\infty\) & 3 & \(2.220 \pm 0.378\) & \(1.939 \pm 0.362\) & \(1.238 \pm 0.243\) & \(1.754 \pm 0.315\) & \(2.116 \pm 0.370\) & \(2.571 \pm 0.441\) & \(3.800 \pm 0.623\) & \(2.563 \pm 0.475\) & .897 \\
    \bottomrule
\end{tabular}
    \label{tab:pr_gamma1}
\end{table}

\setlength{\tabcolsep}{2.8pt}
\begin{table}[h!]
\centering
\fontsize{6}{9.5}\selectfont
\caption{\textbf{Parameter recovery results for \(\gamma\) corresponding to \(\varepsilon^{(Y)}\).} Path coefficients \(\alpha\) and \(\beta\) are both fixed at 0.4, and \(\{\varepsilon_i^{(M)}\}_{i=1}^{n}\) and \(\{\varepsilon_i^{(Y)}\}_{i=1}^{n}\) are drawn from the same distribution. It presents the means ($\pm$ one standard deviation) of posterior means, modes, 2.5th, 25th, 50th, 75th, and 97.5th percentiles, the length of the \(95\%\) credible intervals for \(\gamma\), and the coverage rate of the true value by the credible intervals across \(1\,000\) experiments with sample size \(n = 50\).}
\begin{tabular}{*{12}{c}}
    \toprule
    \textbf{\(\gamma\)} & \textbf{\(\nu\)} & \textbf{True \(\gamma\)} & \textbf{Mean} & \textbf{Mode} & \textbf{2.5th} & \textbf{25th} & \textbf{50th} & \textbf{75th} & \textbf{97.5th} & \textbf{Length} & \textbf{Rate} \\
    \midrule
    0.33 & 3  & 0.33 & \(0.351 \pm 0.103\) & \(0.318 \pm 0.121\) & \(0.118 \pm 0.072\) & \(0.251 \pm 0.099\) & \(0.342 \pm 0.106\) & \(0.439 \pm 0.114\) & \(0.649 \pm 0.140\) & \(0.531 \pm 0.103\) & .976 \\
    0.33 & 10 & 0.33 & \(0.389 \pm 0.119\) & \(0.348 \pm 0.140\) & \(0.123 \pm 0.071\) & \(0.272 \pm 0.109\) & \(0.376 \pm 0.121\) & \(0.489 \pm 0.135\) & \(0.740 \pm 0.183\) & \(0.617 \pm 0.143\) & .977 \\
    0.33 & \(\infty\) & 0.33 & \(0.431 \pm 0.143\) & \(0.383 \pm 0.162\) & \(0.136 \pm 0.078\) & \(0.303 \pm 0.125\) & \(0.416 \pm 0.144\) & \(0.540 \pm 0.166\) & \(0.824 \pm 0.241\) & \(0.688 \pm 0.202\) & .964 \\
    0.5 & 3  & 0.5 & \(0.502 \pm 0.151\) & \(0.477 \pm 0.154\) & \(0.237 \pm 0.138\) & \(0.398 \pm 0.148\) & \(0.494 \pm 0.149\) & \(0.596 \pm 0.156\) & \(0.823 \pm 0.192\) & \(0.587 \pm 0.117\) & .956 \\
    0.5 & 10 & 0.5 & \(0.531 \pm 0.162\) & \(0.496 \pm 0.165\) & \(0.223 \pm 0.131\) & \(0.409 \pm 0.151\) & \(0.520 \pm 0.159\) & \(0.639 \pm 0.174\) & \(0.912 \pm 0.238\) & \(0.689 \pm 0.171\) & .968 \\
    0.5 & \(\infty\) & 0.5 & \(0.543 \pm 0.181\) & \(0.499 \pm 0.183\) & \(0.218 \pm 0.129\) & \(0.410 \pm 0.162\) & \(0.528 \pm 0.177\) & \(0.657 \pm 0.201\) & \(0.965 \pm 0.292\) & \(0.747 \pm 0.223\) & .973 \\
    1 & 3  & 1 & \(1.058 \pm 0.253\) & \(0.991 \pm 0.215\) & \(0.677 \pm 0.174\) & \(0.897 \pm 0.204\) & \(1.033 \pm 0.237\) & \(1.190 \pm 0.286\) & \(1.585 \pm 0.464\) & \(0.909 \pm 0.338\) & .943 \\
    1 & 10 & 1 & \(1.055 \pm 0.252\) & \(0.973 \pm 0.220\) & \(0.629 \pm 0.181\) & \(0.873 \pm 0.207\) & \(1.024 \pm 0.237\) & \(1.202 \pm 0.285\) & \(1.663 \pm 0.462\) & \(1.034 \pm 0.352\) & .960 \\
    1 & \(\infty\) & 1 & \(1.073 \pm 0.289\) & \(0.976 \pm 0.251\) & \(0.607 \pm 0.193\) & \(0.870 \pm 0.231\) & \(1.036 \pm 0.270\) & \(1.233 \pm 0.331\) & \(1.759 \pm 0.551\) & \(1.152 \pm 0.431\) & .955 \\
    2 & 3  & 2 & \(1.993 \pm 0.405\) & \(1.787 \pm 0.342\) & \(1.217 \pm 0.222\) & \(1.631 \pm 0.309\) & \(1.915 \pm 0.380\) & \(2.267 \pm 0.479\) & \(3.224 \pm 0.769\) & \(2.007 \pm 0.592\) & .960 \\
    2 & 10 & 2 & \(1.896 \pm 0.362\) & \(1.674 \pm 0.311\) & \(1.101 \pm 0.210\) & \(1.521 \pm 0.280\) & \(1.812 \pm 0.342\) & \(2.176 \pm 0.427\) & \(3.183 \pm 0.685\) & \(2.082 \pm 0.542\) & .962 \\
    2 & \(\infty\) & 2 & \(1.843 \pm 0.368\) & \(1.616 \pm 0.328\) & \(1.041 \pm 0.218\) & \(1.463 \pm 0.290\) & \(1.757 \pm 0.351\) & \(2.124 \pm 0.435\) & \(3.147 \pm 0.678\) & \(2.105 \pm 0.536\) & .964 \\
    3 & 3  & 3 & \(2.409 \pm 0.363\) & \(2.128 \pm 0.325\) & \(1.419 \pm 0.205\) & \(1.940 \pm 0.286\) & \(2.306 \pm 0.349\) & \(2.763 \pm 0.431\) & \(3.993 \pm 0.644\) & \(2.574 \pm 0.494\) & .909 \\
    3 & 10 & 3 & \(2.240 \pm 0.364\) & \(1.951 \pm 0.340\) & \(1.261 \pm 0.225\) & \(1.772 \pm 0.298\) & \(2.134 \pm 0.356\) & \(2.589 \pm 0.429\) & \(3.831 \pm 0.602\) & \(2.571 \pm 0.455\) & .901 \\
    3 & \(\infty\) & 3 & \(2.162 \pm 0.364\) & \(1.877 \pm 0.344\) & \(1.192 \pm 0.224\) & \(1.698 \pm 0.298\) & \(2.056 \pm 0.356\) & \(2.507 \pm 0.428\) & \(3.743 \pm 0.614\) & \(2.550 \pm 0.470\) & .872 \\
    \bottomrule
    \end{tabular}
    \label{tab:pr_gamma2}
\end{table}

\setlength{\tabcolsep}{1.8pt}
\begin{table}[h!]
\centering
\fontsize{6}{10}\selectfont
\caption{\textbf{Parameter recovery results for \(\nu\) corresponding to \(\varepsilon^{(M)}\).} Path coefficients \(\alpha\) and \(\beta\) are both fixed at 0.4, and \(\{\varepsilon_i^{(M)}\}_{i=1}^{n}\) and \(\{\varepsilon_i^{(Y)}\}_{i=1}^{n}\) are drawn from the same distribution. It presents the means ($\pm$ one standard deviation) of posterior means, modes, 2.5th, 25th, 50th, 75th, and 97.5th percentiles, the length of the \(95\%\) credible intervals for \(\nu\), and the coverage rate of the true value by the credible intervals across \(1\,000\) experiments with sample size \(n = 50\).}
\begin{tabular}{*{12}{c}}
    \toprule
    \textbf{\(\gamma\)} & \textbf{\(\nu\)} & \textbf{True \(\nu\)} & \textbf{Mean} & \textbf{Mode} & \textbf{2.5th} & \textbf{25th} & \textbf{50th} & \textbf{75th} & \textbf{97.5th} & \textbf{Length} & \textbf{Rate} \\
    \midrule
    0.33 & 3  & 3   & \(7.599 \pm 3.951\) & \(3.716 \pm 1.644\) & \(2.348 \pm 0.416\) & \(3.943 \pm 1.674\) & \(5.896 \pm 3.064\) & \(9.314 \pm 5.338\) & \(22.823 \pm 12.772\) & \(20.475 \pm 12.431\) & .905 \\
    0.33 & 10 & 10  & \(13.008 \pm 3.097\) & \(6.191 \pm 1.970\) & \(3.014 \pm 0.599\) & \(6.447 \pm 1.766\) & \(10.328 \pm 2.835\) & \(16.650 \pm 4.225\) & \(38.214 \pm 7.219\) & \(35.200 \pm 6.732\) & .997 \\
    0.33 & \(\infty\) & \(\infty\) & \(14.812 \pm 2.398\) & \(7.390 \pm 1.964\) & \(3.377 \pm 0.622\) & \(7.518 \pm 1.583\) & \(12.017 \pm 2.339\) & \(19.092 \pm 3.187\) & \(42.010 \pm 4.589\) & \(38.633 \pm 4.053\) & \(\backslash\) \\
    0.5 & 3  & 3   & \(7.891 \pm 3.897\) & \(3.760 \pm 1.590\) & \(2.359 \pm 0.414\) & \(4.024 \pm 1.656\) & \(6.080 \pm 3.045\) & \(9.690 \pm 5.295\) & \(24.027 \pm 12.426\) & \(21.668 \pm 12.094\) & .911 \\
    0.5 & 10 & 10  & \(13.416 \pm 3.112\) & \(6.456 \pm 1.998\) & \(3.108 \pm 0.620\) & \(6.698 \pm 1.812\) & \(10.707 \pm 2.883\) & \(17.192 \pm 4.242\) & \(39.073 \pm 6.931\) & \(35.965 \pm 6.406\) & .996 \\
    0.5 & \(\infty\) & \(\infty\) & \(14.973 \pm 2.413\) & \(7.488 \pm 1.909\) & \(3.410 \pm 0.617\) & \(7.619 \pm 1.594\) & \(12.168 \pm 2.362\) & \(19.299 \pm 3.208\) & \(42.349 \pm 4.565\) & \(38.940 \pm 4.025\) & \(\backslash\) \\
    1 & 3  & 3   & \(8.078 \pm 4.085\) & \(3.864 \pm 1.732\) & \(2.388 \pm 0.451\) & \(4.122 \pm 1.773\) & \(6.241 \pm 3.223\) & \(9.943 \pm 5.547\) & \(24.500 \pm 12.832\) & \(22.112 \pm 12.464\) & .891 \\
    1 & 10 & 10  & \(13.325 \pm 3.071\) & \(6.364 \pm 2.010\) & \(3.070 \pm 0.625\) & \(6.618 \pm 1.810\) & \(10.600 \pm 2.860\) & \(17.079 \pm 4.190\) & \(39.001 \pm 6.698\) & \(35.931 \pm 6.171\) & 1 \\
    1 & \(\infty\) & \(\infty\) & \(15.004 \pm 2.309\) & \(7.483 \pm 1.827\) & \(3.419 \pm 0.598\) & \(7.637 \pm 1.535\) & \(12.193 \pm 2.265\) & \(19.342 \pm 3.058\) & \(42.422 \pm 4.338\) & \(39.003 \pm 3.808\) & \(\backslash\) \\
    2 & 3  & 3   & \(7.619 \pm 3.917\) & \(3.678 \pm 1.566\) & \(2.341 \pm 0.403\) & \(3.931 \pm 1.645\) & \(5.892 \pm 3.035\) & \(9.333 \pm 5.301\) & \(23.024 \pm 12.683\) & \(20.683 \pm 12.352\) & .914 \\
    2 & 10 & 10  & \(13.022 \pm 3.123\) & \(6.179 \pm 1.983\) & \(3.012 \pm 0.603\) & \(6.445 \pm 1.800\) & \(10.326 \pm 2.878\) & \(16.668 \pm 4.263\) & \(38.318 \pm 7.086\) & \(35.306 \pm 6.583\) & .998 \\
    2 & \(\infty\) & \(\infty\) & \(14.979 \pm 2.179\) & \(7.445 \pm 1.747\) & \(3.402 \pm 0.583\) & \(7.609 \pm 1.469\) & \(12.162 \pm 2.152\) & \(19.317 \pm 2.877\) & \(42.415 \pm 4.012\) & \(39.013 \pm 3.494\) & \(\backslash\) \\
    3 & 3  & 3   & \(7.394 \pm 3.837\) & \(3.606 \pm 1.549\) & \(2.327 \pm 0.408\) & \(3.849 \pm 1.622\) & \(5.725 \pm 2.974\) & \(9.019 \pm 5.181\) & \(22.283 \pm 12.430\) & \(19.956 \pm 12.099\) & .913 \\
    3 & 10 & 10  & \(12.731 \pm 3.214\) & \(6.039 \pm 1.939\) & \(2.963 \pm 0.584\) & \(6.304 \pm 1.785\) & \(10.083 \pm 2.904\) & \(16.267 \pm 4.394\) & \(37.555 \pm 7.668\) & \(34.593 \pm 7.190\) & .996 \\
    3 & \(\infty\) & \(\infty\) & \(14.569 \pm 2.542\) & \(7.186 \pm 1.926\) & \(3.321 \pm 0.623\) & \(7.367 \pm 1.645\) & \(11.775 \pm 2.467\) & \(18.757 \pm 3.399\) & \(41.578 \pm 4.916\) & \(38.257 \pm 4.368\) & \(\backslash\) \\
    \bottomrule
    \end{tabular}
    \label{tab:pr_v1}
\end{table}

\setlength{\tabcolsep}{1.8pt}
\begin{table}[h!]
\centering
\fontsize{6}{10}\selectfont
\caption{\textbf{Parameter recovery results for \(\nu\) corresponding to \(\varepsilon^{(Y)}\).} Path coefficients \(\alpha\) and \(\beta\) are both fixed at 0.4, and \(\{\varepsilon_i^{(M)}\}_{i=1}^{n}\) and \(\{\varepsilon_i^{(Y)}\}_{i=1}^{n}\) are drawn from the same distribution. It presents the means ($\pm$ one standard deviation) of posterior means, modes, 2.5th, 25th, 50th, 75th, and 97.5th percentiles, the length of the \(95\%\) credible intervals for \(\nu\), and the coverage rate of the true value by the credible intervals across \(1\,000\) experiments with sample size \(n = 50\).}
\begin{tabular}{*{12}{c}}
    \toprule
    \textbf{\(\gamma\)} & \textbf{\(\nu\)} & \textbf{True \(\nu\)} & \textbf{Mean} & \textbf{Mode} & \textbf{2.5th} & \textbf{25th} & \textbf{50th} & \textbf{75th} & \textbf{97.5th} & \textbf{Length} & \textbf{Rate} \\
    \midrule
    0.33 & 3  & 3   & \(7.712 \pm 3.984\) & \(3.719 \pm 1.636\) & \(2.350 \pm 0.422\) & \(3.973 \pm 1.693\) & \(5.973 \pm 3.105\) & \(9.473 \pm 5.400\) & \(23.246 \pm 12.824\) & \(20.896 \pm 12.483\) & .907 \\
    0.33 & 10 & 10  & \(13.070 \pm 3.045\) & \(6.216 \pm 1.972\) & \(3.011 \pm 0.583\) & \(6.468 \pm 1.750\) & \(10.368 \pm 2.806\) & \(16.733 \pm 4.167\) & \(38.467 \pm 6.887\) & \(35.456 \pm 6.401\) & .999 \\
    0.33 & \(\infty\) & \(\infty\) & \(14.779 \pm 2.304\) & \(7.367 \pm 1.885\) & \(3.345 \pm 0.583\) & \(7.483 \pm 1.516\) & \(11.976 \pm 2.253\) & \(19.057 \pm 3.070\) & \(41.997 \pm 4.414\) & \(38.652 \pm 3.912\) & \(\backslash\) \\
    0.5 & 3  & 3   & \(8.123 \pm 4.090\) & \(3.882 \pm 1.777\) & \(2.393 \pm 0.466\) & \(4.137 \pm 1.798\) & \(6.272 \pm 3.246\) & \(10.006 \pm 5.554\) & \(24.672 \pm 12.757\) & \(22.279 \pm 12.384\) & .887 \\
    0.5 & 10 & 10  & \(13.137 \pm 3.175\) & \(6.258 \pm 1.986\) & \(3.023 \pm 0.598\) & \(6.516 \pm 1.803\) & \(10.448 \pm 2.908\) & \(16.832 \pm 4.346\) & \(38.491 \pm 7.296\) & \(35.468 \pm 6.802\) & .995 \\
    0.5 & \(\infty\) & \(\infty\) & \(14.939 \pm 2.327\) & \(7.413 \pm 1.868\) & \(3.381 \pm 0.593\) & \(7.574 \pm 1.541\) & \(12.126 \pm 2.283\) & \(19.272 \pm 3.099\) & \(42.342 \pm 4.374\) & \(38.960 \pm 3.859\) & \(\backslash\) \\
    1 & 3  & 3   & \(7.940 \pm 3.868\) & \(3.740 \pm 1.572\) & \(2.351 \pm 0.405\) & \(4.015 \pm 1.651\) & \(6.084 \pm 3.036\) & \(9.744 \pm 5.267\) & \(24.415 \pm 12.295\) & \(22.065 \pm 11.969\) & .916 \\
    1 & 10 & 10  & \(13.250 \pm 3.125\) & \(6.263 \pm 1.981\) & \(3.035 \pm 0.608\) & \(6.556 \pm 1.815\) & \(10.526 \pm 2.898\) & \(16.984 \pm 4.266\) & \(38.897 \pm 6.984\) & \(35.861 \pm 6.477\) & .997 \\
    1 & \(\infty\) & \(\infty\) & \(15.098 \pm 2.161\) & \(7.506 \pm 1.757\) & \(3.406 \pm 0.572\) & \(7.667 \pm 1.458\) & \(12.276 \pm 2.133\) & \(19.492 \pm 2.860\) & \(42.677 \pm 3.965\) & \(39.270 \pm 3.462\) & \(\backslash\) \\
    2 & 3  & 3   & \(7.833 \pm 4.012\) & \(3.761 \pm 1.688\) & \(2.361 \pm 0.439\) & \(4.016 \pm 1.732\) & \(6.051 \pm 3.151\) & \(9.613 \pm 5.436\) & \(23.745 \pm 12.729\) & \(21.384 \pm 12.377\) & .903 \\
    2 & 10 & 10  & \(13.073 \pm 3.268\) & \(6.228 \pm 2.025\) & \(3.023 \pm 0.609\) & \(6.492 \pm 1.853\) & \(10.394 \pm 2.991\) & \(16.732 \pm 4.469\) & \(38.337 \pm 7.490\) & \(35.314 \pm 6.976\) & .998 \\
    2 & \(\infty\) & \(\infty\) & \(14.741 \pm 2.305\) & \(7.255 \pm 1.776\) & \(3.325 \pm 0.575\) & \(7.443 \pm 1.511\) & \(11.932 \pm 2.252\) & \(19.009 \pm 3.076\) & \(41.988 \pm 4.373\) & \(38.662 \pm 3.871\) & \(\backslash\) \\
    3 & 3  & 3   & \(7.547 \pm 3.795\) & \(3.590 \pm 1.463\) & \(2.319 \pm 0.374\) & \(3.864 \pm 1.558\) & \(5.805 \pm 2.905\) & \(9.245 \pm 5.134\) & \(22.964 \pm 12.528\) & \(20.645 \pm 12.228\) & .931 \\
    3 & 10 & 10  & \(12.649 \pm 3.440\) & \(6.018 \pm 2.061\) & \(2.962 \pm 0.617\) & \(6.278 \pm 1.905\) & \(10.024 \pm 3.099\) & \(16.155 \pm 4.696\) & \(37.245 \pm 8.269\) & \(34.283 \pm 7.759\) & .993 \\
    3 & \(\infty\) & \(\infty\) & \(14.452 \pm 2.535\) & \(7.082 \pm 1.864\) & \(3.272 \pm 0.590\) & \(7.272 \pm 1.614\) & \(11.653 \pm 2.452\) & \(18.613 \pm 3.411\) & \(41.401 \pm 4.959\) & \(38.129 \pm 4.443\) & \(\backslash\) \\
    \bottomrule
    \end{tabular}
    \label{tab:pr_v2}
\end{table}

\subsubsection{Parameters \(\sigma^{(M)}\), \(\sigma^{(Y)}\) and \(\tau\)}

For simplicity, we present the results only for \(\alpha=\beta=0.4\), with \(\{\varepsilon_i^{(M)}\}_{i=1}^{n}\) and \(\{\varepsilon_i^{(Y)}\}_{i=1}^{n}\) drawn from the same distribution. Tables~\ref{tab:pr_sigmaM} and \ref{tab:pr_sigmaY} present the parameter recovery results for \(\sigma^{(M)}\) in Equation~(\ref{eq:my_med_1}) and \(\sigma^{(Y)}\) in Equation~(\ref{eq:my_med_2}). In addition, Table~\ref{tab:pr_tau} gives the results for \(\tau\) in Equation~(\ref{eq:my_med_1}). Collectively, these findings indicate good empirical identifiability of the model.

\setlength{\tabcolsep}{1.8pt}
\begin{table}[h!]
\centering
\fontsize{6}{10}\selectfont
\caption{\textbf{Parameter recovery results for \(\sigma^{(M)}\).} Path coefficients \(\alpha\) and \(\beta\) are both fixed at 0.4, and \(\{\varepsilon_i^{(M)}\}_{i=1}^{n}\) and \(\{\varepsilon_i^{(Y)}\}_{i=1}^{n}\) are drawn from the same distribution. It presents the means ($\pm$ one standard deviation) of posterior means, modes, 2.5th, 25th, 50th, 75th, and 97.5th percentiles, the length of the \(95\%\) credible intervals for \(\sigma^{(M)}\), and the coverage rate of the true value by the credible intervals across \(1\,000\) experiments with sample size \(n = 50\).}
\begin{tabular}{*{12}{c}}
    \toprule
    \textbf{\(\gamma\)} & \textbf{\(\nu\)} & \textbf{True \(\sigma^{(M)}\)} & \textbf{Mean} & \textbf{Mode} & \textbf{2.5th} & \textbf{25th} & \textbf{50th} & \textbf{75th} & \textbf{97.5th} & \textbf{Length} & \textbf{Rate} \\
    \midrule
    0.33 & 3  & 1 & \(1.123 \pm 0.321\) & \(1.063 \pm 0.366\) & \(0.451 \pm 0.267\) & \(0.852 \pm 0.319\) & \(1.105 \pm 0.329\) & \(1.369 \pm 0.342\) & \(1.921 \pm 0.382\) & \(1.470 \pm 0.273\) & .946 \\
    0.33 & 10 & 1 & \(1.042 \pm 0.243\) & \(11 \pm 0.318\) & \(0.407 \pm 0.208\) & \(0.792 \pm 0.260\) & \(1.033 \pm 0.256\) & \(1.275 \pm 0.254\) & \(1.747 \pm 0.260\) & \(1.340 \pm 0.224\) & .989 \\
    0.33 & \(\infty\) & 1 & \(0.999 \pm 0.233\) & \(0.975 \pm 0.315\) & \(0.386 \pm 0.201\) & \(0.765 \pm 0.254\) & \(0.996 \pm 0.249\) & \(1.223 \pm 0.243\) & \(1.654 \pm 0.238\) & \(1.268 \pm 0.209\) & .990 \\
    0.5 & 3  & 1 & \(1.074 \pm 0.262\) & \(1.043 \pm 0.272\) & \(0.575 \pm 0.254\) & \(0.890 \pm 0.260\) & \(1.065 \pm 0.262\) & \(1.247 \pm 0.271\) & \(1.629 \pm 0.300\) & \(1.055 \pm 0.204\) & .939 \\
    0.5 & 10 & 1 & \(0.938 \pm 0.193\) & \(0.932 \pm 0.222\) & \(0.488 \pm 0.222\) & \(0.783 \pm 0.211\) & \(0.939 \pm 0.197\) & \(1.092 \pm 0.188\) & \(1.395 \pm 0.189\) & \(0.907 \pm 0.186\) & .975 \\
    0.5 & \(\infty\) & 1 & \(0.895 \pm 0.181\) & \(0.897 \pm 0.204\) & \(0.464 \pm 0.209\) & \(0.751 \pm 0.199\) & \(0.898 \pm 0.185\) & \(1.040 \pm 0.176\) & \(1.316 \pm 0.173\) & \(0.852 \pm 0.178\) & .960 \\
    1 & 3  & 1 & \(1.080 \pm 0.182\) & \(1.050 \pm 0.187\) & \(0.745 \pm 0.146\) & \(0.951 \pm 0.169\) & \(1.070 \pm 0.182\) & \(1.198 \pm 0.197\) & \(1.473 \pm 0.231\) & \(0.728 \pm 0.123\) & .947 \\
    1 & 10 & 1 & \(0.943 \pm 0.126\) & \(0.934 \pm 0.124\) & \(0.666 \pm 0.124\) & \(0.846 \pm 0.122\) & \(0.940 \pm 0.126\) & \(1.036 \pm 0.132\) & \(1.238 \pm 0.151\) & \(0.572 \pm 0.094\) & .948 \\
    1 & \(\infty\) & 1 & \(0.878 \pm 0.107\) & \(0.869 \pm 0.105\) & \(0.629 \pm 0.113\) & \(0.792 \pm 0.106\) & \(0.875 \pm 0.106\) & \(0.961 \pm 0.110\) & \(1.141 \pm 0.124\) & \(0.512 \pm 0.082\) & .867 \\
    2 & 3  & 1 & \(1.150 \pm 0.227\) & \(1.104 \pm 0.232\) & \(0.725 \pm 0.175\) & \(0.980 \pm 0.210\) & \(1.134 \pm 0.228\) & \(1.301 \pm 0.247\) & \(1.668 \pm 0.287\) & \(0.943 \pm 0.155\) & .929 \\
    2 & 10 & 1 & \(1.030 \pm 0.164\) & \(1.005 \pm 0.173\) & \(0.659 \pm 0.142\) & \(0.890 \pm 0.160\) & \(1.021 \pm 0.166\) & \(1.159 \pm 0.173\) & \(1.449 \pm 0.190\) & \(0.790 \pm 0.111\) & .980 \\
    2 & \(\infty\) & 1 & \(0.972 \pm 0.136\) & \(0.955 \pm 0.146\) & \(0.627 \pm 0.123\) & \(0.844 \pm 0.134\) & \(0.966 \pm 0.138\) & \(1.092 \pm 0.143\) & \(1.352 \pm 0.156\) & \(0.725 \pm 0.101\) & .986 \\
    3 & 3  & 1 & \(1.357 \pm 0.259\) & \(1.279 \pm 0.264\) & \(0.790 \pm 0.181\) & \(1.122 \pm 0.232\) & \(1.330 \pm 0.260\) & \(1.562 \pm 0.289\) & \(2.076 \pm 0.350\) & \(1.286 \pm 0.211\) & .882 \\
    3 & 10 & 1 & \(1.246 \pm 0.191\) & \(1.198 \pm 0.206\) & \(0.738 \pm 0.148\) & \(1.044 \pm 0.180\) & \(1.229 \pm 0.194\) & \(1.428 \pm 0.208\) & \(1.848 \pm 0.237\) & \(1.110 \pm 0.150\) & .940 \\
    3 & \(\infty\) & 1 & \(1.189 \pm 0.179\) & \(1.155 \pm 0.197\) & \(0.709 \pm 0.140\) & \(1.004 \pm 0.172\) & \(1.177 \pm 0.184\) & \(1.361 \pm 0.194\) & \(1.740 \pm 0.213\) & \(1.031 \pm 0.135\) & .968 \\
    \bottomrule
\end{tabular}
    \label{tab:pr_sigmaM}
\end{table}

\setlength{\tabcolsep}{1.8pt}
\begin{table}[h!]
\centering
\fontsize{6}{10}\selectfont
\caption{\textbf{Parameter recovery results for \(\sigma^{(Y)}\).} Path coefficients \(\alpha\) and \(\beta\) are both fixed at 0.4, and \(\{\varepsilon_i^{(M)}\}_{i=1}^{n}\) and \(\{\varepsilon_i^{(Y)}\}_{i=1}^{n}\) are drawn from the same distribution. It presents the means ($\pm$ one standard deviation) of posterior means, modes, 2.5th, 25th, 50th, 75th, and 97.5th percentiles, the length of the \(95\%\) credible intervals for \(\sigma^{(M)}\), and the coverage rate of the true value by the credible intervals across \(1\,000\) experiments with sample size \(n = 50\).}
\begin{tabular}{*{12}{c}}
    \toprule
    \textbf{\(\gamma\)} & \textbf{\(\nu\)} & \textbf{True \(\sigma^{(Y)}\)} & \textbf{Mean} & \textbf{Mode} & \textbf{2.5th} & \textbf{25th} & \textbf{50th} & \textbf{75th} & \textbf{97.5th} & \textbf{Length} & \textbf{Rate} \\
    \midrule
    0.33 & 3  & 1 & \(1.111 \pm 0.292\) & \(1.050 \pm 0.349\) & \(0.410 \pm 0.223\) & \(0.824 \pm 0.290\) & \(1.093 \pm 0.303\) & \(1.372 \pm 0.318\) & \(1.948 \pm 0.358\) & \(1.538 \pm 0.277\) & .971 \\
    0.33 & 10 & 1 & \(1.033 \pm 0.244\) & \(0.998 \pm 0.323\) & \(0.376 \pm 0.199\) & \(0.774 \pm 0.259\) & \(1.027 \pm 0.259\) & \(1.278 \pm 0.257\) & \(1.760 \pm 0.266\) & \(1.384 \pm 0.228\) & .992 \\
    0.33 & \(\infty\) & 1 & \(1.027 \pm 0.238\) & \(1.013 \pm 0.324\) & \(0.389 \pm 0.203\) & \(0.789 \pm 0.260\) & \(1.027 \pm 0.253\) & \(1.258 \pm 0.245\) & \(1.691 \pm 0.239\) & \(1.303 \pm 0.212\) & .984 \\
    0.5 & 3  & 1 & \(1.051 \pm 0.259\) & \(1.026 \pm 0.275\) & \(0.537 \pm 0.259\) & \(0.859 \pm 0.263\) & \(1.043 \pm 0.260\) & \(1.232 \pm 0.265\) & \(1.622 \pm 0.292\) & \(1.085 \pm 0.215\) & .956 \\
    0.5 & 10 & 1 & \(0.937 \pm 0.203\) & \(0.936 \pm 0.233\) & \(0.463 \pm 0.227\) & \(0.776 \pm 0.222\) & \(0.939 \pm 0.207\) & \(1.098 \pm 0.198\) & \(1.408 \pm 0.198\) & \(0.945 \pm 0.191\) & .979 \\
    0.5 & \(\infty\) & 1 & \(0.877 \pm 0.186\) & \(0.882 \pm 0.214\) & \(0.427 \pm 0.214\) & \(0.724 \pm 0.210\) & \(0.880 \pm 0.192\) & \(1.030 \pm 0.178\) & \(1.316 \pm 0.171\) & \(0.889 \pm 0.179\) & .962 \\
    1 & 3  & 1 & \(1.089 \pm 0.180\) & \(1.058 \pm 0.184\) & \(0.743 \pm 0.140\) & \(0.956 \pm 0.165\) & \(1.079 \pm 0.181\) & \(1.211 \pm 0.198\) & \(1.495 \pm 0.237\) & \(0.752 \pm 0.137\) & .957 \\
    1 & 10 & 1 & \(0.941 \pm 0.125\) & \(0.933 \pm 0.124\) & \(0.660 \pm 0.125\) & \(0.843 \pm 0.121\) & \(0.938 \pm 0.124\) & \(1.036 \pm 0.131\) & \(1.241 \pm 0.150\) & \(0.581 \pm 0.097\) & .947 \\
    1 & \(\infty\) & 1 & \(0.872 \pm 0.111\) & \(0.866 \pm 0.113\) & \(0.612 \pm 0.121\) & \(0.784 \pm 0.110\) & \(0.870 \pm 0.110\) & \(0.958 \pm 0.114\) & \(1.142 \pm 0.129\) & \(0.530 \pm 0.095\) & .862 \\
    2 & 3  & 1 & \(1.153 \pm 0.230\) & \(1.106 \pm 0.234\) & \(0.717 \pm 0.173\) & \(0.979 \pm 0.210\) & \(1.136 \pm 0.231\) & \(1.308 \pm 0.252\) & \(1.683 \pm 0.298\) & \(0.966 \pm 0.170\) & .932 \\
    2 & 10 & 1 & \(1.035 \pm 0.164\) & \(1.012 \pm 0.173\) & \(0.654 \pm 0.138\) & \(0.892 \pm 0.158\) & \(1.027 \pm 0.166\) & \(1.168 \pm 0.175\) & \(1.461 \pm 0.196\) & \(0.807 \pm 0.117\) & .986 \\
    2 & \(\infty\) & 1 & \(0.971 \pm 0.139\) & \(0.956 \pm 0.148\) & \(0.618 \pm 0.125\) & \(0.842 \pm 0.138\) & \(0.966 \pm 0.141\) & \(1.094 \pm 0.146\) & \(1.357 \pm 0.160\) & \(0.739 \pm 0.106\) & .986 \\
    3 & 3  & 1 & \(1.388 \pm 0.263\) & \(1.310 \pm 0.272\) & \(0.803 \pm 0.181\) & \(1.147 \pm 0.235\) & \(1.361 \pm 0.264\) & \(1.600 \pm 0.295\) & \(2.126 \pm 0.357\) & \(1.323 \pm 0.222\) & .858 \\
    3 & 10 & 1 & \(1.272 \pm 0.197\) & \(1.226 \pm 0.215\) & \(0.748 \pm 0.148\) & \(1.066 \pm 0.186\) & \(1.256 \pm 0.202\) & \(1.459 \pm 0.216\) & \(1.885 \pm 0.248\) & \(1.137 \pm 0.159\) & .939 \\
    3 & \(\infty\) & 1 & \(1.211 \pm 0.180\) & \(1.177 \pm 0.200\) & \(0.716 \pm 0.140\) & \(1.021 \pm 0.173\) & \(1.199 \pm 0.185\) & \(1.388 \pm 0.195\) & \(1.773 \pm 0.217\) & \(1.057 \pm 0.138\) & .965 \\
    \bottomrule
\end{tabular}
    \label{tab:pr_sigmaY}
\end{table}

\setlength{\tabcolsep}{1.8pt}
\begin{table}[h!]
\centering
\fontsize{6}{8.5}\selectfont
\caption{\textbf{Parameter recovery results for \(\tau\).} Path coefficients \(\alpha\) and \(\beta\) are both fixed at 0.4, and \(\{\varepsilon_i^{(M)}\}_{i=1}^{n}\) and \(\{\varepsilon_i^{(Y)}\}_{i=1}^{n}\) are drawn from the same distribution. It presents the means ($\pm$ one standard deviation) of posterior means, modes, 2.5th, 25th, 50th, 75th, and 97.5th percentiles, the length of the \(95\%\) credible intervals for \(\tau\), and the coverage rate of the true value by the credible intervals across \(1\,000\) experiments with sample size \(n = 50\).}
\begin{tabular}{*{12}{c}}
    \toprule
    \textbf{\(\gamma\)} & \textbf{\(\nu\)} & \textbf{True \(\tau\)} & \textbf{Mean} & \textbf{Mode} & \textbf{2.5th} & \textbf{25th} & \textbf{50th} & \textbf{75th} & \textbf{97.5th} & \textbf{Length} & \textbf{Rate} \\
    \midrule
    0.33 & 3  & 0.2 & \(0.205 \pm 0.237\) & \(0.203 \pm 0.246\) & \(-0.285 \pm 0.275\) & \(0.045 \pm 0.242\) & \(0.204 \pm 0.237\) & \(0.364 \pm 0.243\) & \(0.701 \pm 0.281\) & \(0.985 \pm 0.269\) & .957 \\
    0.33 & 10 & 0.2 & \(0.187 \pm 0.201\) & \(0.185 \pm 0.204\) & \(-0.243 \pm 0.238\) & \(0.047 \pm 0.205\) & \(0.186 \pm 0.200\) & \(0.326 \pm 0.203\) & \(0.617 \pm 0.232\) & \(0.861 \pm 0.215\) & .964 \\
    0.33 & \(\infty\) & 0.2 & \(0.189 \pm 0.204\) & \(0.190 \pm 0.205\) & \(-0.227 \pm 0.235\) & \(0.054 \pm 0.209\) & \(0.189 \pm 0.203\) & \(0.324 \pm 0.205\) & \(0.602 \pm 0.231\) & \(0.829 \pm 0.199\) & .952 \\
    0.5 & 3  & 0.2 & \(0.189 \pm 0.217\) & \(0.188 \pm 0.221\) & \(-0.232 \pm 0.244\) & \(0.049 \pm 0.221\) & \(0.189 \pm 0.217\) & \(0.328 \pm 0.219\) & \(0.611 \pm 0.238\) & \(0.844 \pm 0.203\) & .935 \\
    0.5 & 10 & 0.2 & \(0.191 \pm 0.187\) & \(0.190 \pm 0.193\) & \(-0.173 \pm 0.207\) & \(0.070 \pm 0.191\) & \(0.190 \pm 0.187\) & \(0.311 \pm 0.188\) & \(0.555 \pm 0.200\) & \(0.728 \pm 0.161\) & .949 \\
    0.5 & \(\infty\) & 0.2 & \(0.189 \pm 0.176\) & \(0.191 \pm 0.179\) & \(-0.148 \pm 0.193\) & \(0.078 \pm 0.179\) & \(0.190 \pm 0.176\) & \(0.301 \pm 0.178\) & \(0.526 \pm 0.191\) & \(0.675 \pm 0.144\) & .934 \\
    1 & 3  & 0.2 & \(0.200 \pm 0.202\) & \(0.200 \pm 0.204\) & \(-0.204 \pm 0.221\) & \(0.065 \pm 0.205\) & \(0.200 \pm 0.202\) & \(0.336 \pm 0.203\) & \(0.604 \pm 0.215\) & \(0.807 \pm 0.160\) & .952 \\
    1 & 10 & 0.2 & \(0.204 \pm 0.177\) & \(0.205 \pm 0.179\) & \(-0.135 \pm 0.184\) & \(0.090 \pm 0.178\) & \(0.204 \pm 0.177\) & \(0.319 \pm 0.179\) & \(0.544 \pm 0.185\) & \(0.678 \pm 0.112\) & .946 \\
    1 & \(\infty\) & 0.2 & \(0.207 \pm 0.174\) & \(0.206 \pm 0.176\) & \(-0.113 \pm 0.179\) & \(0.099 \pm 0.174\) & \(0.207 \pm 0.174\) & \(0.315 \pm 0.176\) & \(0.525 \pm 0.183\) & \(0.638 \pm 0.109\) & .923 \\
    2 & 3  & 0.2 & \(0.195 \pm 0.209\) & \(0.196 \pm 0.209\) & \(-0.237 \pm 0.234\) & \(0.052 \pm 0.212\) & \(0.195 \pm 0.208\) & \(0.338 \pm 0.211\) & \(0.628 \pm 0.233\) & \(0.864 \pm 0.199\) & .954 \\
    2 & 10 & 0.2 & \(0.189 \pm 0.185\) & \(0.190 \pm 0.185\) & \(-0.185 \pm 0.202\) & \(0.065 \pm 0.187\) & \(0.189 \pm 0.185\) & \(0.313 \pm 0.187\) & \(0.562 \pm 0.204\) & \(0.747 \pm 0.155\) & .945 \\
    2 & \(\infty\) & 0.2 & \(0.207 \pm 0.176\) & \(0.206 \pm 0.176\) & \(-0.144 \pm 0.190\) & \(0.090 \pm 0.177\) & \(0.207 \pm 0.176\) & \(0.324 \pm 0.179\) & \(0.559 \pm 0.194\) & \(0.703 \pm 0.139\) & .957 \\
    3 & 3  & 0.2 & \(0.204 \pm 0.244\) & \(0.205 \pm 0.241\) & \(-0.326 \pm 0.280\) & \(0.031 \pm 0.247\) & \(0.204 \pm 0.242\) & \(0.377 \pm 0.247\) & \(0.732 \pm 0.280\) & \(1.059 \pm 0.242\) & .963 \\
    3 & 10 & 0.2 & \(0.195 \pm 0.221\) & \(0.195 \pm 0.217\) & \(-0.265 \pm 0.254\) & \(0.045 \pm 0.225\) & \(0.195 \pm 0.219\) & \(0.346 \pm 0.221\) & \(0.656 \pm 0.244\) & \(0.921 \pm 0.197\) & .969 \\
    3 & \(\infty\) & 0.2 & \(0.207 \pm 0.225\) & \(0.206 \pm 0.224\) & \(-0.226 \pm 0.245\) & \(0.064 \pm 0.225\) & \(0.207 \pm 0.223\) & \(0.349 \pm 0.227\) & \(0.642 \pm 0.250\) & \(0.868 \pm 0.179\) & .948 \\
    \bottomrule
\end{tabular}
    \label{tab:pr_tau}
\end{table}

\newpage

\vspace{-0.5cm}
\subsubsection{Estimated Mediation Effects under the Misspecified Normal Model} \label{sm:pr_mis}

\vspace{-0.5cm}

For simplicity, we present the results only for \(\alpha=\beta=0.4\) and \(\{\varepsilon_i^{(M)}\}_{i=1}^{n}\), with \(\{\varepsilon_i^{(Y)}\}_{i=1}^{n}\) drawn from the same distribution. Table~\ref{tab:pr_norm3} presents the estimated mediation effect \(\alpha\beta\) under the Normal Model, even though the errors may be skewed and/or heavy-tailed. The \(95\%\) credible intervals are generally wider than those obtained with the Full Model.

\setlength{\tabcolsep}{2.4pt}
\begin{table}[h!]
\centering
\fontsize{6}{8.5}\selectfont
\caption{\textbf{Estimated mediation effects under the Normal Model}. Path coefficients \(\alpha\) and \(\beta\) are both fixed at 0.4, and \(\{\varepsilon_i^{(M)}\}_{i=1}^{n}\) and \(\{\varepsilon_i^{(Y)}\}_{i=1}^{n}\) are drawn from the same distribution. It presents the means ($\pm$ one standard deviation) of posterior means, modes, 2.5th, 25th, 50th, 75th, and 97.5th percentiles, the length of the \(95\%\) credible intervals for \(\alpha\beta\), and the coverage rate of the true value by the credible intervals across \(1\,000\) experiments with sample size \(n = 50\).}
\begin{tabular}{*{12}{c}}
    \toprule
    \textbf{\(\gamma\)} & \textbf{\(\nu\)} & \textbf{True \(\alpha\beta\)} & \textbf{Mean} & \textbf{Mode} & \textbf{2.5th} & \textbf{25th} & \textbf{50th} & \textbf{75th} & \textbf{97.5th} & \textbf{Length} & \textbf{Rate} \\
    \midrule
    0.33 & 3  & 0.16 & \(0.164 \pm 0.238\) & \(0.113 \pm 0.204\) & \(-0.315 \pm 0.337\) & \(0.000 \pm 0.229\) & \(0.148 \pm 0.226\) & \(0.314 \pm 0.260\) & \(0.722 \pm 0.380\) & \(1.036 \pm 0.463\) & .973 \\
    0.33 & 10 & 0.16 & \(0.169 \pm 0.151\) & \(0.126 \pm 0.137\) & \(-0.118 \pm 0.140\) & \(0.063 \pm 0.126\) & \(0.155 \pm 0.145\) & \(0.263 \pm 0.174\) & \(0.525 \pm 0.238\) & \(0.643 \pm 0.192\) & .962 \\
    0.33 & \(\infty\) & 0.16 & \(0.164 \pm 0.135\) & \(0.125 \pm 0.122\) & \(-0.086 \pm 0.119\) & \(0.069 \pm 0.111\) & \(0.151 \pm 0.130\) & \(0.247 \pm 0.156\) & \(0.479 \pm 0.212\) & \(0.565 \pm 0.164\) & .948 \\
    0.5 & 3  & 0.16 & \(0.158 \pm 0.176\) & \(0.116 \pm 0.157\) & \(-0.179 \pm 0.288\) & \(0.037 \pm 0.161\) & \(0.144 \pm 0.169\) & \(0.266 \pm 0.224\) & \(0.562 \pm 0.394\) & \(0.741 \pm 0.564\) & .958 \\
    0.5 & 10 & 0.16 & \(0.157 \pm 0.117\) & \(0.122 \pm 0.109\) & \(-0.050 \pm 0.098\) & \(0.076 \pm 0.097\) & \(0.146 \pm 0.114\) & \(0.227 \pm 0.135\) & \(0.423 \pm 0.181\) & \(0.473 \pm 0.135\) & .953 \\
    0.5 & \(\infty\) & 0.16 & \(0.167 \pm 0.105\) & \(0.133 \pm 0.101\) & \(-0.021 \pm 0.075\) & \(0.091 \pm 0.086\) & \(0.156 \pm 0.103\) & \(0.232 \pm 0.122\) & \(0.414 \pm 0.162\) & \(0.434 \pm 0.116\) & .948 \\
    1 & 3  & 0.16 & \(0.156 \pm 0.118\) & \(0.120 \pm 0.109\) & \(-0.069 \pm 0.122\) & \(0.070 \pm 0.100\) & \(0.145 \pm 0.114\) & \(0.232 \pm 0.137\) & \(0.442 \pm 0.196\) & \(0.511 \pm 0.195\) & .960 \\
    1 & 10 & 0.16 & \(0.164 \pm 0.091\) & \(0.135 \pm 0.089\) & \(0.001 \pm 0.064\) & \(0.097 \pm 0.076\) & \(0.155 \pm 0.090\) & \(0.222 \pm 0.105\) & \(0.380 \pm 0.136\) & \(0.380 \pm 0.096\) & .953 \\
    1 & \(\infty\) & 0.16 & \(0.159 \pm 0.087\) & \(0.132 \pm 0.085\) & \(0.006 \pm 0.060\) & \(0.096 \pm 0.072\) & \(0.150 \pm 0.085\) & \(0.213 \pm 0.100\) & \(0.360 \pm 0.130\) & \(0.354 \pm 0.089\) & .943 \\
    2 & 3  & 0.16 & \(0.162 \pm 0.175\) & \(0.120 \pm 0.157\) & \(-0.166 \pm 0.201\) & \(0.045 \pm 0.150\) & \(0.148 \pm 0.168\) & \(0.267 \pm 0.206\) & \(0.556 \pm 0.303\) & \(0.722 \pm 0.331\) & .961 \\
    2 & 10 & 0.16 & \(0.162 \pm 0.115\) & \(0.126 \pm 0.109\) & \(-0.045 \pm 0.095\) & \(0.081 \pm 0.096\) & \(0.151 \pm 0.112\) & \(0.233 \pm 0.132\) & \(0.430 \pm 0.174\) & \(0.475 \pm 0.130\) & .951 \\
    2 & \(\infty\) & 0.16 & \(0.161 \pm 0.103\) & \(0.128 \pm 0.099\) & \(-0.023 \pm 0.079\) & \(0.087 \pm 0.085\) & \(0.150 \pm 0.101\) & \(0.225 \pm 0.119\) & \(0.404 \pm 0.156\) & \(0.427 \pm 0.112\) & .958 \\
    3 & 3  & 0.16 & \(0.158 \pm 0.257\) & \(0.113 \pm 0.216\) & \(-0.320 \pm 0.359\) & \(-0.004 \pm 0.245\) & \(0.144 \pm 0.242\) & \(0.308 \pm 0.280\) & \(0.708 \pm 0.412\) & \(1.028 \pm 0.490\) & .964 \\
    3 & 10 & 0.16 & \(0.164 \pm 0.150\) & \(0.121 \pm 0.135\) & \(-0.118 \pm 0.142\) & \(0.059 \pm 0.126\) & \(0.150 \pm 0.144\) & \(0.256 \pm 0.172\) & \(0.515 \pm 0.234\) & \(0.634 \pm 0.186\) & .961 \\
    3 & \(\infty\) & 0.16 & \(0.156 \pm 0.133\) & \(0.116 \pm 0.121\) & \(-0.089 \pm 0.121\) & \(0.063 \pm 0.111\) & \(0.144 \pm 0.128\) & \(0.237 \pm 0.153\) & \(0.466 \pm 0.205\) & \(0.555 \pm 0.157\) & .959 \\
    \bottomrule
\end{tabular}
    \label{tab:pr_norm3}
\end{table}

\newpage

\subsubsection{Parameter Recovery Results Using Uncentred Two-Piece Distributions} \label{sm:equiv_model}

We also present parameter recovery results in Tables~\ref{tab:pr_alpha_uncentred}--~\ref{tab:pr_sigmaY_uncentred} for the model obtained with uncentred errors whose distribution is specified in (\ref{eq:tpd}), under the same prior specification described in Sections~\ref{sec:3.2} and~\ref{sec:4} and the simulation settings described in Section~\ref{sec:5.1}. Again, for simplicity, we present the results only for \(\alpha=\beta=0.4\), with \(\{\varepsilon_i^{(M)}\}_{i=1}^{n}\) and \(\{\varepsilon_i^{(Y)}\}_{i=1}^{n}\) drawn from the same uncentred two-piece distribution. The posterior summaries are very similar to those from the model with centred errors, empirically supporting the equivalence of the two models.

\vspace{0.5cm}

\setlength{\tabcolsep}{2.3pt}
\begin{table}[h!]
\centering
\fontsize{6}{10}\selectfont
\caption{\textbf{Parameter recovery results for \(\alpha\) in the equivalent model with uncentred errors.} Path coefficients \(\alpha\) and \(\beta\) are both fixed at 0.4, and \(\{\varepsilon_i^{(M)}\}_{i=1}^{n}\) and \(\{\varepsilon_i^{(Y)}\}_{i=1}^{n}\) are drawn from the same uncentred two-piece distribution. It presents the means ($\pm$ one standard deviation) of posterior means, modes, 2.5th, 25th, 50th, 75th, and 97.5th percentiles, the length of the \(95\%\) credible intervals for \(\alpha\), and the coverage rate of the true value by the credible intervals across \(1\,000\) experiments with sample size \(n = 50\).}
\begin{tabular}{*{12}{c}}
    \toprule
    \textbf{\(\gamma\)} & \textbf{\(\nu\)} & \textbf{True \(\alpha\)} & \textbf{Mean} & \textbf{Mode} & \textbf{2.5th} & \textbf{25th} & \textbf{50th} & \textbf{75th} & \textbf{97.5th} & \textbf{Length} & \textbf{Rate} \\
    \midrule
    0.33 & 3  & 0.4 & \(0.413 \pm 0.243\) & \(0.413 \pm 0.250\) & \(-0.055 \pm 0.269\) & \(0.262 \pm 0.244\) & \(0.413 \pm 0.243\) & \(0.564 \pm 0.251\) & \(0.879 \pm 0.288\) & \(0.934 \pm 0.254\) & .954 \\
    0.33 & 10 & 0.4 & \(0.401 \pm 0.196\) & \(0.399 \pm 0.201\) & \(-0.006 \pm 0.226\) & \(0.269 \pm 0.200\) & \(0.400 \pm 0.196\) & \(0.532 \pm 0.199\) & \(0.812 \pm 0.225\) & \(0.818 \pm 0.202\) & .960 \\
    0.33 & \(\infty\) & 0.4 & \(0.405 \pm 0.187\) & \(0.404 \pm 0.189\) & \(0.022 \pm 0.216\) & \(0.281 \pm 0.188\) & \(0.405 \pm 0.186\) & \(0.529 \pm 0.192\) & \(0.791 \pm 0.221\) & \(0.769 \pm 0.192\) & .953 \\
    0.5 & 3  & 0.4 & \(0.396 \pm 0.206\) & \(0.394 \pm 0.213\) & \(-0.014 \pm 0.224\) & \(0.259 \pm 0.207\) & \(0.395 \pm 0.207\) & \(0.532 \pm 0.212\) & \(0.808 \pm 0.232\) & \(0.822 \pm 0.199\) & .949 \\
    0.5 & 10 & 0.4 & \(0.398 \pm 0.173\) & \(0.399 \pm 0.173\) & \(0.056 \pm 0.189\) & \(0.285 \pm 0.174\) & \(0.398 \pm 0.173\) & \(0.511 \pm 0.175\) & \(0.738 \pm 0.191\) & \(0.682 \pm 0.147\) & .943 \\
    0.5 & \(\infty\) & 0.4 & \(0.408 \pm 0.163\) & \(0.407 \pm 0.166\) & \(0.089 \pm 0.179\) & \(0.302 \pm 0.165\) & \(0.408 \pm 0.163\) & \(0.514 \pm 0.164\) & \(0.728 \pm 0.177\) & \(0.639 \pm 0.139\) & .934 \\
    1 & 3  & 0.4 & \(0.396 \pm 0.193\) & \(0.396 \pm 0.193\) & \(0.016 \pm 0.204\) & \(0.268 \pm 0.193\) & \(0.396 \pm 0.193\) & \(0.523 \pm 0.196\) & \(0.775 \pm 0.212\) & \(0.759 \pm 0.155\) & .947 \\
    1 & 10 & 0.4 & \(0.403 \pm 0.160\) & \(0.402 \pm 0.164\) & \(0.089 \pm 0.167\) & \(0.296 \pm 0.161\) & \(0.403 \pm 0.161\) & \(0.509 \pm 0.162\) & \(0.718 \pm 0.168\) & \(0.630 \pm 0.106\) & .947 \\
    1 & \(\infty\) & 0.4 & \(0.399 \pm 0.154\) & \(0.399 \pm 0.157\) & \(0.107 \pm 0.160\) & \(0.300 \pm 0.155\) & \(0.399 \pm 0.154\) & \(0.498 \pm 0.155\) & \(0.693 \pm 0.161\) & \(0.586 \pm 0.097\) & .945 \\
    2 & 3  & 0.4 & \(0.396 \pm 0.194\) & \(0.395 \pm 0.196\) & \(-0.019 \pm 0.218\) & \(0.258 \pm 0.197\) & \(0.396 \pm 0.194\) & \(0.533 \pm 0.197\) & \(0.811 \pm 0.217\) & \(0.830 \pm 0.193\) & .970 \\
    2 & 10 & 0.4 & \(0.392 \pm 0.164\) & \(0.394 \pm 0.164\) & \(0.038 \pm 0.186\) & \(0.275 \pm 0.167\) & \(0.393 \pm 0.163\) & \(0.510 \pm 0.165\) & \(0.743 \pm 0.179\) & \(0.704 \pm 0.147\) & .968 \\
    2 & \(\infty\) & 0.4 & \(0.401 \pm 0.162\) & \(0.402 \pm 0.161\) & \(0.071 \pm 0.177\) & \(0.291 \pm 0.164\) & \(0.401 \pm 0.161\) & \(0.511 \pm 0.163\) & \(0.729 \pm 0.177\) & \(0.658 \pm 0.128\) & .957 \\
    3 & 3  & 0.4 & \(0.401 \pm 0.229\) & \(0.401 \pm 0.225\) & \(-0.099 \pm 0.271\) & \(0.238 \pm 0.234\) & \(0.401 \pm 0.227\) & \(0.563 \pm 0.229\) & \(0.899 \pm 0.259\) & \(0.998 \pm 0.235\) & .966 \\
    3 & 10 & 0.4 & \(0.408 \pm 0.201\) & \(0.409 \pm 0.198\) & \(-0.023 \pm 0.226\) & \(0.268 \pm 0.202\) & \(0.409 \pm 0.199\) & \(0.549 \pm 0.203\) & \(0.838 \pm 0.226\) & \(0.861 \pm 0.177\) & .964 \\
    3 & \(\infty\) & 0.4 & \(0.398 \pm 0.190\) & \(0.397 \pm 0.186\) & \(-0.010 \pm 0.215\) & \(0.264 \pm 0.191\) & \(0.397 \pm 0.188\) & \(0.531 \pm 0.192\) & \(0.807 \pm 0.217\) & \(0.817 \pm 0.168\) & .966 \\
    \bottomrule
\end{tabular}
    \label{tab:pr_alpha_uncentred}
\end{table}

\setlength{\tabcolsep}{1.8pt}
\begin{table}[h!]
\centering
\fontsize{6}{10}\selectfont
\caption{\textbf{Parameter recovery results for \(\beta\) in the equivalent model with uncentred errors.} Path coefficients \(\alpha\) and \(\beta\) are both fixed at 0.4, and \(\{\varepsilon_i^{(M)}\}_{i=1}^{n}\) and \(\{\varepsilon_i^{(Y)}\}_{i=1}^{n}\) are drawn from the same uncentred two-piece distribution. It presents the means ($\pm$ one standard deviation) of posterior means, modes, 2.5th, 25th, 50th, 75th, and 97.5th percentiles, the length of the \(95\%\) credible intervals for \(\beta\), and the coverage rate of the true value by the credible intervals across \(1\,000\) experiments with sample size \(n = 50\).}
\begin{tabular}{*{12}{c}}
    \toprule
    \textbf{\(\gamma\)} & \textbf{\(\nu\)} & \textbf{True \(\beta\)} & \textbf{Mean} & \textbf{Mode} & \textbf{2.5th} & \textbf{25th} & \textbf{50th} & \textbf{75th} & \textbf{97.5th} & \textbf{Length} & \textbf{Rate} \\
    \midrule
    0.33 & 3  & 0.4 & \(0.413 \pm 0.243\) & \(0.413 \pm 0.250\) & \( -0.055 \pm 0.269\) & \(0.262 \pm 0.244\) & \(0.413 \pm 0.243\) & \(0.564 \pm 0.251\) & \(0.879 \pm 0.288\) & \(0.934 \pm 0.254\) & .954 \\
    0.33 & 10 & 0.4 & \(0.405 \pm 0.092\) & \(0.418 \pm 0.093\) & \(0.199 \pm 0.107\) & \(0.343 \pm 0.094\) & \(0.409 \pm 0.092\) & \(0.470 \pm 0.094\) & \(0.590 \pm 0.110\) & \(0.391 \pm 0.103\) & .959 \\
    0.33 & \(\infty\) & 0.4 & \(0.398 \pm 0.102\) & \(0.405 \pm 0.100\) & \(0.178 \pm 0.118\) & \(0.330 \pm 0.104\) & \(0.401 \pm 0.101\) & \(0.469 \pm 0.102\) & \(0.607 \pm 0.115\) & \(0.429 \pm 0.096\) & .965 \\
    0.5 & 3  & 0.4 & \(0.395 \pm 0.100\) & \(0.406 \pm 0.103\) & \(0.192 \pm 0.123\) & \(0.333 \pm 0.104\) & \(0.399 \pm 0.100\) & \(0.460 \pm 0.102\) & \(0.577 \pm 0.118\) & \(0.385 \pm 0.133\) & .946 \\
    0.5 & 10 & 0.4 & \(0.407 \pm 0.127\) & \(0.415 \pm 0.130\) & \(0.164 \pm 0.137\) & \(0.330 \pm 0.128\) & \(0.409 \pm 0.127\) & \(0.486 \pm 0.129\) & \(0.637 \pm 0.141\) & \(0.473 \pm 0.109\) & .932 \\
    0.5 & \(\infty\) & 0.4 & \(0.406 \pm 0.121\) & \(0.413 \pm 0.124\) & \(0.161 \pm 0.131\) & \(0.328 \pm 0.122\) & \(0.408 \pm 0.121\) & \(0.487 \pm 0.123\) & \(0.642 \pm 0.134\) & \(0.482 \pm 0.106\) & .942 \\
    1 & 3  & 0.4 & \(0.398 \pm 0.132\) & \(0.399 \pm 0.134\) & \(0.142 \pm 0.149\) & \(0.313 \pm 0.134\) & \(0.398 \pm 0.133\) & \(0.483 \pm 0.135\) & \(0.652 \pm 0.150\) & \(0.510 \pm 0.141\) & .952 \\
    1 & 10 & 0.4 & \(0.404 \pm 0.150\) & \(0.404 \pm 0.151\) & \(0.117 \pm 0.158\) & \(0.307 \pm 0.151\) & \(0.404 \pm 0.150\) & \(0.501 \pm 0.151\) & \(0.692 \pm 0.158\) & \(0.575 \pm 0.108\) & .943 \\
    1 & \(\infty\) & 0.4 & \(0.395 \pm 0.152\) & \(0.396 \pm 0.154\) & \(0.102 \pm 0.158\) & \(0.296 \pm 0.153\) & \(0.395 \pm 0.153\) & \(0.494 \pm 0.154\) & \(0.688 \pm 0.160\) & \(0.587 \pm 0.100\) & .942 \\
    2 & 3  & 0.4 & \(0.399 \pm 0.091\) & \(0.410 \pm 0.092\) & \(0.196 \pm 0.113\) & \(0.337 \pm 0.094\) & \(0.403 \pm 0.091\) & \(0.464 \pm 0.094\) & \(0.583 \pm 0.115\) & \(0.387 \pm 0.130\) & .962 \\
    2 & 10 & 0.4 & \(0.407 \pm 0.120\) & \(0.412 \pm 0.121\) & \(0.164 \pm 0.132\) & \(0.329 \pm 0.122\) & \(0.409 \pm 0.120\) & \(0.487 \pm 0.121\) & \(0.640 \pm 0.132\) & \(0.476 \pm 0.103\) & .947 \\
    2 & \(\infty\) & 0.4 & \(0.400 \pm 0.120\) & \(0.405 \pm 0.121\) & \(0.147 \pm 0.130\) & \(0.318 \pm 0.122\) & \(0.402 \pm 0.120\) & \(0.483 \pm 0.121\) & \(0.645 \pm 0.131\) & \(0.497 \pm 0.092\) & .955 \\
    3 & 3  & 0.4 & \(0.399 \pm 0.080\) & \(0.412 \pm 0.080\) & \(0.218 \pm 0.097\) & \(0.346 \pm 0.081\) & \(0.403 \pm 0.080\) & \(0.456 \pm 0.084\) & \(0.558 \pm 0.106\) & \(0.341 \pm 0.114\) & .978 \\
    3 & 10 & 0.4 & \(0.400 \pm 0.101\) & \(0.409 \pm 0.100\) & \(0.180 \pm 0.113\) & \(0.332 \pm 0.102\) & \(0.403 \pm 0.100\) & \(0.471 \pm 0.103\) & \(0.604 \pm 0.117\) & \(0.424 \pm 0.094\) & .963 \\
    3 & \(\infty\) & 0.4 & \(0.402 \pm 0.111\) & \(0.409 \pm 0.110\) & \(0.171 \pm 0.123\) & \(0.330 \pm 0.112\) & \(0.405 \pm 0.111\) & \(0.477 \pm 0.113\) & \(0.622 \pm 0.127\) & \(0.451 \pm 0.097\) & .951 \\
    \bottomrule
\end{tabular}
    \label{tab:pr_beta_uncentred}
\end{table}

\setlength{\tabcolsep}{1.8pt}
\begin{table}[h!]
\centering
\fontsize{6}{10}\selectfont
\caption{\textbf{Parameter recovery results for \(\alpha\beta\) in the equivalent model with uncentred errors.} Path coefficients \(\alpha\) and \(\beta\) are both fixed at 0.4, and \(\{\varepsilon_i^{(M)}\}_{i=1}^{n}\) and \(\{\varepsilon_i^{(Y)}\}_{i=1}^{n}\) are drawn from the same uncentred two-piece distribution. It presents the means ($\pm$ one standard deviation) of posterior means, modes, 2.5th, 25th, 50th, 75th, and 97.5th percentiles, the length of the \(95\%\) credible intervals for \(\alpha\beta\), and the coverage rate of the true value by the credible intervals across \(1\,000\) experiments with sample size \(n = 50\).}
\begin{tabular}{*{12}{c}}
    \toprule
    \textbf{\(\gamma\)} & \textbf{\(\nu\)} & \textbf{True \(\alpha\beta\)} & \textbf{Mean} & \textbf{Mode} & \textbf{2.5th} & \textbf{25th} & \textbf{50th} & \textbf{75th} & \textbf{97.5th} & \textbf{Length} & \textbf{Rate} \\
    \midrule
    0.33 & 3  & 0.16 & \(0.165 \pm 0.102\) & \(0.151 \pm 0.100\) & \(-0.030 \pm 0.104\) & \(0.096 \pm 0.094\) & \(0.160 \pm 0.101\) & \(0.229 \pm 0.112\) & \(0.384 \pm 0.146\) & \(0.413 \pm 0.129\) & .967 \\
    0.33 & 10 & 0.16 & \(0.161 \pm 0.088\) & \(0.143 \pm 0.086\) & \(-0.015 \pm 0.086\) & \(0.097 \pm 0.079\) & \(0.155 \pm 0.086\) & \(0.220 \pm 0.097\) & \(0.371 \pm 0.129\) & \(0.386 \pm 0.109\) & .973 \\
    0.33 & \(\infty\) & 0.16 & \(0.161 \pm 0.088\) & \(0.142 \pm 0.085\) & \(-0.008 \pm 0.081\) & \(0.098 \pm 0.077\) & \(0.155 \pm 0.086\) & \(0.218 \pm 0.099\) & \(0.367 \pm 0.133\) & \(0.375 \pm 0.110\) & .960 \\
    0.5 & 3  & 0.16 & \(0.157 \pm 0.093\) & \(0.138 \pm 0.092\) & \(-0.018 \pm 0.084\) & \(0.091 \pm 0.083\) & \(0.150 \pm 0.092\) & \(0.216 \pm 0.105\) & \(0.364 \pm 0.137\) & \(0.382 \pm 0.113\) & .963 \\
    0.5 & 10 & 0.16 & \(0.162 \pm 0.091\) & \(0.142 \pm 0.090\) & \(0.003 \pm 0.072\) & \(0.100 \pm 0.078\) & \(0.155 \pm 0.090\) & \(0.217 \pm 0.104\) & \(0.360 \pm 0.136\) & \(0.357 \pm 0.104\) & .946 \\
    0.5 & \(\infty\) & 0.16 & \(0.167 \pm 0.087\) & \(0.146 \pm 0.087\) & \(0.014 \pm 0.065\) & \(0.107 \pm 0.075\) & \(0.160 \pm 0.086\) & \(0.220 \pm 0.100\) & \(0.359 \pm 0.129\) & \(0.345 \pm 0.094\) & .938 \\
    1 & 3  & 0.16 & \(0.157 \pm 0.096\) & \(0.131 \pm 0.092\) & \(-0.016 \pm 0.079\) & \(0.089 \pm 0.082\) & \(0.148 \pm 0.094\) & \(0.217 \pm 0.109\) & \(0.377 \pm 0.144\) & \(0.393 \pm 0.108\) & .949 \\
    1 & 10 & 0.16 & \(0.163 \pm 0.091\) & \(0.136 \pm 0.090\) & \(0.003 \pm 0.065\) & \(0.098 \pm 0.077\) & \(0.154 \pm 0.090\) & \(0.219 \pm 0.106\) & \(0.373 \pm 0.137\) & \(0.370 \pm 0.098\) & .950 \\
    1 & \(\infty\) & 0.16 & \(0.159 \pm 0.090\) & \(0.132 \pm 0.089\) & \(0.006 \pm 0.063\) & \(0.096 \pm 0.076\) & \(0.150 \pm 0.089\) & \(0.212 \pm 0.104\) & \(0.359 \pm 0.135\) & \(0.354 \pm 0.094\) & .935 \\
    2 & 3  & 0.16 & \(0.158 \pm 0.087\) & \(0.140 \pm 0.085\) & \(-0.018 \pm 0.082\) & \(0.092 \pm 0.078\) & \(0.152 \pm 0.086\) & \(0.218 \pm 0.097\) & \(0.367 \pm 0.126\) & \(0.385 \pm 0.106\) & .974 \\
    2 & 10 & 0.16 & \(0.160 \pm 0.085\) & \(0.137 \pm 0.083\) & \(-0.001 \pm 0.069\) & \(0.097 \pm 0.073\) & \(0.153 \pm 0.083\) & \(0.216 \pm 0.096\) & \(0.363 \pm 0.127\) & \(0.364 \pm 0.097\) & .963 \\
    2 & \(\infty\) & 0.16 & \(0.161 \pm 0.085\) & \(0.138 \pm 0.083\) & \(0.006 \pm 0.065\) & \(0.099 \pm 0.072\) & \(0.153 \pm 0.083\) & \(0.214 \pm 0.097\) & \(0.357 \pm 0.127\) & \(0.351 \pm 0.093\) & .959 \\
    3 & 3  & 0.16 & \(0.161 \pm 0.100\) & \(0.144 \pm 0.094\) & \(-0.046 \pm 0.106\) & \(0.087 \pm 0.091\) & \(0.155 \pm 0.097\) & \(0.228 \pm 0.108\) & \(0.396 \pm 0.143\) & \(0.442 \pm 0.122\) & .970 \\
    3 & 10 & 0.16 & \(0.163 \pm 0.092\) & \(0.142 \pm 0.087\) & \(-0.022 \pm 0.085\) & \(0.094 \pm 0.080\) & \(0.156 \pm 0.089\) & \(0.225 \pm 0.103\) & \(0.387 \pm 0.139\) & \(0.408 \pm 0.112\) & .975 \\
    3 & \(\infty\) & 0.16 & \(0.160 \pm 0.091\) & \(0.137 \pm 0.086\) & \(-0.019 \pm 0.083\) & \(0.092 \pm 0.079\) & \(0.152 \pm 0.088\) & \(0.220 \pm 0.102\) & \(0.379 \pm 0.137\) & \(0.398 \pm 0.110\) & .972 \\
    \bottomrule
\end{tabular}
    \label{tab:pr_med_uncentred}
\end{table}

\setlength{\tabcolsep}{1.8pt}
\begin{table}[h!]
\centering
\fontsize{6}{10}\selectfont
\caption{\textbf{Parameter recovery results for \(\tau\) in the equivalent model with uncentred errors.} Path coefficients \(\alpha\) and \(\beta\) are both fixed at 0.4, and \(\{\varepsilon_i^{(M)}\}_{i=1}^{n}\) and \(\{\varepsilon_i^{(Y)}\}_{i=1}^{n}\) are drawn from the same uncentred two-piece distribution. It presents the means ($\pm$ one standard deviation) of posterior means, modes, 2.5th, 25th, 50th, 75th, and 97.5th percentiles, the length of the \(95\%\) credible intervals for \(\tau\), and the coverage rate of the true value by the credible intervals across \(1\,000\) experiments with sample size \(n = 50\).}
\begin{tabular}{*{12}{c}}
    \toprule
    \textbf{\(\gamma\)} & \textbf{\(\nu\)} & \textbf{True \(\tau\)} & \textbf{Mean} & \textbf{Mode} & \textbf{2.5th} & \textbf{25th} & \textbf{50th} & \textbf{75th} & \textbf{97.5th} & \textbf{Length} & \textbf{Rate} \\
    \midrule
    0.33 & 3 & 0.2 & \(0.205 \pm 0.237\) & \(0.202 \pm 0.243\) & \(-0.285 \pm 0.275\) & \(0.045 \pm 0.242\) & \(0.204 \pm 0.237\) & \(0.364 \pm 0.243\) & \(0.701 \pm 0.282\) & \(0.986 \pm 0.269\) & .954 \\
    0.33 & 10 & 0.2 & \(0.186 \pm 0.201\) & \(0.184 \pm 0.204\) & \(-0.244 \pm 0.238\) & \(0.047 \pm 0.205\) & \(0.186 \pm 0.200\) & \(0.325 \pm 0.203\) & \(0.617 \pm 0.232\) & \(0.861 \pm 0.215\) & .960 \\
    0.33 & \(\infty\) & 0.2 & \(0.189 \pm 0.204\) & \(0.190 \pm 0.205\) & \(-0.228 \pm 0.236\) & \(0.054 \pm 0.209\) & \(0.189 \pm 0.203\) & \(0.324 \pm 0.205\) & \(0.603 \pm 0.231\) & \(0.830 \pm 0.200\) & .952 \\
    0.5 & 3 & 0.2 & \(0.189 \pm 0.217\) & \(0.189 \pm 0.221\) & \(-0.233 \pm 0.244\) & \(0.050 \pm 0.221\) & \(0.189 \pm 0.218\) & \(0.328 \pm 0.219\) & \(0.612 \pm 0.239\) & \(0.844 \pm 0.204\) & .940 \\
    0.5 & 10 & 0.2 & \(0.190 \pm 0.186\) & \(0.190 \pm 0.191\) & \(-0.173 \pm 0.207\) & \(0.070 \pm 0.191\) & \(0.190 \pm 0.187\) & \(0.311 \pm 0.188\) & \(0.555 \pm 0.199\) & \(0.728 \pm 0.161\) & .947 \\
    0.5 & \(\infty\) & 0.2 & \(0.189 \pm 0.176\) & \(0.189 \pm 0.178\) & \(-0.149 \pm 0.192\) & \(0.078 \pm 0.178\) & \(0.189 \pm 0.176\) & \(0.301 \pm 0.178\) & \(0.526 \pm 0.191\) & \(0.675 \pm 0.144\) & .935 \\
    1 & 3 & 0.2 & \(0.200 \pm 0.202\) & \(0.200 \pm 0.204\) & \(-0.204 \pm 0.221\) & \(0.065 \pm 0.205\) & \(0.200 \pm 0.202\) & \(0.336 \pm 0.203\) & \(0.604 \pm 0.215\) & \(0.807 \pm 0.160\) & .952 \\
    1 & 10 & 0.2 & \(0.204 \pm 0.177\) & \(0.205 \pm 0.179\) & \(-0.135 \pm 0.184\) & \(0.090 \pm 0.178\) & \(0.204 \pm 0.177\) & \(0.319 \pm 0.179\) & \(0.544 \pm 0.185\) & \(0.678 \pm 0.112\) & .946 \\
    1 & \(\infty\) & 0.2 & \(0.207 \pm 0.174\) & \(0.206 \pm 0.176\) & \(-0.113 \pm 0.179\) & \(0.099 \pm 0.174\) & \(0.207 \pm 0.174\) & \(0.315 \pm 0.176\) & \(0.525 \pm 0.183\) & \(0.638 \pm 0.109\) & .923 \\
    2 & 3 & 0.2 & \(0.195 \pm 0.209\) & \(0.196 \pm 0.210\) & \(-0.237 \pm 0.234\) & \(0.052 \pm 0.211\) & \(0.195 \pm 0.208\) & \(0.339 \pm 0.211\) & \(0.628 \pm 0.233\) & \(0.864 \pm 0.200\) & .958 \\
    2 & 10 & 0.2 & \(0.189 \pm 0.185\) & \(0.191 \pm 0.186\) & \(-0.185 \pm 0.202\) & \(0.065 \pm 0.187\) & \(0.189 \pm 0.185\) & \(0.313 \pm 0.187\) & \(0.562 \pm 0.204\) & \(0.747 \pm 0.155\) & .946 \\
    2 & \(\infty\) & 0.2 & \(0.207 \pm 0.176\) & \(0.206 \pm 0.176\) & \(-0.144 \pm 0.190\) & \(0.090 \pm 0.177\) & \(0.207 \pm 0.176\) & \(0.324 \pm 0.179\) & \(0.559 \pm 0.194\) & \(0.703 \pm 0.139\) & .956 \\
    3 & 3 & 0.2 & \(0.204 \pm 0.244\) & \(0.203 \pm 0.243\) & \(-0.326 \pm 0.280\) & \(0.031 \pm 0.247\) & \(0.204 \pm 0.242\) & \(0.377 \pm 0.247\) & \(0.733 \pm 0.280\) & \(1.059 \pm 0.242\) & .966 \\
    3 & 10 & 0.2 & \(0.196 \pm 0.221\) & \(0.195 \pm 0.217\) & \(-0.265 \pm 0.254\) & \(0.045 \pm 0.225\) & \(0.196 \pm 0.219\) & \(0.347 \pm 0.221\) & \(0.656 \pm 0.244\) & \(0.921 \pm 0.196\) & .966 \\
    3 & \(\infty\) & 0.2 & \(0.207 \pm 0.225\) & \(0.206 \pm 0.222\) & \(-0.226 \pm 0.245\) & \(0.064 \pm 0.225\) & \(0.207 \pm 0.223\) & \(0.349 \pm 0.227\) & \(0.642 \pm 0.250\) & \(0.868 \pm 0.178\) & .947 \\
    \bottomrule
\end{tabular}
    \label{tab:pr_tau_uncentred}
\end{table}

\setlength{\tabcolsep}{1.8pt}
\begin{table}[h!]
\centering
\fontsize{6}{10}\selectfont
\caption{\textbf{Parameter recovery results for \(\gamma\) corresponding to \(\varepsilon^{(M)}\) in the equivalent model with uncentred errors.} Path coefficients \(\alpha\) and \(\beta\) are both fixed at 0.4, and \(\{\varepsilon_i^{(M)}\}_{i=1}^{n}\) and \(\{\varepsilon_i^{(Y)}\}_{i=1}^{n}\) are drawn from the same uncentred two-piece distribution. It presents the means ($\pm$ one standard deviation) of posterior means, modes, 2.5th, 25th, 50th, 75th, and 97.5th percentiles, the length of the \(95\%\) credible intervals for \(\gamma\), and the coverage rate of the true value by the credible intervals across \(1\,000\) experiments with sample size \(n = 50\).}
\begin{tabular}{*{12}{c}}
    \toprule
    \textbf{\(\gamma\)} & \textbf{\(\nu\)} & \textbf{True \(\gamma\)} & \textbf{Mean} & \textbf{Mode} & \textbf{2.5th} & \textbf{25th} & \textbf{50th} & \textbf{75th} & \textbf{97.5th} & \textbf{Length} & \textbf{Rate}\\
    \midrule
    0.33 & 3  & 0.33 & \(0.357 \pm 0.109\) & \(0.326 \pm 0.123\) & \(0.133 \pm 0.084\) & \(0.263 \pm 0.105\) & \(0.348 \pm 0.110\) & \(0.440 \pm 0.116\) & \(0.640 \pm 0.137\) & \(0.506 \pm 0.088\) & .960\\
    0.33 & 10 & 0.33 & \(0.387 \pm 0.116\) & \(0.344 \pm 0.135\) & \(0.132 \pm 0.078\) & \(0.277 \pm 0.109\) & \(0.375 \pm 0.118\) & \(0.482 \pm 0.131\) & \(0.720 \pm 0.169\) & \(0.588 \pm 0.131\) & .969\\
    0.33 & \(\infty\) & 0.33 & \(0.410 \pm 0.130\) & \(0.363 \pm 0.152\) & \(0.133 \pm 0.075\) & \(0.289 \pm 0.114\) & \(0.395 \pm 0.131\) & \(0.512 \pm 0.151\) & \(0.779 \pm 0.216\) & \(0.646 \pm 0.181\) & .973\\
    0.5 & 3  & 0.5 & \(0.509 \pm 0.150\) & \(0.482 \pm 0.153\) & \(0.255 \pm 0.138\) & \(0.410 \pm 0.146\) & \(0.501 \pm 0.148\) & \(0.597 \pm 0.155\) & \(0.812 \pm 0.186\) & \(0.557 \pm 0.105\) & .939\\
    0.5 & 10 & 0.5 & \(0.527 \pm 0.149\) & \(0.492 \pm 0.152\) & \(0.237 \pm 0.125\) & \(0.411 \pm 0.140\) & \(0.515 \pm 0.146\) & \(0.629 \pm 0.160\) & \(0.892 \pm 0.211\) & \(0.655 \pm 0.147\) & .983\\
    0.5 & \(\infty\) & 0.5 & \(0.549 \pm 0.167\) & \(0.507 \pm 0.171\) & \(0.236 \pm 0.129\) & \(0.422 \pm 0.152\) & \(0.535 \pm 0.164\) & \(0.659 \pm 0.185\) & \(0.947 \pm 0.261\) & \(0.712 \pm 0.198\) & .971\\
    1 & 3  & 1 & \(1.046 \pm 0.240\) & \(0.983 \pm 0.213\) & \(0.679 \pm 0.173\) & \(0.892 \pm 0.201\) & \(1.023 \pm 0.228\) & \(1.174 \pm 0.268\) & \(1.543 \pm 0.402\) & \(0.864 \pm 0.275\) & .935\\
    1 & 10 & 1 & \(1.041 \pm 0.252\) & \(0.964 \pm 0.223\) & \(0.632 \pm 0.180\) & \(0.866 \pm 0.209\) & \(1.013 \pm 0.239\) & \(1.184 \pm 0.285\) & \(1.617 \pm 0.442\) & \(0.985 \pm 0.329\) & .960\\
    1 & \(\infty\) & 1 & \(1.063 \pm 0.262\) & \(0.975 \pm 0.233\) & \(0.621 \pm 0.178\) & \(0.872 \pm 0.213\) & \(1.030 \pm 0.247\) & \(1.216 \pm 0.300\) & \(1.696 \pm 0.484\) & \(1.075 \pm 0.379\) & .969\\
    2 & 3  & 2 & \(1.989 \pm 0.410\) & \(1.799 \pm 0.345\) & \(1.234 \pm 0.222\) & \(1.642 \pm 0.312\) & \(1.918 \pm 0.386\) & \(2.256 \pm 0.485\) & \(3.161 \pm 0.770\) & \(1.926 \pm 0.588\) & .954\\
    2 & 10 & 2 & \(1.922 \pm 0.385\) & \(1.713 \pm 0.338\) & \(1.138 \pm 0.218\) & \(1.557 \pm 0.299\) & \(1.843 \pm 0.366\) & \(2.198 \pm 0.456\) & \(3.168 \pm 0.720\) & \(2.029 \pm 0.567\) & .968\\
    2 & \(\infty\) & 2 & \(1.863 \pm 0.369\) & \(1.650 \pm 0.334\) & \(1.070 \pm 0.220\) & \(1.493 \pm 0.293\) & \(1.783 \pm 0.355\) & \(2.141 \pm 0.436\) & \(3.122 \pm 0.668\) & \(2.051 \pm 0.530\) & .967\\
    3 & 3  & 3 & \(2.467 \pm 0.375\) & \(2.192 \pm 0.338\) & \(1.460 \pm 0.211\) & \(1.993 \pm 0.295\) & \(2.367 \pm 0.360\) & \(2.829 \pm 0.444\) & \(4.053 \pm 0.671\) & \(2.594 \pm 0.515\) & .922\\
    3 & 10 & 3 & \(2.316 \pm 0.365\) & \(2.034 \pm 0.338\) & \(1.324 \pm 0.216\) & \(1.846 \pm 0.294\) & \(2.213 \pm 0.354\) & \(2.671 \pm 0.430\) & \(3.899 \pm 0.627\) & \(2.575 \pm 0.476\) & .900\\
    3 & \(\infty\) & 3 & \(2.219 \pm 0.377\) & \(1.939 \pm 0.361\) & \(1.238 \pm 0.241\) & \(1.754 \pm 0.314\) & \(2.116 \pm 0.370\) & \(2.570 \pm 0.441\) & \(3.796 \pm 0.621\) & \(2.558 \pm 0.473\) & .894\\
    \bottomrule
\end{tabular}
    \label{tab:pr_gamma_1_uncentred}
\end{table}

\setlength{\tabcolsep}{1.8pt}
\begin{table}[h!]
\centering
\fontsize{6}{10}\selectfont
\caption{\textbf{Parameter recovery results for \(\gamma\) corresponding to \(\varepsilon^{(Y)}\) in the equivalent model with uncentred errors.} Path coefficients \(\alpha\) and \(\beta\) are both fixed at 0.4, and \(\{\varepsilon_i^{(M)}\}_{i=1}^{n}\) and \(\{\varepsilon_i^{(Y)}\}_{i=1}^{n}\) are drawn from the same uncentred two-piece distribution. It presents the means ($\pm$ one standard deviation) of posterior means, modes, 2.5th, 25th, 50th, 75th, and 97.5th percentiles, the length of the \(95\%\) credible intervals for \(\gamma\), and the coverage rate of the true value by the credible intervals across \(1\,000\) experiments with sample size \(n = 50\).}
\begin{tabular}{*{12}{c}}
    \toprule
    \textbf{\(\gamma\)} & \textbf{\(\nu\)} & \textbf{True \(\gamma\)} & \textbf{Mean} & \textbf{Mode} & \textbf{2.5th} & \textbf{25th} & \textbf{50th} & \textbf{75th} & \textbf{97.5th} & \textbf{Length} & \textbf{Rate} \\
    \midrule
    0.33 & 3  & 0.33 & \(0.352 \pm 0.104\) & \(0.320 \pm 0.119\) & \(0.120 \pm 0.072\) & \(0.252 \pm 0.099\) & \(0.342 \pm 0.106\) & \(0.439 \pm 0.115\) & \(0.650 \pm 0.141\) & \(0.530 \pm 0.102\) & .975 \\
    0.33 & 10 & 0.33 & \(0.389 \pm 0.119\) & \(0.347 \pm 0.141\) & \(0.125 \pm 0.071\) & \(0.272 \pm 0.109\) & \(0.376 \pm 0.120\) & \(0.489 \pm 0.135\) & \(0.740 \pm 0.181\) & \(0.615 \pm 0.142\) & .975 \\
    0.33 & \(\infty\) & 0.33 & \(0.432 \pm 0.143\) & \(0.363 \pm 0.152\) & \(0.139 \pm 0.079\) & \(0.304 \pm 0.125\) & \(0.417 \pm 0.143\) & \(0.540 \pm 0.166\) & \(0.823 \pm 0.239\) & \(0.685 \pm 0.196\) & .964 \\
    0.5 & 3  & 0.5 & \(0.503 \pm 0.151\) & \(0.476 \pm 0.155\) & \(0.237 \pm 0.138\) & \(0.398 \pm 0.148\) & \(0.494 \pm 0.149\) & \(0.596 \pm 0.156\) & \(0.823 \pm 0.192\) & \(0.587 \pm 0.116\) & .954 \\
    0.5 & 10 & 0.5 & \(0.531 \pm 0.161\) & \(0.492 \pm 0.171\) & \(0.226 \pm 0.129\) & \(0.408 \pm 0.152\) & \(0.519 \pm 0.159\) & \(0.639 \pm 0.174\) & \(0.911 \pm 0.236\) & \(0.686 \pm 0.171\) & .970 \\
    0.5 & \(\infty\) & 0.5 & \(0.543 \pm 0.180\) & \(0.497 \pm 0.186\) & \(0.220 \pm 0.129\) & \(0.410 \pm 0.162\) & \(0.528 \pm 0.177\) & \(0.657 \pm 0.200\) & \(0.964 \pm 0.290\) & \(0.744 \pm 0.223\) & .971 \\
    1 & 3  & 1 & \(1.058 \pm 0.253\) & \(0.991 \pm 0.215\) & \(0.677 \pm 0.174\) & \(0.897 \pm 0.204\) & \(1.033 \pm 0.237\) & \(1.190 \pm 0.286\) & \(1.585 \pm 0.464\) & \(0.909 \pm 0.338\) & .943 \\
    1 & 10 & 1 & \(1.055 \pm 0.252\) & \(0.973 \pm 0.220\) & \(0.629 \pm 0.181\) & \(0.873 \pm 0.207\) & \(1.024 \pm 0.237\) & \(1.202 \pm 0.285\) & \(1.663 \pm 0.462\) & \(1.034 \pm 0.352\) & .960 \\
    1 & \(\infty\) & 1 & \(1.073 \pm 0.289\) & \(0.976 \pm 0.233\) & \(0.607 \pm 0.193\) & \(0.872 \pm 0.213\) & \(1.036 \pm 0.270\) & \(1.233 \pm 0.331\) & \(1.759 \pm 0.551\) & \(1.152 \pm 0.431\) & .955 \\
    2 & 3  & 2 & \(1.993 \pm 0.405\) & \(1.787 \pm 0.345\) & \(1.217 \pm 0.224\) & \(1.631 \pm 0.309\) & \(1.915 \pm 0.380\) & \(2.267 \pm 0.479\) & \(3.224 \pm 0.768\) & \(2.007 \pm 0.590\) & .960 \\
    2 & 10 & 2 & \(1.896 \pm 0.361\) & \(1.673 \pm 0.311\) & \(1.101 \pm 0.210\) & \(1.521 \pm 0.280\) & \(1.812 \pm 0.341\) & \(2.176 \pm 0.427\) & \(3.180 \pm 0.683\) & \(2.080 \pm 0.540\) & .964 \\
    2 & \(\infty\) & 2 & \(1.842 \pm 0.368\) & \(1.615 \pm 0.327\) & \(1.042 \pm 0.217\) & \(1.463 \pm 0.289\) & \(1.756 \pm 0.351\) & \(2.123 \pm 0.435\) & \(3.142 \pm 0.677\) & \(2.100 \pm 0.534\) & .963 \\
    3 & 3  & 3 & \(2.408 \pm 0.363\) & \(2.129 \pm 0.326\) & \(1.420 \pm 0.206\) & \(1.940 \pm 0.286\) & \(2.306 \pm 0.349\) & \(2.762 \pm 0.430\) & \(3.989 \pm 0.639\) & \(2.569 \pm 0.487\) & .909 \\
    3 & 10 & 3 & \(2.240 \pm 0.365\) & \(1.954 \pm 0.341\) & \(1.260 \pm 0.225\) & \(1.772 \pm 0.298\) & \(2.134 \pm 0.356\) & \(2.590 \pm 0.429\) & \(3.831 \pm 0.610\) & \(2.571 \pm 0.462\) & .898 \\
    3 & \(\infty\) & 3 & \(2.161 \pm 0.363\) & \(1.876 \pm 0.344\) & \(1.192 \pm 0.224\) & \(1.697 \pm 0.298\) & \(2.055 \pm 0.355\) & \(2.507 \pm 0.428\) & \(3.743 \pm 0.611\) & \(2.551 \pm 0.469\) & .876 \\
    \bottomrule
\end{tabular}
    \label{tab:pr_gamma_2_centred}
\end{table}

\setlength{\tabcolsep}{1.8pt}
\begin{table}[h!]
\centering
\fontsize{6}{9}\selectfont
\caption{\textbf{Parameter recovery results for \(\nu\) corresponding to \(\varepsilon^{(M)}\) in the equivalent model with uncentred errors.} Path coefficients \(\alpha\) and \(\beta\) are both fixed at 0.4, and \(\{\varepsilon_i^{(M)}\}_{i=1}^{n}\) and \(\{\varepsilon_i^{(Y)}\}_{i=1}^{n}\) are drawn from the same uncentred two-piece distribution. It presents the means ($\pm$ one standard deviation) of posterior means, modes, 2.5th, 25th, 50th, 75th, and 97.5th percentiles, the length of the \(95\%\) credible intervals for \(\nu\), and the coverage rate of the true value by the credible intervals across \(1\,000\) experiments with sample size \(n = 50\).}
\begin{tabular}{*{12}{c}}
    \toprule
    \textbf{\(\gamma\)} & \textbf{\(\nu\)} & \textbf{True \(\nu\)} & \textbf{Mean} & \textbf{Mode} & \textbf{2.5th} & \textbf{25th} & \textbf{50th} & \textbf{75th} & \textbf{97.5th} & \textbf{Length} & \textbf{Rate} \\
    \midrule
    0.33 & 3           & 3   & \(7.601 \pm 3.960\) & \(3.695 \pm 1.619\) & \(2.348 \pm 0.419\) & \(3.943 \pm 1.677\) & \(5.904 \pm 3.080\) & \(9.317 \pm 5.350\) & \(22.817 \pm 12.786\) & \(20.469 \pm 12.445\) & .904 \\
    0.33 & 10          & 10  & \(12.998 \pm 3.098\) & \(6.180 \pm 1.953\) & \(3.012 \pm 0.603\) & \(6.441 \pm 1.762\) & \(10.317 \pm 2.827\) & \(16.637 \pm 4.223\) & \(38.198 \pm 7.216\) & \(35.186 \pm 6.725\) & .997 \\
    0.33 & \(\infty\)  & \(\infty\) & \(14.830 \pm 2.391\) & \(7.392 \pm 1.978\) & \(3.373 \pm 0.618\) & \(7.524 \pm 1.586\) & \(12.028 \pm 2.341\) & \(19.110 \pm 3.180\) & \(42.085 \pm 4.549\) & \(38.712 \pm 4.025\) & .000 \\
    0.5 & 3           & 3   & \(7.889 \pm 3.898\) & \(3.759 \pm 1.597\) & \(2.358 \pm 0.409\) & \(4.023 \pm 1.655\) & \(6.077 \pm 3.042\) & \(9.690 \pm 5.296\) & \(24.030 \pm 12.442\) & \(21.673 \pm 12.110\) & .911 \\
    0.5 & 10          & 10  & \(13.413 \pm 3.121\) & \(6.473 \pm 2.020\) & \(3.107 \pm 0.623\) & \(6.703 \pm 1.818\) & \(10.711 \pm 2.893\) & \(17.182 \pm 4.249\) & \(39.055 \pm 6.963\) & \(35.947 \pm 6.437\) & .996 \\
    0.5 & \(\infty\)  & \(\infty\) & \(14.978 \pm 2.417\) & \(7.512 \pm 2.019\) & \(3.412 \pm 0.621\) & \(7.626 \pm 1.597\) & \(12.173 \pm 2.370\) & \(19.306 \pm 3.225\) & \(42.377 \pm 4.555\) & \(38.966 \pm 4.014\) & .000 \\
    1 & 3           & 3   & \(8.078 \pm 4.085\) & \(3.864 \pm 1.732\) & \(2.388 \pm 0.451\) & \(4.122 \pm 1.773\) & \(6.241 \pm 3.223\) & \(9.943 \pm 5.547\) & \(24.500 \pm 12.832\) & \(22.112 \pm 12.464\) & .891 \\
    1 & 10          & 10  & \(13.325 \pm 3.071\) & \(6.364 \pm 2.010\) & \(3.070 \pm 0.625\) & \(6.618 \pm 1.810\) & \(10.600 \pm 2.860\) & \(17.079 \pm 4.190\) & \(39.001 \pm 6.698\) & \(35.931 \pm 6.171\) & 1 \\
    1 & \(\infty\)  & \(\infty\) & \(15.004 \pm 2.309\) & \(7.483 \pm 1.827\) & \(3.419 \pm 0.598\) & \(7.637 \pm 1.535\) & \(12.193 \pm 2.265\) & \(19.342 \pm 3.058\) & \(42.422 \pm 4.338\) & \(39.003 \pm 3.808\) & .000 \\
    2 & 3           & 3   & \(7.619 \pm 3.916\) & \(3.682 \pm 1.577\) & \(2.339 \pm 0.398\) & \(3.932 \pm 1.645\) & \(5.892 \pm 3.033\) & \(9.333 \pm 5.303\) & \(23.016 \pm 12.671\) & \(20.678 \pm 12.344\) & .919 \\
    2 & 10          & 10  & \(13.023 \pm 3.125\) & \(6.168 \pm 1.968\) & \(3.015 \pm 0.604\) & \(6.446 \pm 1.800\) & \(10.326 \pm 2.881\) & \(16.666 \pm 4.263\) & \(38.328 \pm 7.076\) & \(35.314 \pm 6.572\) & .998 \\
    2 & \(\infty\)  & \(\infty\) & \(14.977 \pm 2.178\) & \(7.447 \pm 1.761\) & \(3.399 \pm 0.581\) & \(7.608 \pm 1.469\) & \(12.161 \pm 2.149\) & \(19.315 \pm 2.877\) & \(42.401 \pm 3.998\) & \(39.002 \pm 3.483\) & .000 \\
    3 & 3           & 3   & \(7.393 \pm 3.836\) & \(3.615 \pm 1.581\) & \(2.326 \pm 0.405\) & \(3.847 \pm 1.623\) & \(5.721 \pm 2.971\) & \(9.017 \pm 5.181\) & \(22.283 \pm 12.417\) & \(19.956 \pm 12.088\) & .918 \\
    3 & 10          & 10  & \(12.729 \pm 3.219\) & \(6.024 \pm 1.933\) & \(2.963 \pm 0.584\) & \(6.300 \pm 1.785\) & \(10.079 \pm 2.906\) & \(16.269 \pm 4.397\) & \(37.545 \pm 7.680\) & \(34.581 \pm 7.200\) & .997 \\
    3 & \(\infty\)  & \(\infty\) & \(14.567 \pm 2.541\) & \(7.188 \pm 1.928\) & \(3.323 \pm 0.625\) & \(7.365 \pm 1.647\) & \(11.774 \pm 2.469\) & \(18.755 \pm 3.394\) & \(41.574 \pm 4.910\) & \(38.251 \pm 4.360\) & .000 \\
    \bottomrule
\end{tabular}
    \label{tab:pr_nu_1_uncentred}
\end{table}

\setlength{\tabcolsep}{1.8pt}
\begin{table}[h!]
\centering
\fontsize{6}{9}\selectfont
\caption{\textbf{Parameter recovery results for \(\nu\) corresponding to \(\varepsilon^{(Y)}\) in the equivalent model with uncentred errors.} Path coefficients \(\alpha\) and \(\beta\) are both fixed at 0.4, and \(\{\varepsilon_i^{(M)}\}_{i=1}^{n}\) and \(\{\varepsilon_i^{(Y)}\}_{i=1}^{n}\) are drawn from the same uncentred two-piece distribution. It presents the means ($\pm$ one standard deviation) of posterior means, modes, 2.5th, 25th, 50th, 75th, and 97.5th percentiles, the length of the \(95\%\) credible intervals for \(\nu\), and the coverage rate of the true value by the credible intervals across \(1\,000\) experiments with sample size \(n = 50\).}
\begin{tabular}{*{12}{c}}
    \toprule
    \textbf{\(\gamma\)} & \textbf{\(\nu\)} & \textbf{True \(\nu\)} & \textbf{Mean} & \textbf{Mode} & \textbf{2.5th} & \textbf{25th} & \textbf{50th} & \textbf{75th} & \textbf{97.5th} & \textbf{Length} & \textbf{Rate} \\
    \midrule
    0.33 & 3   & 3   & \(7.711 \pm 3.982\) & \(3.717 \pm 1.634\) & \(2.348 \pm 0.421\) & \(3.968 \pm 1.689\) & \(5.967 \pm 3.095\) & \(9.474 \pm 5.392\) & \(23.253 \pm 12.838\) & \(20.905 \pm 12.501\) & .913 \\
    0.33 & 10  & 10  & \(13.079 \pm 3.042\) & \(6.180 \pm 1.915\) & \(3.004 \pm 0.582\) & \(6.464 \pm 1.741\) & \(10.375 \pm 2.801\) & \(16.755 \pm 4.164\) & \(38.490 \pm 6.906\) & \(35.485 \pm 6.423\) & .999 \\
    0.33 & \(\infty\) &  \(\infty\) & \(14.763 \pm 2.295\) & \(7.314 \pm 1.819\) & \(3.341 \pm 0.581\) & \(7.467 \pm 1.503\) & \(11.955 \pm 2.241\) & \(19.032 \pm 3.067\) & \(41.996 \pm 4.408\) & \(38.654 \pm 3.910\) & .000 \\
    0.5 & 3   & 3   & \(8.126 \pm 4.092\) & \(3.871 \pm 1.755\) & \(2.393 \pm 0.464\) & \(4.135 \pm 1.796\) & \(6.273 \pm 3.247\) & \(10.010 \pm 5.555\) & \(24.673 \pm 12.780\) & \(22.280 \pm 12.407\) & .888 \\
    0.5 & 10  & 10  & \(13.136 \pm 3.176\) & \(6.250 \pm 1.995\) & \(3.023 \pm 0.599\) & \(6.519 \pm 1.820\) & \(10.446 \pm 2.910\) & \(16.819 \pm 4.341\) & \(38.521 \pm 7.308\) & \(35.497 \pm 6.815\) & .995 \\
    0.5 & \(\infty\) &  \(\infty\) & \(14.945 \pm 2.344\) & \(7.485 \pm 2.177\) & \(3.379 \pm 0.599\) & \(7.581 \pm 1.556\) & \(12.124 \pm 2.290\) & \(19.280 \pm 3.133\) & \(42.370 \pm 4.441\) & \(38.990 \pm 3.934\) & .000 \\
    1 & 3   & 3   & \(7.940 \pm 3.868\) & \(3.740 \pm 1.572\) & \(2.351 \pm 0.405\) & \(4.015 \pm 1.651\) & \(6.084 \pm 3.036\) & \(9.744 \pm 5.267\) & \(24.415 \pm 12.295\) & \(22.065 \pm 11.969\) & .916 \\
    1 & 10  & 10  & \(13.250 \pm 3.125\) & \(6.263 \pm 1.981\) & \(3.035 \pm 0.608\) & \(6.556 \pm 1.815\) & \(10.526 \pm 2.898\) & \(16.984 \pm 4.266\) & \(38.897 \pm 6.984\) & \(35.861 \pm 6.477\) & .997 \\
    1 & \(\infty\) &  \(\infty\) & \(15.098 \pm 2.161\) & \(7.506 \pm 1.757\) & \(3.406 \pm 0.572\) & \(7.667 \pm 1.458\) & \(12.276 \pm 2.133\) & \(19.492 \pm 2.860\) & \(42.677 \pm 3.965\) & \(39.270 \pm 3.462\) & .000 \\
    2 & 3   & 3   & \(7.837 \pm 4.016\) & \(3.767 \pm 1.694\) & \(2.363 \pm 0.441\) & \(4.019 \pm 1.734\) & \(6.055 \pm 3.155\) & \(9.622 \pm 5.445\) & \(23.732 \pm 12.723\) & \(21.369 \pm 12.369\) & .900 \\
    2 & 10  & 10  & \(13.067 \pm 3.269\) & \(6.224 \pm 2.029\) & \(3.026 \pm 0.611\) & \(6.491 \pm 1.855\) & \(10.387 \pm 2.991\) & \(16.722 \pm 4.470\) & \(38.322 \pm 7.497\) & \(35.297 \pm 6.981\) & .998 \\
    2 & \(\infty\) &  \(\infty\) & \(14.745 \pm 2.304\) & \(7.252 \pm 1.781\) & \(3.325 \pm 0.574\) & \(7.442 \pm 1.511\) & \(11.933 \pm 2.253\) & \(19.013 \pm 3.071\) & \(42.012 \pm 4.376\) & \(38.687 \pm 3.875\) & .000 \\
    3 & 3   & 3   & \(7.545 \pm 3.794\) & \(3.602 \pm 1.478\) & \(2.319 \pm 0.375\) & \(3.866 \pm 1.560\) & \(5.803 \pm 2.904\) & \(9.240 \pm 5.134\) & \(22.967 \pm 12.526\) & \(20.648 \pm 12.225\) & .924 \\
    3 & 10  & 10  & \(12.651 \pm 3.445\) & \(6.019 \pm 2.066\) & \(2.966 \pm 0.624\) & \(6.279 \pm 1.909\) & \(10.027 \pm 3.104\) & \(16.157 \pm 4.702\) & \(37.256 \pm 8.286\) & \(34.291 \pm 7.772\) & .993 \\
    3 & \(\infty\) &  \(\infty\) & \(14.456 \pm 2.537\) & \(7.068 \pm 1.882\) & \(3.273 \pm 0.592\) & \(7.275 \pm 1.618\) & \(11.661 \pm 2.454\) & \(18.620 \pm 3.408\) & \(41.416 \pm 4.986\) & \(38.144 \pm 4.468\) & .000 \\
    \bottomrule
\end{tabular}
    \label{tab:pr_nu_2_uncentred}
\end{table}

\setlength{\tabcolsep}{1.8pt}
\begin{table}[h!]
\centering
\fontsize{6}{9.5}\selectfont
\caption{\textbf{Parameter recovery results for \(\sigma^{(M)}\) in the equivalent model with uncentred errors.} Path coefficients \(\alpha\) and \(\beta\) are both fixed at 0.4, and \(\{\varepsilon_i^{(M)}\}_{i=1}^{n}\) and \(\{\varepsilon_i^{(Y)}\}_{i=1}^{n}\) are drawn from the same uncentred two-piece distribution. It presents the means ($\pm$ one standard deviation) of posterior means, modes, 2.5th, 25th, 50th, 75th, and 97.5th percentiles, the length of the \(95\%\) credible intervals for \(\sigma^{(M)}\), and the coverage rate of the true value by the credible intervals across \(1\,000\) experiments with sample size \(n = 50\).}
\begin{tabular}{*{12}{c}}
    \toprule
    \textbf{\(\gamma\)} & \textbf{\(\nu\)} & \textbf{True \(\sigma^{(M)}\)} & \textbf{Mean} & \textbf{Mode} & \textbf{2.5th} & \textbf{25th} & \textbf{50th} & \textbf{75th} & \textbf{97.5th} & \textbf{Length} & \textbf{Rate} \\
    \midrule
    0.33 & 3  & 1 & \(1.125 \pm 0.320\) & \(1.064 \pm 0.363\) & \(0.456 \pm 0.265\) & \(0.854 \pm 0.318\) & \(1.106 \pm 0.328\) & \(1.370 \pm 0.342\) & \(1.923 \pm 0.383\) & \(1.467 \pm 0.274\) & .948 \\
    0.33 & 10 & 1 & \(1.040 \pm 0.243\) & \(1.004 \pm 0.321\) & \(0.403 \pm 0.212\) & \(0.792 \pm 0.260\) & \(1.033 \pm 0.257\) & \(1.275 \pm 0.254\) & \(1.747 \pm 0.259\) & \(1.344 \pm 0.233\) & .986 \\
    0.33 & \(\infty\) & 1 & \(0.999 \pm 0.231\) & \(0.979 \pm 0.308\) & \(0.381 \pm 0.195\) & \(0.766 \pm 0.251\) & \(0.997 \pm 0.247\) & \(1.224 \pm 0.242\) & \(1.655 \pm 0.238\) & \(1.273 \pm 0.211\) & .989 \\
    0.5 & 3  & 1 & \(1.075 \pm 0.261\) & \(1.045 \pm 0.272\) & \(0.577 \pm 0.257\) & \(0.889 \pm 0.262\) & \(1.065 \pm 0.262\) & \(1.248 \pm 0.269\) & \(1.630 \pm 0.298\) & \(1.053 \pm 0.204\) & .942 \\
    0.5 & 10 & 1 & \(0.940 \pm 0.192\) & \(0.936 \pm 0.214\) & \(0.492 \pm 0.221\) & \(0.785 \pm 0.208\) & \(0.940 \pm 0.195\) & \(1.093 \pm 0.188\) & \(1.395 \pm 0.189\) & \(0.903 \pm 0.180\) & .977 \\
    0.5 & \(\infty\) & 1 & \(0.894 \pm 0.182\) & \(0.896 \pm 0.207\) & \(0.461 \pm 0.211\) & \(0.749 \pm 0.202\) & \(0.897 \pm 0.186\) & \(1.040 \pm 0.176\) & \(1.316 \pm 0.173\) & \(0.855 \pm 0.179\) & .957 \\
    1 & 3  & 1 & \(1.080 \pm 0.182\) & \(1.050 \pm 0.187\) & \(0.745 \pm 0.146\) & \(0.951 \pm 0.169\) & \(1.070 \pm 0.182\) & \(1.198 \pm 0.197\) & \(1.473 \pm 0.231\) & \(0.728 \pm 0.123\) & .947 \\
    1 & 10 & 1 & \(0.943 \pm 0.126\) & \(0.934 \pm 0.124\) & \(0.666 \pm 0.124\) & \(0.846 \pm 0.122\) & \(0.940 \pm 0.126\) & \(1.036 \pm 0.132\) & \(1.238 \pm 0.151\) & \(0.572 \pm 0.094\) & .948 \\
    1 & \(\infty\) & 1 & \(0.878 \pm 0.107\) & \(0.869 \pm 0.105\) & \(0.629 \pm 0.113\) & \(0.792 \pm 0.106\) & \(0.875 \pm 0.106\) & \(0.961 \pm 0.110\) & \(1.141 \pm 0.124\) & \(0.512 \pm 0.082\) & .867 \\
    2 & 3  & 1 & \(1.150 \pm 0.227\) & \(1.102 \pm 0.232\) & \(0.725 \pm 0.175\) & \(0.980 \pm 0.209\) & \(1.134 \pm 0.228\) & \(1.301 \pm 0.246\) & \(1.668 \pm 0.287\) & \(0.943 \pm 0.155\) & .929 \\
    2 & 10 & 1 & \(1.030 \pm 0.164\) & \(1.006 \pm 0.173\) & \(0.660 \pm 0.142\) & \(0.890 \pm 0.160\) & \(1.021 \pm 0.166\) & \(1.159 \pm 0.173\) & \(1.449 \pm 0.190\) & \(0.789 \pm 0.111\) & .981 \\
    2 & \(\infty\) & 1 & \(0.972 \pm 0.136\) & \(0.955 \pm 0.144\) & \(0.627 \pm 0.123\) & \(0.845 \pm 0.134\) & \(0.966 \pm 0.138\) & \(1.092 \pm 0.143\) & \(1.352 \pm 0.156\) & \(0.725 \pm 0.100\) & .986 \\
    3 & 3  & 1 & \(1.357 \pm 0.259\) & \(1.280 \pm 0.264\) & \(0.790 \pm 0.181\) & \(1.122 \pm 0.232\) & \(1.330 \pm 0.260\) & \(1.562 \pm 0.289\) & \(2.076 \pm 0.350\) & \(1.286 \pm 0.212\) & .879 \\
    3 & 10 & 1 & \(1.246 \pm 0.190\) & \(1.198 \pm 0.207\) & \(0.739 \pm 0.147\) & \(1.044 \pm 0.179\) & \(1.229 \pm 0.194\) & \(1.428 \pm 0.208\) & \(1.848 \pm 0.236\) & \(1.110 \pm 0.149\) & .941 \\
    3 & \(\infty\) & 1 & \(1.190 \pm 0.179\) & \(1.155 \pm 0.199\) & \(0.710 \pm 0.140\) & \(1.004 \pm 0.172\) & \(1.178 \pm 0.184\) & \(1.361 \pm 0.194\) & \(1.740 \pm 0.214\) & \(1.030 \pm 0.135\) & .965 \\
    \bottomrule
\end{tabular}
    \label{tab:pr_sigmaM_uncentred}
\end{table}

\setlength{\tabcolsep}{1.8pt}
\begin{table}[h!]
\centering
\fontsize{6}{10}\selectfont
\caption{\textbf{Parameter recovery results for \(\sigma^{(Y)}\) in the equivalent model with uncentred errors.} Path coefficients \(\alpha\) and \(\beta\) are both fixed at 0.4, and \(\{\varepsilon_i^{(M)}\}_{i=1}^{n}\) and \(\{\varepsilon_i^{(Y)}\}_{i=1}^{n}\) are drawn from the same uncentred two-piece distribution. It presents the means ($\pm$ one standard deviation) of posterior means, modes, 2.5th, 25th, 50th, 75th, and 97.5th percentiles, the length of the \(95\%\) credible intervals for \(\sigma^{(Y)}\), and the coverage rate of the true value by the credible intervals across \(1\,000\) experiments with sample size \(n = 50\).}
\begin{tabular}{*{12}{c}}
    \toprule
    \textbf{\(\gamma\)} & \textbf{\(\nu\)} & \textbf{True \(\sigma^{(Y)}\)} &
    \textbf{Mean} & \textbf{Mode} &
    \textbf{2.5th} & \textbf{25th} & \textbf{50th} &
    \textbf{75th} & \textbf{97.5th} & \textbf{Length} & \textbf{Rate} \\
    \midrule
    0.33 & 3   & 1 & \(1.113 \pm 0.293\) & \(1.055 \pm 0.348\) & \(0.415 \pm 0.223\) & \(0.827 \pm 0.291\) & \(1.094 \pm 0.304\) & \(1.372 \pm 0.319\) & \(1.949 \pm 0.360\) & \(1.534 \pm 0.274\) & .971 \\
    0.33 & 10  & 1 & \(1.033 \pm 0.243\) & \(0.995 \pm 0.326\) & \(0.380 \pm 0.201\) & \(0.773 \pm 0.260\) & \(1.027 \pm 0.257\) & \(1.278 \pm 0.257\) & \(1.760 \pm 0.267\) & \(1.380 \pm 0.229\) & .985 \\
    0.33 & \(\infty\) & 1 & \(1.028 \pm 0.236\) & \(1.019 \pm 0.307\) & \(0.395 \pm 0.205\) & \(0.791 \pm 0.257\) & \(1.028 \pm 0.252\) & \(1.258 \pm 0.244\) & \(1.692 \pm 0.238\) & \(1.297 \pm 0.210\) & .984 \\
    0.5 & 3   & 1 & \(1.052 \pm 0.257\) & \(1.022 \pm 0.277\) & \(0.536 \pm 0.259\) & \(0.859 \pm 0.262\) & \(1.044 \pm 0.259\) & \(1.232 \pm 0.265\) & \(1.622 \pm 0.292\) & \(1.087 \pm 0.220\) & .957 \\
    0.5 & 10  & 1 & \(0.937 \pm 0.203\) & \(0.933 \pm 0.235\) & \(0.469 \pm 0.222\) & \(0.775 \pm 0.224\) & \(0.939 \pm 0.208\) & \(1.098 \pm 0.198\) & \(1.408 \pm 0.198\) & \(0.939 \pm 0.186\) & .981 \\
    0.5 & \(\infty\) & 1 & \(0.877 \pm 0.187\) & \(0.879 \pm 0.223\) & \(0.431 \pm 0.214\) & \(0.723 \pm 0.213\) & \(0.880 \pm 0.192\) & \(1.030 \pm 0.178\) & \(1.316 \pm 0.171\) & \(0.885 \pm 0.179\) & .961 \\
    1 & 3   & 1 & \(1.089 \pm 0.180\) & \(1.058 \pm 0.184\) & \(0.743 \pm 0.140\) & \(0.956 \pm 0.165\) & \(1.079 \pm 0.181\) & \(1.211 \pm 0.198\) & \(1.495 \pm 0.237\) & \(0.752 \pm 0.137\) & .957 \\
    1 & 10  & 1 & \(0.941 \pm 0.125\) & \(0.933 \pm 0.124\) & \(0.660 \pm 0.125\) & \(0.843 \pm 0.121\) & \(0.938 \pm 0.124\) & \(1.036 \pm 0.131\) & \(1.241 \pm 0.150\) & \(0.581 \pm 0.097\) & .947 \\
    1 & \(\infty\) & 1 & \(0.872 \pm 0.111\) & \(0.866 \pm 0.113\) & \(0.612 \pm 0.121\) & \(0.784 \pm 0.110\) & \(0.870 \pm 0.110\) & \(0.958 \pm 0.114\) & \(1.142 \pm 0.129\) & \(0.530 \pm 0.095\) & .862 \\
    2 & 3   & 1 & \(1.153 \pm 0.230\) & \(1.106 \pm 0.234\) & \(0.716 \pm 0.173\) & \(0.979 \pm 0.210\) & \(1.136 \pm 0.231\) & \(1.308 \pm 0.252\) & \(1.683 \pm 0.299\) & \(0.967 \pm 0.170\) & .932 \\
    2 & 10  & 1 & \(1.035 \pm 0.164\) & \(1.013 \pm 0.174\) & \(0.654 \pm 0.138\) & \(0.892 \pm 0.158\) & \(1.027 \pm 0.166\) & \(1.168 \pm 0.175\) & \(1.461 \pm 0.195\) & \(0.807 \pm 0.115\) & .984 \\
    2 & \(\infty\) & 1 & \(0.971 \pm 0.139\) & \(0.955 \pm 0.148\) & \(0.619 \pm 0.124\) & \(0.842 \pm 0.137\) & \(0.966 \pm 0.141\) & \(1.094 \pm 0.146\) & \(1.357 \pm 0.160\) & \(0.737 \pm 0.105\) & .985 \\
    3 & 3   & 1 & \(1.388 \pm 0.263\) & \(1.312 \pm 0.272\) & \(0.803 \pm 0.180\) & \(1.147 \pm 0.235\) & \(1.361 \pm 0.265\) & \(1.600 \pm 0.295\) & \(2.126 \pm 0.357\) & \(1.323 \pm 0.220\) & .857 \\
    3 & 10  & 1 & \(1.272 \pm 0.197\) & \(1.227 \pm 0.215\) & \(0.748 \pm 0.150\) & \(1.066 \pm 0.186\) & \(1.256 \pm 0.201\) & \(1.459 \pm 0.216\) & \(1.885 \pm 0.248\) & \(1.137 \pm 0.160\) & .939 \\
    3 & \(\infty\) & 1 & \(1.211 \pm 0.180\) & \(1.179 \pm 0.2\) & \(0.717 \pm 0.140\) & \(1.021 \pm 0.173\) & \(1.200 \pm 0.185\) & \(1.388 \pm 0.195\) & \(1.774 \pm 0.217\) & \(1.057 \pm 0.138\) & .963 \\
    \bottomrule
\end{tabular}
    \label{tab:pr_sigmaY_uncentred}
\end{table}

\clearpage
\newpage
\subsection{Additional Hypothesis Testing Results} \label{sm:add_ht}

\vspace{-0.25cm}
\subsubsection{Centred Two-Piece Student \(t\) Distribution} \label{sm:tp_ht}

\vspace{-0.5cm}
As stated in Section~\ref{sec:5.2}, we calculate the true positive rate and the false positive rate of our Full Model and three sub-models using a cutoff of \(10\) for the Bayes factor under each combination of \(\gamma\) and \(\nu\) values. Then, they are compared with the true positive rate and the false positive rate of the OLS bootstrap method and the robust bootstrap method under the nominal \(0.05\) significance level. Table~\ref{tab:bf_comparison_3} presents the result.

\setlength{\tabcolsep}{8pt}
\begin{table}[h!]
    \centering
    \fontsize{6}{8}\selectfont
    \caption{\textbf{Comparison of true positive and false positive rates for four models (with Bayes factor cutoff of 10), OLS bootstrap, and robust bootstrap methods when the errors follow the same centred two-piece Student \(t\) distribution}: The first two columns specify the skewness and tail parameters in the error distribution. Results are based on \(1\,000\) experiments with sample size \(n = 50\). \textbf{Bold} and \underline{underlined} denote the best and second-best results, respectively.}
    \begin{tabular}{cc|cccccccccccc}
        \toprule
        \multicolumn{2}{c}{Settings} & \multicolumn{2}{c}{OLS Bootstrap} & \multicolumn{2}{c}{Robust Bootstrap} & \multicolumn{2}{c}{Full} & \multicolumn{2}{c}{\(\gamma\)-Only} & \multicolumn{2}{c}{\(\nu\)-Only} & \multicolumn{2}{c}{Normal} \\
        \cmidrule(lr){1-2} \cmidrule(lr){3-4} \cmidrule(lr){5-6} \cmidrule(lr){7-8} \cmidrule(lr){9-10} \cmidrule(lr){11-12} \cmidrule(lr){13-14}
        \(\gamma\) & \(\nu\) & TPR & FPR & TPR & FPR & TPR & FPR & TPR & FPR & TPR & FPR & TPR & FPR \\
        \midrule
        \multicolumn{14}{c}{Path Coefficients: \(\alpha = 0.4, \beta = 0.4\)} \\
        0.33  & 3  & .109 & .004 & .229 & .010 & .468 & .010 & .370 & .010 & .274 & .014 & .175 & .024 \\ 
        0.33  & 10  & .204 & .009 & .222 & .009 & .525 & .012 & .505 & .012 & .228 & .010 & .221 & .009 \\ 
        0.33     & \(\infty\)  & .252 & .012 & .227 & .010 & .514 & .013 & .516 & .011 & .258 & .010 & .246 & .006 \\ 
        0.5     & 3  & .191 & .007 & .354 & .019 & .460 & .008 & .362 & .006 & .346 & .013 & .203 & .018 \\ 
        0.5   & 10  & .325 & .013 & .322 & .010 & .511 & .011 & .493 & .014 & .340 & .009 & .298 & .009 \\ 
        0.5   & \(\infty\)  & .399 & .016 & .349 & .014 & .570 & .018 & .584 & .015 & .365 & .011 & .347 & .011 \\ 
        1   & 3  & .301 & .017 & .422 & .017 & .403 & .005 & .300 & .012 & .403 & .005 & .264 & .009 \\ 
        1   & 10  & .511 & .019 & .436 & .015 & .469 & .011 & .448 & .012 & .469 & .009 & .447 & .009 \\ 
        1   & \(\infty\)  & .558 & .017 & .422 & .012 & .476 & .004 & .480 & .006 & .477 & .006 & .482 & .005 \\ 
        \midrule
        \multicolumn{14}{c}{Path Coefficients: \(\alpha = 0.4, \beta = 0.7\)} \\
        0.33  & 3  & .166 & .013 & .253 & .008 & .509 & .019 & .434 & .019 & .337 & .025 & .279 & .043 \\ 
        0.33  & 10  & .285 & .025 & .286 & .020 & .572 & .027 & .565 & .030 & .346 & .027 & .336 & .030 \\ 
        0.33     & \(\infty\)  & .301 & .017 & .302 & .018 & .576 & .020 & .576 & .023 & .355 & .022 & .337 & .019 \\  
        0.5     & 3  & .299 & .019 & .409 & .019 & .558 & .021 & .484 & .036 & .443 & .026 & .333 & .041 \\ 
        0.5   & 10  & .462 & .028 & .430 & .021 & .643 & .023 & .642 & .027 & .489 & .021 & .466 & .027 \\ 
        0.5   & \(\infty\)  & .546 & .027 & .492 & .019 & .685 & .027 & .692 & .028 & .548 & .017 & .527 & .014 \\ 
        1   & 3  & .456 & .032 & .574 & .032 & .578 & .032 & .469 & .032 & .574 & .034 & .443 & .036 \\ 
        1   & 10  & .728 & .039 & .634 & .027 & .698 & .031 & .685 & .036 & .701 & .029 & .681 & .028 \\ 
        1   & \(\infty\)  & .774 & .037 & .657 & .028 & .717 & .029 & .721 & .032 & .713 & .029 & .724 & .024 \\ 
        \midrule
        \multicolumn{14}{c}{Path Coefficients: \(\alpha = 0.7, \beta = 0.4\)} \\
        0.33  & 3  & .232 & .011 & .509 & .015 & .849 & .022 & .735 & .022 & .554 & .015 & .304 & .038 \\ 
        0.33  & 10  & .462 & .013 & .492 & .011 & .868 & .022 & .857 & .018 & .546 & .014 & .465 & .012 \\ 
        0.33     & \(\infty\)  & .531 & .012 & .468 & .026 & .849 & .028 & .855 & .026 & .548 & .015 & .526 & .014 \\ 
        0.5     & 3  & .440 & .012 & .677 & .023 & .836 & .022 & .696 & .032 & .725 & .023 & .438 & .025 \\ 
        0.5   & 10  & .636 & .020 & .630 & .024 & .823 & .025 & .819 & .032 & .662 & .018 & .600 & .018 \\ 
        0.5   & \(\infty\)  & .726 & .033 & .628 & .020 & .826 & .021 & .820 & .028 & .678 & .025 & .655 & .027 \\ 
        1   & 3  & .578 & .026 & .727 & .027 & .753 & .022 & .613 & .020 & .776 & .015 & .600 & .022 \\ 
        1   & 10  & .740 & .022 & .673 & .019 & .714 & .014 & .694 & .017 & .726 & .013 & .712 & .013 \\ 
        1   & \(\infty\)  & .742 & .027 & .628 & .022 & .705 & .017 & .703 & .022 & .697 & .014 & .701 & .015 \\
        \midrule
        \multicolumn{14}{c}{Path Coefficients: \(\alpha = 0.7, \beta = 0.7\)} \\
        0.33  & 3  & .357 & .017 & .564 & .031 & .885 & .047 & .833 & .050 & .676 & .050 & .498 & .052 \\ 
        0.33  & 10  & .599 & .025 & .618 & .021 & .923 & .045 & .930 & .049 & .733 & .032 & .678 & .034 \\ 
        0.33     & \(\infty\)  & .716 & .031 & .672 & .031 & .936 & .054 & .943 & .064 & .781 & .040 & .766 & .038 \\ 
        0.5     & 3  & .557 & .026 & .783 & .027 & .928 & .032 & .852 & .048 & .847 & .038 & .645 & .047 \\ 
        0.5   & 10  & .852 & .052 & .853 & .039 & .971 & .060 & .963 & .068 & .917 & .052 & .880 & .051 \\ 
        0.5   & \(\infty\)  & .935 & .042 & .886 & .029 & .980 & .055 & .982 & .063 & .949 & .039 & .946 & .044 \\ 
        1   & 3  & .776 & .049 & .896 & .046 & .931 & .049 & .827 & .060 & .940 & .047 & .829 & .044 \\ 
        1   & 10  & .975 & .053 & .945 & .034 & .979 & .052 & .976 & .053 & .984 & .042 & .978 & .043 \\ 
        1   & \(\infty\)  & .984 & .083 & .944 & .046 & .976 & .057 & .979 & .060 & .977 & .056 & .982 & .058 \\
        \bottomrule
    \end{tabular}
    \label{tab:bf_comparison_3}
\end{table}

\newpage

Next, we present the hypothesis testing results where the \(\varepsilon^{(M)}\) and \(\varepsilon^{(Y)}\) come from different centred two-piece Student \(t\) distribution with opposing skewness and different tail-heaviness. In Table~\ref{tab:oppo_skew}, the cutoff values of the Bayes factor for the four Bayesian models are estimated from null data so that the resulting false positive rate matches that of the OLS bootstrap or the robust bootstrap. In Table~\ref{tab:opposing_skewness_10}, we set the cutoff values of the Bayes factor at 10.

\vspace{1cm}

\setlength{\tabcolsep}{4.5pt}
\begin{table}[h!]
    \centering
    \caption{\textbf{Comparison of true positive rates for four Bayesian models, OLS bootstrap, and robust bootstrap at matched false positive rates when the errors follow distinct centred two-piece Student \(t\) distribution with opposing skewness and different tail-heaviness.} The first two columns specify the distributions of \(\varepsilon^{(M)}\) and \(\varepsilon^{(Y)}\). Cutoff values of the Bayes factor are estimated from null data so that the resulting false positive rate matches that of the OLS bootstrap or the robust bootstrap under comparison. Results are based on \(1\,000\) experiments with sample size \(n = 50\). \textbf{Bold} and \underline{underlined} denote the best and second-best results, respectively.}

    \fontsize{6}{10}\selectfont
    \begin{tabular}{cc|cccccc|cccccc}
        \toprule
        \multicolumn{2}{c}{Settings} & \multicolumn{2}{c}{OLS Bootstrap}  & \multicolumn{1}{c}{Full} & \multicolumn{1}{c}{\(\gamma\)-Only} & \multicolumn{1}{c}{\(\nu\)-Only} & \multicolumn{1}{c}{Normal} & \multicolumn{2}{c}{Robust Bootstrap}  & \multicolumn{1}{c}{Full} & \multicolumn{1}{c}{\(\gamma\)-Only} & \multicolumn{1}{c}{\(\nu\)-Only} & \multicolumn{1}{c}{Normal}\\
        \cmidrule(lr){1-2} \cmidrule(lr){3-4} \cmidrule(lr){9-10}
        \(\varepsilon^{(M)}\) & \(\varepsilon^{(Y)}\) & TPR & FPR &  &  & &  & TPR & FPR &  &  &  &  \\
        \midrule
        \multicolumn{14}{c}{Path Coefficients: \(\alpha = 0.4, \beta = 0.4\)} \\
        ctpt(0.33, 3)  & ctpt(2, 10)  & .192 & .027 & \textbf{.432} & \underline{.341} & .324 & .179 & .255 & .027 & \textbf{.432} & \underline{.341} & .324 & .179 \\
        ctpt(0.33, 10)  & ctpt(2, 3)  & .167 & .016 & \textbf{.554} & \underline{.471} & .361 & .253 & .233 & .013 & \textbf{.519} & \underline{.433} & .359 & .236  \\ 
        ctpt(2, 3)  & ctpt(0.33, 10)  & .170 & .006 & \textbf{.323} & .197 & \underline{.264} & .158 & .270 & .010 & \textbf{.359} & .242 & \underline{.354} & .188 \\ 
        ctpt(2, 10)  & ctpt(0.33, 3) & .137 & .008 & \textbf{.284} & .243 & \underline{.250} & .131 & .174 & .011 & \textbf{.365} & \underline{.299} & .266 & .151 \\ 
        \midrule
        \multicolumn{14}{c}{Path Coefficients: \(\alpha = 0.4, \beta = 0.7\)} \\
        ctpt(0.33, 3)  & ctpt(2, 10) & .215 & .045 & \textbf{.455} & \underline{.396} & .318 & .192 & .257 & .028 & \textbf{.380} & \underline{.320} & .252 & .153 \\ 
        ctpt(0.33, 10)  & ctpt(2, 3) & .260 & .019 & \textbf{.466} & \underline{.409} & .325 & .287 & .293 & .015 & \textbf{.419} & \underline{.365} & .269 & .265 \\ 
        ctpt(2, 3)  & ctpt(0.33, 10)  & .305 & .023 & \textbf{.461} & .383 & \underline{.430} & .273 & .369 & .017 & \textbf{.416} & .299 & \underline{.406} & .259 \\ 
        ctpt(2, 10)  & ctpt(0.33, 3)  & .262 & .021 & \textbf{.511} & \underline{.475} & .446 & .283 & .363 & .014 & \textbf{.478} & \underline{.431} & .383 & .222 \\ 
        \midrule
        \multicolumn{14}{c}{Path Coefficients: \(\alpha = 0.7, \beta = 0.4\)} \\
        ctpt(0.33, 3)  & ctpt(2, 10) & .373 & .033 & \textbf{.784} & \underline{.682} & .564 & .383 & .516 & .029 & \textbf{.762} & \underline{.665} & .554 & .366 \\ 
        ctpt(0.33, 10)  & ctpt(2, 3) & .434 & .013 & \textbf{.733} & \underline{.708} & .652 & .539 & .549 & .013 & \textbf{.733} & \underline{.708} & .652 & .539\\ 
        ctpt(2, 3)  & ctpt(0.33, 10)  & .410 & .012 & \textbf{.665} & \underline{.565} & .483 & .300 & .552 & .018 & \textbf{.714} & \underline{.651} & .598 & .409 \\ 
        ctpt(2, 10)  & ctpt(0.33, 3) & .259 & .011 & \textbf{.549} & \underline{.480} & .390 & .318 & .379 & .020 & \textbf{.626} & \underline{.566} & .467 & .415 \\ 
        \midrule
        \multicolumn{14}{c}{Path Coefficients: \(\alpha = 0.7, \beta = 0.7\)} \\
        ctpt(0.33, 3)  & ctpt(2, 10)  & .394 & .044 & \textbf{.835} & \underline{.679} & .666 & .398 & .568 & .035 & \textbf{.811} & \underline{.652} & .636 & .375 \\ 
        ctpt(0.33, 10)  & ctpt(2, 3)  & .571 & .039 & \textbf{.885} & \underline{.833} & .763 & .688 & .661 & .027 & \textbf{.852} & .729 & \underline{.732} & .631 \\ 
        ctpt(2, 3)  & ctpt(0.33, 10)  & .576 & .039 & \textbf{.855} & .754 & \underline{.830} & .648 & .751 & .031 & \textbf{.842} & .714 & \underline{.798} & .624 \\ 
        ctpt(2, 10)  & ctpt(0.33, 3) & .512 & .028 & \textbf{.876} & \underline{.809} & .751 & .585 & .670 & .021 & \textbf{.852} & \underline{.780} & .738 & .547 \\ 
        \bottomrule
    \end{tabular}
    \label{tab:oppo_skew}
\end{table}

\setlength{\tabcolsep}{6.5pt}
\begin{table}[h!]
    \centering
    \fontsize{6}{10}\selectfont
    \caption{\textbf{Comparison of true positive and false positive rates for four models (with Bayes factor cutoff of 10), OLS bootstrap, and robust bootstrap methods when the errors follow distinct centred two-piece Student \(t\) distribution with opposing skewness and different tail-heaviness.} The first two columns specify the distributions of \(\varepsilon^{(M)}\) and \(\varepsilon^{(Y)}\). Results are based on \(1\,000\) experiments with sample size \(n = 50\). \textbf{Bold} and \underline{underlined} denote the best and second-best results, respectively.}
    \begin{tabular}{cc|cccccccccccc}
        \toprule
        \multicolumn{2}{c}{Settings} & \multicolumn{2}{c}{OLS Bootstrap} & \multicolumn{2}{c}{Robust Bootstrap} & \multicolumn{2}{c}{Full} & \multicolumn{2}{c}{\(\gamma\)-Only} & \multicolumn{2}{c}{\(\nu\)-Only} & \multicolumn{2}{c}{Normal} \\
        \cmidrule(lr){1-2} \cmidrule(lr){3-4} \cmidrule(lr){5-6} \cmidrule(lr){7-8} \cmidrule(lr){9-10} \cmidrule(lr){11-12} \cmidrule(lr){13-14}
        \(\varepsilon^{(M)}\) & \(\varepsilon^{(Y)}\) & TPR & FPR & TPR & FPR & TPR & FPR & TPR & FPR & TPR & FPR & TPR & FPR \\
        \midrule
        \multicolumn{14}{c}{Path Coefficients: \(\alpha = 0.4, \beta = 0.4\)} \\
        ctpt(0.33, 3)  & ctpt(2, 10)  & .192 & .027 & .255 & .027 & .494 & .038 & .430 & .057 & .334 & .029 & .326 & .055 \\ 
        ctpt(0.33, 10)  & ctpt(2, 3)  & .167 & .016 & .233 & .013 & .516 & .011 & .474 & .016 & .272 & .010 & .198 & .008 \\ 
        ctpt(2, 3)  & ctpt(0.33, 10)  & .170 & .006 & .270 & .010 & .457 & .018 & .404 & .019 & .285 & .006 & .205 & .011 \\ 
        ctpt(2, 10)  & ctpt(0.33, 3)  & .137 & .008 & .174 & .011 & .381 & .013 & .359 & .015 & .170 & .004 & .128 & .007 \\ 
        \midrule
        \multicolumn{14}{c}{Path Coefficients: \(\alpha = 0.4, \beta = 0.7\)} \\
        ctpt(0.33, 3)  & ctpt(2, 10)  & .215 & .045 & .257 & .028 & .518 & .052 & .467 & .060 & .341 & .052 & .344 & .102 \\ 
        ctpt(0.33, 10)  & ctpt(2, 3)  & .260 & .019 & .293 & .015 & .578 & .035 & .557 & .052 & .369 & .027 & .325 & .025 \\ 
        ctpt(2, 3)  & ctpt(0.33, 10)  & .305 & .023 & .369 & .017 & .513 & .045 & .464 & .046 & .412 & .018 & .344 & .037 \\ 
        ctpt(2, 10)  & ctpt(0.33, 3)  & .262 & .021 & .363 & .014 & .596 & .031 & .564 & .035 & .380 & .012 & .271 & .019 \\ 
        \midrule
        \multicolumn{14}{c}{Path Coefficients: \(\alpha = 0.7, \beta = 0.4\)} \\
        ctpt(0.33, 3)  & ctpt(2, 10)  & .373 & .033 & .516 & .029 & .876 & .070 & .827 & .078 & .622 & .040 & .526 & .062 \\ 
        ctpt(0.33, 10)  & ctpt(2, 3)  & .434 & .013 & .549 & .013 & .871 & .028 & .830 & .034 & .603 & .009 & .464 & .009 \\ 
        ctpt(2, 3)  & ctpt(0.33, 10)  & .410 & .012 & .552 & .018 & .835 & .049 & .785 & .054 & .605 & .018 & .474 & .025 \\ 
        ctpt(2, 10)  & ctpt(0.33, 3)  & .259 & .011 & .379 & .020 & .698 & .029 & .626 & .025 & .373 & .010 & .297 & .008 \\ 
        \midrule
        \multicolumn{14}{c}{Path Coefficients: \(\alpha = 0.7, \beta = 0.7\)} \\
        ctpt(0.33, 3)  & ctpt(2, 10)  & .394 & .044 & .568 & .035 & .889 & .082 & .849 & .113 & .678 & .047 & .567 & .088 \\
        ctpt(0.33, 10)  & ctpt(2, 3)  & .571 & .039 & .661 & .027 & .931 & .068 & .935 & .097 & .742 & .031 & .662 & .033 \\ 
        ctpt(2, 3)  & ctpt(0.33, 10)  & .576 & .039 & .751 & .031 & .918 & .077 & .853 & .103 & .816 & .036 & .657 & .043 \\ 
        ctpt(2, 10)  & ctpt(0.33, 3)  & .512 & .028 & .670 & .021 & .917 & .061 & .897 & .075 & .733 & .018 & .588 & .029 \\ 
        \bottomrule
    \end{tabular}
    \label{tab:opposing_skewness_10}
\end{table}

\newpage

\subsubsection{Tukey \(g\)-and-\(h\) Distribution}

As stated in Section~\ref{sec:5.2}, we calculate the true positive rate and the false positive rate of our Full Model and three sub-models using a cutoff of \(10\) for the Bayes factor under each combination of \(g\) and \(h\) values. Then, they are compared with the true positive rate and the false positive rate of the OLS bootstrap method and the robust bootstrap method under the nominal \(0.05\) significance level. Table~\ref{tab:bf_comparison_3} presents the result.

\setlength{\tabcolsep}{8pt}
\begin{table}[h!]
    \centering
    \fontsize{6}{9}\selectfont
    \caption{\textbf{Comparison of true positive and false positive rates for four models (with Bayes factor cutoff of 10), OLS bootstrap, and robust bootstrap methods when the errors follow the same Tukey \(g\)-and-\(h\) distribution.} The first two columns specify the values of \(g\) and \(h\) in the error distribution. Results are based on \(1\,000\) experiments with sample size \(n = 50\). \textbf{Bold} and \underline{underlined} denote the best and second-best results, respectively.}
    \begin{tabular}{cc|cccccccccccc}
        \toprule
        \multicolumn{2}{c}{Settings} & \multicolumn{2}{c}{OLS Bootstrap} & \multicolumn{2}{c}{Robust Bootstrap} & \multicolumn{2}{c}{Full} & \multicolumn{2}{c}{\(\gamma\)-Only} & \multicolumn{2}{c}{\(\nu\)-Only} & \multicolumn{2}{c}{Normal} \\
        \cmidrule(lr){1-2} \cmidrule(lr){3-4} \cmidrule(lr){5-6} \cmidrule(lr){7-8} \cmidrule(lr){9-10} \cmidrule(lr){11-12} \cmidrule(lr){13-14}
        \(g\) & \(h\) & TPR & FPR & TPR & FPR & TPR & FPR & TPR & FPR & TPR & FPR & TPR & FPR \\
        \midrule
        \multicolumn{14}{c}{Path Coefficients: \(\alpha = 0.4, \beta = 0.4\)} \\
        0.2  & 0  & .583 & .015 & .488 & .012 & .531 & .009 & .543 & .009 & .507 & .004 & .479 & .005 \\ 
        0.2  & 0.2  & .364 & .011 & .514 & .009 & .489 & .005 & .342 & .007 & .485 & .005 & .326 & .011 \\ 
        0.5  & 0  & .485 & .013 & .597 & .010 & .777 & .010 & .735 & .010 & .548 & .005 & .394 & .008 \\ 
        0.5  & 0.2  & .311 & .014 & .572 & .013 & .603 & .009 & .385 & .010 & .538 & .008 & .274 & .017 \\ 
        \midrule
        \multicolumn{14}{c}{Path Coefficients: \(\alpha = 0.4, \beta = 0.7\)} \\
        0.2  & 0  & .746 & .033 & .668 & .025 & .744 & .020 & .752 & .016 & .708 & .019 & .707 & .017 \\ 
        0.2  & 0.2  & .496 & .025 & .610 & .023 & .608 & .020 & .508 & .023 & .609 & .017 & .479 & .021 \\ 
        0.5 & 0  & .632 & .028 & .723 & .028 & .858 & .024 & .854 & .024 & .692 & .019 & .588 & .024 \\ 
        0.5 & 0.2  & .423 & .025 & .668 & .029 & .680 & .027 & .535 & .022 & .621 & .023 & .404 & .029 \\ 
        \midrule
        \multicolumn{14}{c}{Path Coefficients: \(\alpha = 0.7, \beta = 0.4\)} \\
        0.2  & 0  & .768 & .025 & .671 & .020 & .756 & .018 & .757 & .020 & .744 & .018 & .727 & .021 \\ 
        0.2  & 0.2  & .639 & .019 & .772 & .023 & .811 & .015 & .683 & .018 & .805 & .024 & .639 & .023 \\ 
        0.5  & 0  & .733 & .020 & .816 & .020 & .918 & .020 & .891 & .023 & .799 & .024 & .697 & .021 \\ 
        0.5  & 0.2  & .570 & .021 & .854 & .028 & .898 & .020 & .726 & .022 & .874 & .030 & .557 & .030 \\ 
        \midrule
        \multicolumn{14}{c}{Path Coefficients: \(\alpha = 0.7, \beta = 0.7\)} \\
        0.2  & 0  & .978 & .062 & .945 & .037 & .977 & .052 & .980 & .057 & .978 & .050 & .980 & .046 \\ 
        0.2  & 0.2  & .831 & .043 & .919 & .040 & .957 & .055 & .896 & 055 & .960 & .050 & .873 & .045 \\ 
        0.5  & 0  & .921 & .052 & .959 & .044 & .994 & .061 & .992 & .076 & .977 & .050 & .943 & .048 \\ 
        0.5  & 0.2  & .729 & .037 & .945 & .044 & .977 & .055 & .915 & .053 & .958 & .050 & .772 & .048 \\ 
        \bottomrule
    \end{tabular}
    \label{tab:bf_comparison_tukey}
\end{table}

\newpage

\subsubsection{Comparing Different Sample Sizes in Hypothesis Testing} \label{sm:diff_sample}

In Tables~\ref{tab:bf_comparison_3} and \ref{tab:bf_comparison_match} the sample size was fixed at \(50\). We now enlarge the sample size to \(100\) and \(200\) and repeat the same simulation studies, with the path coefficients \(\alpha = 0.4\) and \(\beta = 0.4\); all other settings remain unchanged. For each combination of \(\gamma\) and \(\nu\) we compute the true positive and false positive rates of the Full Model and its three sub-models, using a Bayes-factor cutoff of 3 as before. These rates are compared with those of the OLS bootstrap and the robust bootstrap at the nominal 0.05 significance level, and the results are summarised in Table~\ref{tab:sample_bf_comparison_3}. In Table~\ref{tab:sample_bf_comparison_match}, we additionally select cutoff values for the Full Model and the sub-models such that their false positive rates match those of the OLS and robust bootstraps, thereby enabling a direct comparison.

\vspace{0.6cm}
\setlength{\tabcolsep}{8pt}
\begin{table}[h!]
    \centering
    \fontsize{6}{8}\selectfont
    \caption{\textbf{Comparison of true positive rates for four Bayesian models, OLS bootstrap, and robust bootstrap at matched false positive rates when the errors follow the same centred two-piece Student \(t\) distribution under sample size \(n = 50\), \(100\) and \(200\).} The first two columns specify the skewness and tail parameters in the error distribution. Cutoff values of the Bayes factor are estimated from null data so that the resulting false positive rate matches that of the OLS bootstrap or the robust bootstrap under comparison. Results are based on \(1\,000\) experiments with \(\alpha=\beta=0.4\). \textbf{Bold} and \underline{underlined} denote the best and second-best results, respectively.}
    \begin{tabular}{cc|cccccccccccc}
        \toprule
        \multicolumn{2}{c}{Settings} & \multicolumn{2}{c}{OLS Bootstrap} & \multicolumn{2}{c}{Robust Bootstrap} & \multicolumn{2}{c}{Full} & \multicolumn{2}{c}{\(\gamma\)-Only} & \multicolumn{2}{c}{\(\nu\)-Only} & \multicolumn{2}{c}{Normal} \\
        \cmidrule(lr){1-2} \cmidrule(lr){3-4} \cmidrule(lr){5-6} \cmidrule(lr){7-8} \cmidrule(lr){9-10} \cmidrule(lr){11-12} \cmidrule(lr){13-14}
        \(\gamma\) & \(\nu\) & TPR & FPR & TPR & FPR & TPR & FPR & TPR & FPR & TPR & FPR & TPR & FPR \\
        \midrule
        \multicolumn{14}{c}{Sample Size: \(n=50\)} \\
        0.33  & 3  & .109 & .004 & .229 & .010 & .468 & .010 & .370 & .010 & .274 & .014 & .175 & .024 \\ 
        0.33  & 10  & .204 & .009 & .222 & .009 & .525 & .012 & .505 & .012 & .228 & .010 & .221 & .009 \\ 
        0.33     & \(\infty\)  & .252 & .012 & .227 & .010 & .514 & .013 & .516 & .011 & .258 & .010 & .246 & .006 \\ 
        0.5     & 3  & .191 & .007 & .354 & .019 & .460 & .008 & .362 & .006 & .346 & .013 & .203 & .018 \\ 
        0.5   & 10  & .325 & .013 & .322 & .010 & .511 & .011 & .493 & .014 & .340 & .009 & .298 & .009 \\ 
        0.5   & \(\infty\)  & .399 & .016 & .349 & .014 & .570 & .018 & .584 & .015 & .365 & .011 & .347 & .011 \\ 
        1   & 3  & .301 & .017 & .422 & .017 & .403 & .005 & .300 & .012 & .403 & .005 & .264 & .009 \\ 
        1   & 10  & .511 & .019 & .436 & .015 & .469 & .011 & .448 & .012 & .469 & .009 & .447 & .009 \\ 
        1   & \(\infty\)  & .558 & .017 & .422 & .012 & .476 & .004 & .480 & .006 & .477 & .006 & .482 & .005 \\ 
        \midrule
        \multicolumn{14}{c}{Sample Size: \(n=100\)} \\
        0.33  & 3  & .224 & .018 & .413 & .018 & .754 & .024 & .649 & .021 & .409 & .019 & .265 & .027 \\ 
        0.33  & 10  & .427 & .021 & .469 & .022 & .835 & .028 & .830 & .029 & .479 & .014 & .403 & .015 \\ 
        0.33     & \(\infty\)  & .576 & .018 & .513 & .020 & .878 & .023 & .889 & .022 & .526 & .005 & .508 & .007 \\ 
        0.5     & 3  & .410 & .017 & .668 & .028 & .808 & .014 & .646 & .016 & .624 & .021 & .351 & .017 \\ 
        0.5   & 10  & .713 & .026 & .722 & .024  & .860 & .022 & .846 & .021 & .682 & .015 & .627 & .013 \\
        0.5   & \(\infty\)  & .784 & .028 & .723 & .027 & .886 & .021 & .899 & .020 & .733 & .014 & .712 & .012 \\ 
        1   & 3  & .642 & .025 & .833 & .040 & .779 & .024 & .585 & .023 & .792 & .022 & .581 & .017 \\ 
        1   & 10  & .892 & .038 & .857 & .036 & .857 & .015 & .844 & .014 & .863 & .013 & .843 & .014 \\ 
        1   & \(\infty\)  & .938 & .056 & .858 & .046 & .884 & .024 & .889 & .026 & .886 & .023 & .889 & .023 \\ 
        \midrule
        \multicolumn{14}{c}{Sample Size: \(n=200\)} \\
        0.33  & 3  & .389 & .038 & .721 & .039 & .955 & .048 & .851 & .057 & .659 & .039 & .409 & .041 \\  
        0.33  & 10  & .731 & .027 & .784 & .040 & .982 & .053 & .975 & .061 & .747 & .015 & .699 & .016 \\ 
        0.33     & \(\infty\)  & .835 & .049 & .814 & .040 & .994 & .066 & .994 & .075 & .789 & .026 & .786 & .018 \\ 
        0.5     & 3  & .646 & .048 & .925 & .046 & .977 & .044 & .880 & .040 & .883 & .038 & .592 & .031 \\ 
        0.5   & 10  & .947 & .076 & .967 & .077 & .998 & .057 & .995 & .064 & .949 & .043 & .924 & .036 \\ 
        0.5   & \(\infty\)  & .985 & .100 & .972 & .072 & .999 & .071 & .999 & .074 & .971 & .035 & .972 & .031 \\ 
        1   & 3  & .880 & .083 & .985 & .094 & .984 & .054 & .858 & .039 & .983 & .050 & .842 & .033 \\ 
        1   & 10  & .998 & .112 & .997 & .098 & .995 & .044 & .992 & .039 & .994 & .045 & .991 & .036 \\ 
        1   & \(\infty\)  & .999 & .158 & .997 & .103 & .998 & .055 & .999 & .055 & .999 & .054 & .999 & .059 \\ 
        \bottomrule
    \end{tabular}
    \label{tab:sample_bf_comparison_3}
\end{table}

On the basis of Tables~\ref{tab:sample_bf_comparison_3} and \ref{tab:sample_bf_comparison_match}, we find that when the Tukey \(g\)-and-\(h\) errors are both skewed and heavy-tailed, the Full Model generally outperforms the three sub-models and the two bootstrap-based approaches. When the errors are symmetric (\(g = 1\)), the \(\nu\)-Only Model and the Normal Model perform slightly better, but their advantage over the Full Model is small. We therefore conclude that the Full Model remains robust under various skewness and tail-heaviness.

\setlength{\tabcolsep}{7pt}
\begin{table}[h!]
    \centering
    \caption{\textbf{Comparison of true positive and false positive rates for four models (with Bayes factor cutoff of 10), OLS bootstrap, and robust bootstrap methods when the errors follow the same centred two-piece Student \(t\) distribution under sample size \(n = 50\), \(100\) and \(200\).} The first two columns specify the distributions of \(\varepsilon^{(M)}\) and \(\varepsilon^{(Y)}\). Results are based on \(1\,000\) experiments with \(\alpha=\beta=0.4\). \textbf{Bold} and \underline{underlined} denote the best and second-best results, respectively.}
    \fontsize{6}{7.5}\selectfont
    \begin{tabular}{cc|cccccc|cccccc}
        \toprule
        \multicolumn{2}{c}{Settings} & \multicolumn{2}{c}{OLS Bootstrap}  & \multicolumn{1}{c}{Full} & \multicolumn{1}{c}{\(\gamma\)-Only} & \multicolumn{1}{c}{\(\nu\)-Only} & \multicolumn{1}{c}{Normal} & \multicolumn{2}{c}{Robust Bootstrap}  & \multicolumn{1}{c}{Full} & \multicolumn{1}{c}{\(\gamma\)-Only} & \multicolumn{1}{c}{\(\nu\)-Only} & \multicolumn{1}{c}{Normal}\\
        \cmidrule(lr){1-2} \cmidrule(lr){3-4} \cmidrule(lr){9-10}
        \(\gamma\) & \(\nu\) & TPR & FPR &  &  & &  & TPR & FPR &  &  &  &  \\
        \midrule
        \multicolumn{14}{c}{Sample Size: \(n=50\)} \\
        0.33  & 3  & .109 & .004 & \textbf{.374} & \underline{.225} & .129 & .059 & .229 & .010 & \textbf{.435} & \underline{.368} & .228 & .111 \\ 
        0.33  & 10  & .204 & .009 & \textbf{.466} & \underline{.456} & .212 & .219 & .222 & .009 & \textbf{.466} & \underline{.456} & .212 & .219  \\ 
        0.33     & \(\infty\)  & .252 & .012 & \underline{.503} & \textbf{.516} & .294 & .276 & .227 & .010 & \underline{.472} & \textbf{.500} & .248 & .272 \\ 
        0.5     & 3  & .191 & .007 & \textbf{.413} & \underline{.310} & .177 & .111 & .354 & .019 & \textbf{.571} & \underline{.452} & .427 & .205 \\ 
        0.5   & 10  & .325 & .013 & \textbf{.522} & \underline{.473} & .383 & .355 & .322 & .010 & \textbf{.469} & \underline{.431} & .362 & .316 \\ 
        0.5   & \(\infty\)  & .399 & .016 & \underline{.557} & \textbf{.568} & .388 & .394 & .043 & .013 & \underline{.533} & \textbf{.548} & .379 & .360 \\ 
        1   & 3  & .301 & .017 & \underline{.510} & .324 & \textbf{.553} & .295 & .422 & .017 & \underline{.510} & .324 & \textbf{.553} & .295 \\ 
        1   & 10  & .511 & .019 & \underline{.569} & .534 & \textbf{.613} & .568 & .436 & .015 & \underline{.523} & .475 & \textbf{.586} & .506 \\ 
        1   & \(\infty\)  & .558 & .017 & .606 & .596 & \underline{.630} & \textbf{.639} & .422 & .012 & .553 & \underline{.577} & .571 & \textbf{.586} \\ 
        \midrule
        \multicolumn{14}{c}{Sample Size: \(n=100\)} \\
        0.33  & 3  & .224 & .018 & \textbf{.722} & \underline{.624} & .401 & .187 & .413 & .018 & \textbf{.722} & \underline{.624} & .401 & .187 \\ 
        0.33  & 10  & .427 & .021 & \underline{.781} & \textbf{.796} & .504 & .477 & .469 & .022 & \underline{.783} & \textbf{.799} & .504 & .495 \\ 
        0.33     & \(\infty\)  & .567 & .018 & \underline{.830} & \textbf{.845} & .635 & .602 & .513 & .020 & \underline{.840} & \textbf{.878} & .644 & .614  \\ 
        0.5     & 3  & .410 & .017 & \textbf{.814} & \underline{.646} & .620 & .345 & .668 & .028 & \textbf{.863} & \underline{.721} & .690 & .460 \\ 
        0.5   & 10  & .713 & .026 & \textbf{.881} & \underline{.857} & .768 & .726 & .722 & .024 & \textbf{.874} & \underline{.852} & .753 & .721 \\ 
        0.5   & \(\infty\)  & .784 & .028 & \underline{.908} & \textbf{.918} & .810 & .809 & .723 & .027 & \underline{.906} & \textbf{.913} & .801 & .804 \\ 
        1   & 3  & .642 & .025 & \underline{.781} & .626 & \textbf{.819} & .626 & .833 & .040 & \textbf{.887} & .687 & \textbf{.887} & .722 \\ 
        1   & 10  & .892 & .038 & \underline{.921} & .900 & \textbf{.937} & .914 & .857 & .036 & \underline{.921} & .890 & \textbf{.930} & .907 \\ 
        1   & \(\infty\)  & .938 & .056 & .942 & .941 & \textbf{.953} & \underline{.944} & .858 & .046 & .934 & \underline{.938} & .931 & \textbf{.939} \\ 
        \midrule
        \multicolumn{14}{c}{Sample Size: \(n=200\)} \\
        0.33  & 3  & .389 & .038 & \textbf{.948} & \underline{.808} & .651 & .387 & .721 & .039 & \textbf{.948} & \underline{.813} & .656 & .388 \\ 
        0.33  & 10  & .731 & .027 & \textbf{.965} & \underline{.963} & .789 & .780 & .784 & .040 & \textbf{.976} & \underline{.971} & .831 & .826 \\ 
        0.33  & \(\infty\)  & .835 & .049 & \textbf{.991} & \textbf{.991} & .884 & .865 & .814 & .040 & \underline{.983} & \textbf{.987} & .854 & .853 \\ 
        0.5  & 3  & .646 & .048 & \textbf{.977} & .895 & \underline{.910} & .674 & .925 & .046 & \textbf{.977} & .892 & \underline{.905} & .667 \\ 
        0.5   & 10  & .947 & .076 & \textbf{.999} & \underline{.998} & .976 & .959 & .967 & .077 & \textbf{.999} & \underline{.998} & .976 & .964 \\ 
        0.5   & \(\infty\)  & .985 & .100 & \textbf{1} & \textbf{1} & .989 & .985 & .972 & .072 & \textbf{.999} & \textbf{.999} & .988 & .985 \\ 
        1   & 3  & .880 & .083 & .990 & .927 & \textbf{.991} & .920 & \underline{.985} & .094 & \underline{.990} & .938 & \textbf{.991} & .929 \\ 
        1   & 10  & .998 & .112 & \textbf{.999} & \underline{.998} & \textbf{.999} & \underline{.998} & .997 & .098 & \textbf{.999} & .997 & \textbf{.999} & .998 \\ 
        1   & \(\infty\)  & \textbf{.999} & .158 & \textbf{.999} & \textbf{.999} & \textbf{.999} & \textbf{.999} & .997 & .103 & \textbf{.999} & \textbf{.999} & \textbf{.999} & \textbf{.999} \\ 
        \bottomrule
    \end{tabular}
    \label{tab:sample_bf_comparison_match}
\end{table}

\newpage
\subsection{Additional Figures of Case Study} \label{sm:add_case}

\vspace{-0.1cm}
\begin{figure}[h!]
\vspace{-10pt} 
\centering
\subcaptionbox{OLS residuals from Equation (\ref{eq:case_1}) \label{fig:case1}}{\includegraphics[width=0.4\linewidth]{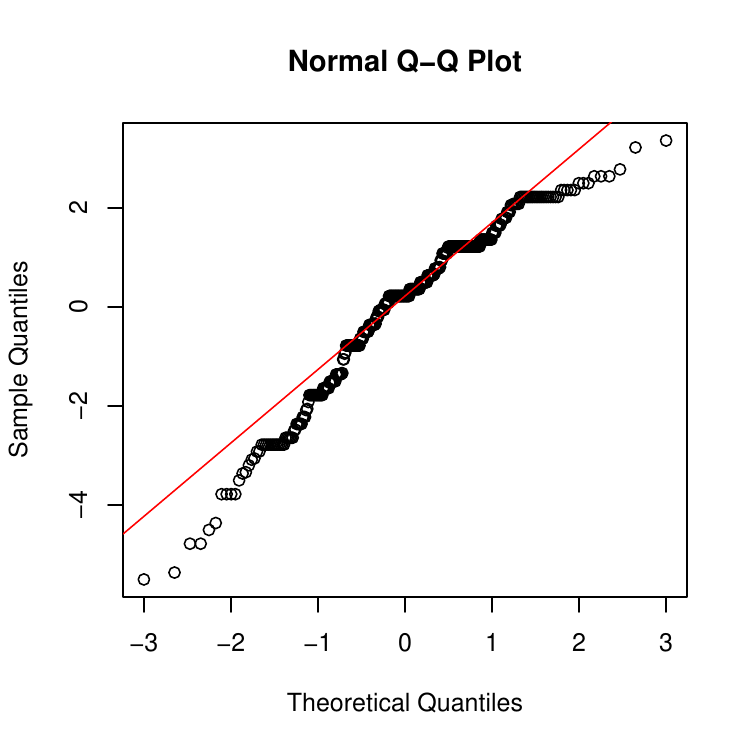}}
\subcaptionbox{OLS residuals from Equation (\ref{eq:case_2})\label{fig:case2}} {\includegraphics[width=0.4\linewidth]{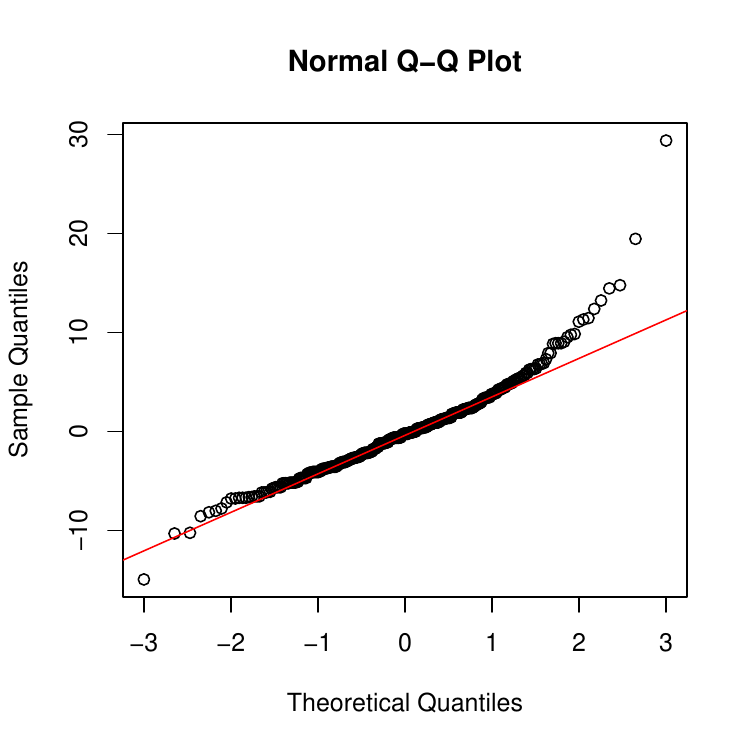}}
\caption{Normal Q-Q plots of the residuals from OLS regressions in (\ref{eq:case_1}) and (\ref{eq:case_2}).}
\label{fig:8}
\end{figure}

\begin{figure}[h!]
\vspace{-10pt} 
\centering
\subcaptionbox{Estimated posterior density of \(\gamma\)  \label{fig:gamma_1}}{\includegraphics[width=0.495\linewidth]{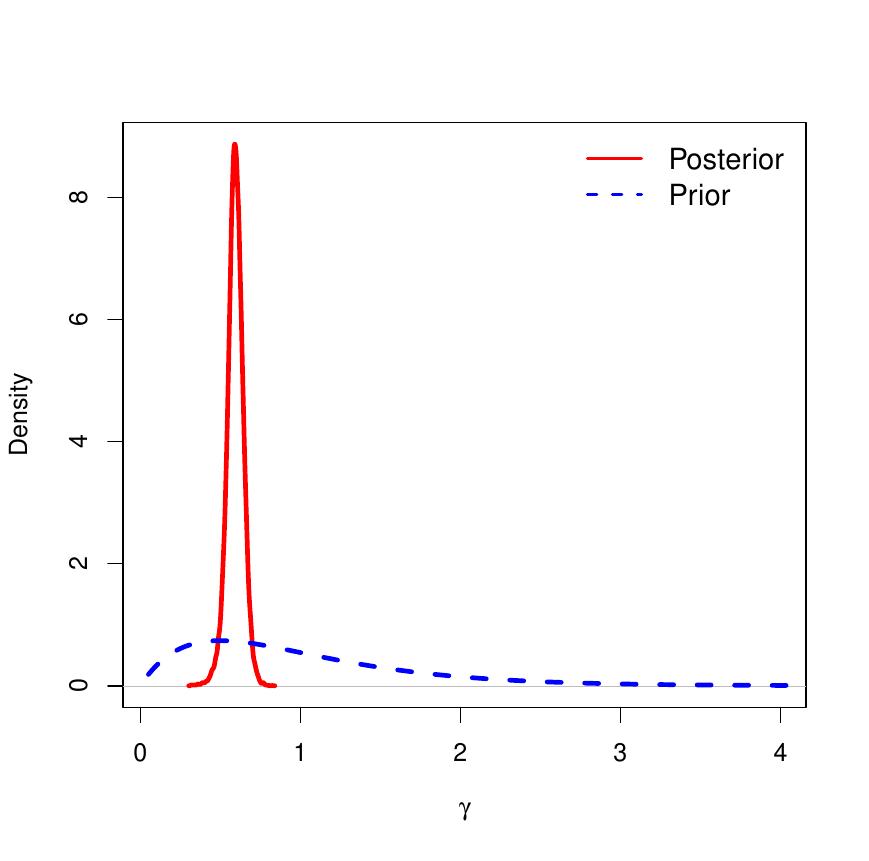}}
\subcaptionbox{Estimated posterior density of \(\nu\)\label{fig:v_1}}{\includegraphics[width=0.495\linewidth]{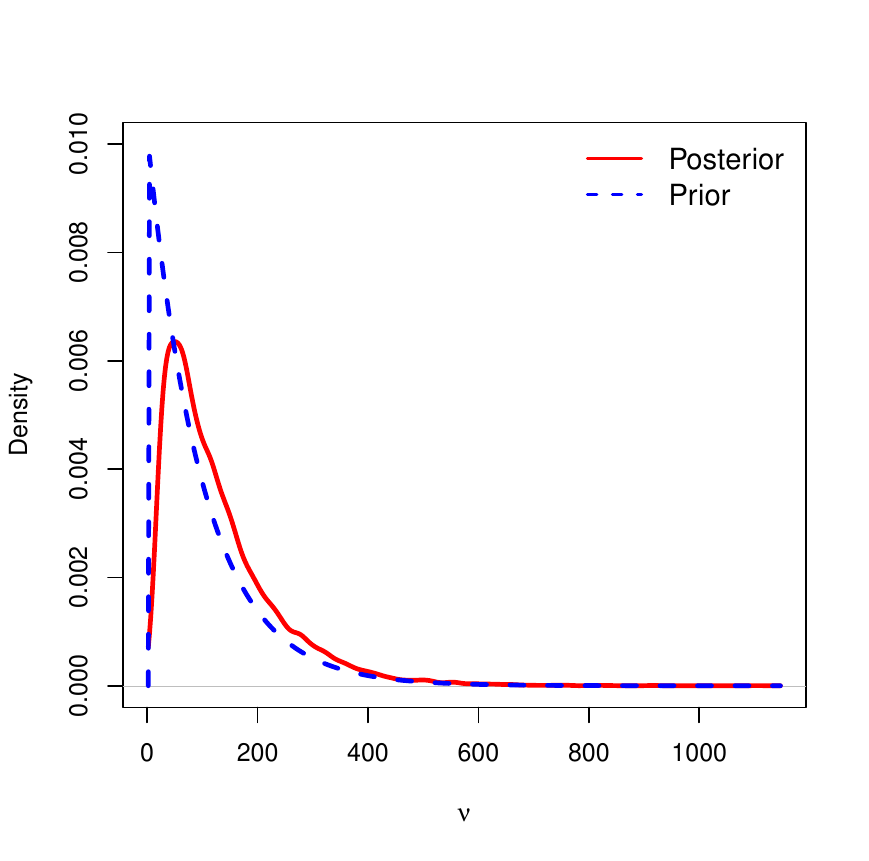}}
\caption{Estimated posterior densities of \(\gamma\) and \(\nu\) for \(\varepsilon^{(M)}\) based on the Full Model.}
\label{fig:full1}
\end{figure}

\begin{figure}[h!]
\centering
\subcaptionbox{Estimated posterior density of \(\gamma\)  \label{fig:gamma_2}}{\includegraphics[width=0.495\linewidth]{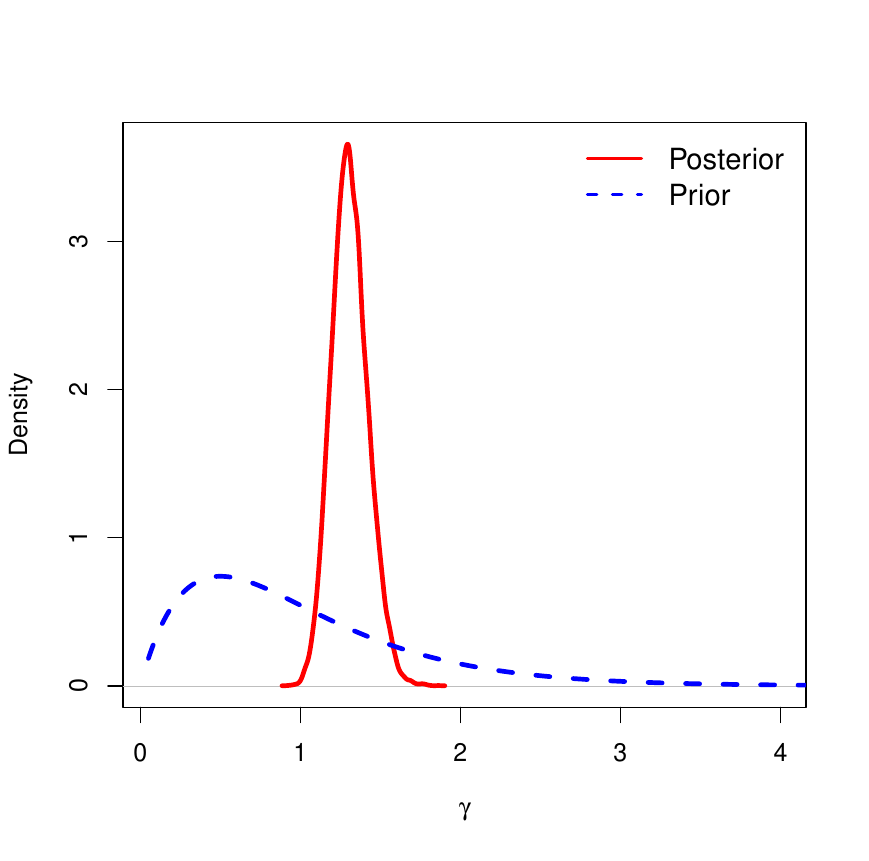}}
\subcaptionbox{Estimated posterior density of \(\nu\)\label{fig:v_2}}{\includegraphics[width=0.495\linewidth]{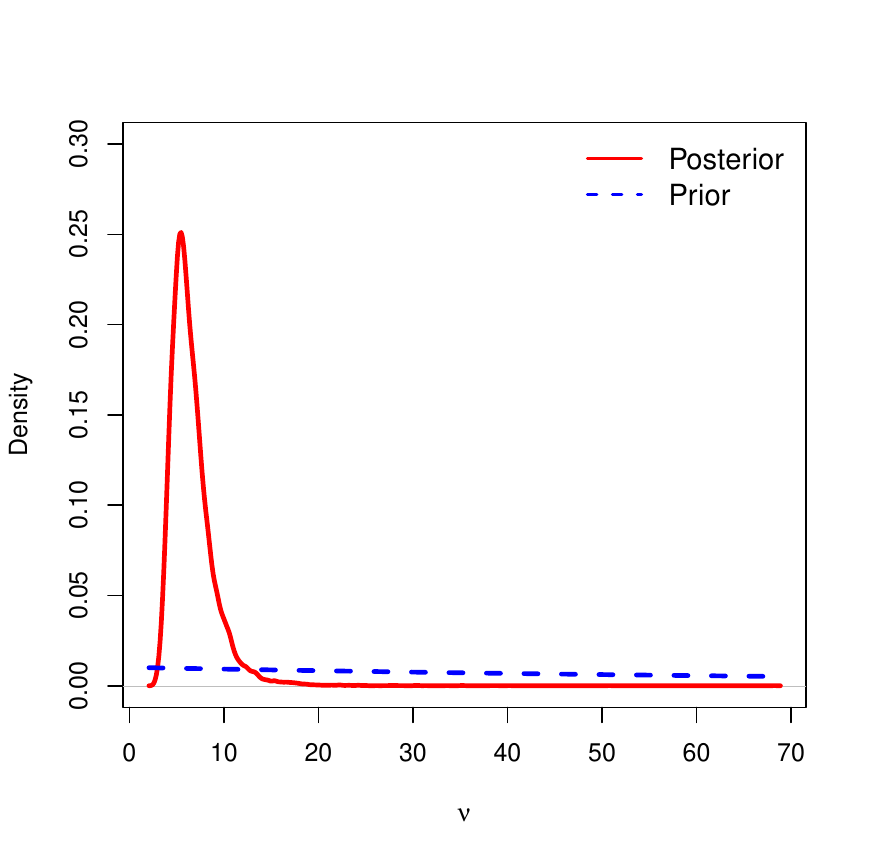}}
\caption{Estimated posterior densities of \(\gamma\) and \(\nu\) for \(\varepsilon^{(M)}\) based on the Full Model.}
\label{fig:full2}
\end{figure}

\end{document}